\documentclass[hyper]{JHEP3} % 10pt is ignored!

\usepackage{epsfig}
\usepackage{amsmath}
\usepackage{cite}

\newcommand{\beq}{\begin{equation}}
\newcommand{\eeq}{\end{equation}}

\title{Precise Calculation of the Relic Density of Kaluza-Klein Dark Matter
in Universal Extra Dimensions}

\author{Kyoungchul Kong, Konstantin T.~Matchev\\
        Physics Department, University of Florida,
        Gainesville, FL 32611, USA\\
        E-mail: \email{kong@phys.ufl.edu, matchev@phys.ufl.edu}}

\received{\today}               %%
\accepted{\today}               %% These are for published papers.

\preprint{UFIFT-HEP-05-22 \\
          September 12, 2005
          } % OR: \preprint{Aaaa/Mm/Yy\\Aaa-aa/Nnnnnn}
\abstract{We revisit the calculation of the relic density of the
lightest Kaluza-Klein particle (LKP) in the model of Universal 
Extra Dimensions. The Kaluza-Klein (KK) particle spectrum at level one
is rather degenerate, and various coannihilation processes may be relevant.
We extend the calculation of hep-ph/0206071 to include coannihilation 
processes with {\em all} level one KK particles. In our computation we 
consider a most general KK particle spectrum, without any simplifying 
assumptions. In particular, we do not assume a completely degenerate 
KK spectrum and instead retain the dependence on each individual KK mass. 
As an application of our results, we calculate the Kaluza-Klein relic 
density in the Minimal UED model, turning on coannihilations with all
level one KK particles. We then go beyond the minimal model and
discuss the size of the coannihilation effects separately for 
each class of level 1 KK particles. 
Our results provide the basis for consistent relic density computations 
in arbitrarily general models with Universal Extra Dimenions.}

\keywords{Dark Matter, Cosmology of Theories beyond the SM, Field Theories in Higher Dimensions, Beyond Standard Model, Compactification and String Models}

\begin{document} 

%%%%%%%%Section #1  %%%%%%%%%%%%%%%%%%%%%%%%%%%%%%%%%%%
%%%%%%%%%%%%%%%%%%%%%%%%%%%%%%%%%%%%%%%%%%%%%%%%%%%%%%%
\section{Introduction}
\label{sec:intro}

The Standard Model (SM) of particle physics has been astonishingly successful in
explaining much of the presently available experimental data. However, 
it still leaves open 
a number of outstanding fundamental questions whose answers are expected to 
emerge in a more general theoretical framework which extends the SM at 
higher energy scales. The main motivation for pushing the energy frontier 
beyond the Terascale comes from two main issues. 
The first is the question, what breaks the 
electroweak gauge symmetry, and if it is a Higgs field, why is the Higgs 
particle so light. The second is the dark matter problem, which has no 
explanation within the SM. By now we have accumulated a wealth of astrophysical 
data which all point to the existence of a dark, non-baryonic component of the 
matter in the Universe. The most recent WMAP data confirm the standard 
cosmological model and pin down the amount of cold dark matter\footnote{From 
now on and throughout the paper we shall use $\Omega\equiv \Omega_{CDM}$
to denote the dark matter relic density.}
as $0.094<\Omega_{CDM}h^2<0.129$, which is consistent with 
earlier indications, but much more precise.
However, the microscopic nature of dark matter is presently unknown, 
as all known particles are excluded as dark matter candidates.
This makes the dark matter problem the most pressing
phenomenological motivation for particles and interactions
beyond the Standard Model.

Among the most attractive explanations of the dark matter 
problem is the ``WIMP'' hypothesis - that dark matter
consists of stable, neutral, weakly interacting massive particles.
Indeed, many theoretically motivated extensions of the Standard Model 
which attempt to resolve the gauge hierarchy problem in the
electroweak sector, already contain particles which can be identified 
as WIMP dark matter candidates. A straightforward calculation of the 
WIMP thermal relic abundance, which we shall review in some detail 
in Section~\ref{sec:rd}, reveals that WIMP particles with masses
in the TeV range may provide most of the dark matter.

Supersymmetry (SUSY) and Extra Dimensions, which appear rather naturally 
in string theory, are the leading candidate theories for physics beyond the SM. 
By now they are both known to possess WIMP dark matter candidates.
The collider and astroparticle signals of SUSY dark matter
have been extensively studied~\cite{Chung:2003fi,Jungman:1995df}.
Recently, WIMPs from extra dimensional theories, have started 
to attract similar interest. A particularly appealing scenario 
is offered by the so called Universal Extra Dimensions (UED),
originally proposed in~\cite{Appelquist:2000nn}, where all SM
particles are allowed to freely propagate into the bulk of one or more 
extra dimensions. The case of UED bears interesting analogies
to supersymmetry, and sometimes has been referred to as  
``bosonic supersymmetry'' \cite{Cheng:2002ab}. Many of the
virtues of supersymmetry remain valid in the UED framework. 
For example, the existence of a dark matter candidate,
the lightest Kaluza-Klein particle (LKP),
is guaranteed by a discrete symmetry, called KK-parity.
KK-parity also eliminates dangerous tree-level contributions to
electroweak precision observables, rendering the model viable for a 
large range of parameters below the TeV scale. 

There are also some important differences between SUSY and UED. 
For example, the mass spectrum of Kaluza-Klein particles 
is rather degenerate, even after radiative corrections~\cite{Cheng:2002iz}. 
This means that the computation of the LKP relic density
gets rather complicated, since there are many possible coannihilation 
processes with other particles in the KK spectrum~\cite{Servant:2002aq}.
This situation should be contrasted with the case of supersymmetric models
where typically the spectrum is widely spread\footnote{Notice that 
the recently proposed little Higgs models with $T$-parity 
\cite{Cheng:2003ju,Cheng:2004yc,Hubisz:2004ft} are reminiscent of UED,
and their spectrum does not have to be degenerate, so they
are four dimensional examples which bear analogies to both UED and SUSY.}, 
and barring fine-tuned coincidences, only one or at most a few 
additional particles participate in coannihilation processes.

A second important distinction between SUSY and UED is the presence 
of a KK tower in the case of extra dimensions. The KK masses roughly scale as
$n/R$, where $R$ is the size of the extra dimension, and $n$ is the KK level.
This has an important implication for the KK relic density calculation,
since the $n=2$ particles naturally have masses twice as large as the
LKP mass, and resonant annihilation processes may become 
relevant~\cite{Kakizaki:2005en,Kakizaki:2005uy}.

A third important difference is encoded in the spins of the new particles.
The superpartners have spins which differ by $1/2$ unit from their SM counterparts, 
while the spins of the KK particles are the same as in the SM.
This has important consequences for the dark matter abundance as well.
For example, the dark matter candidate in SUSY is usually the
lightest neutralino, which is in general some mixture of a Bino, 
a Wino and Higgsinos (the superpartners of the hypercharge 
gauge boson, the neutral $SU(2)_W$ gauge boson, and the
neutral Higgs bosons, respectively). The neutralino 
is a Majorana particle, and its annihilation
to SM fermion final states is helicity suppressed.
In contrast, the dark matter candidate in UED is the KK 
mode of a Higgs or gauge boson, and its annihilation 
cross-sections have no anomalous suppressions. 
For example, in the Minimal UED model (MUED), 
which we review in more detail below in Section~\ref{sec:ued},
the LKP is $B_1$, the KK partner of the hypercharge 
gauge boson, and is a spin one particle.
The preferred mass range for the LKP is therefore 
somewhat larger than in supersymmetry.

The first and only comprehensive calculation of the UED relic density 
to date was performed in~\cite{Servant:2002aq}. 
The authors considered two cases of LKP: 
the KK hypercharge gauge boson $B_1$ and the KK neutrino $\nu_1$. 
The case of $B_1$ LKP is naturally obtained in MUED, where the
radiative corrections to $B_1$ are the smallest in size,
since they are only due to hypercharge interactions.
The authors of \cite{Servant:2002aq} also realized the
importance of coannihilation processes and included
in their analysis coannihilations with the
$SU(2)_W$-singlet KK leptons, which in MUED 
are the lightest among the remaining $n=1$ KK particles.
It was therefore expected that their coannihilations will be most important.
Subsequently, Refs.~\cite{Kakizaki:2005en,Kakizaki:2005uy} analyzed
the resonant enhancement of the $n=1$ (co)annihilation cross-sections
due to $n=2$ KK particles.

Our goal in this paper will be to complete the LKP relic density calculation 
of Ref.~\cite{Servant:2002aq}. We will attempt to improve in three different aspects:
\begin{itemize}
\item We will include coannihilation effects with {\em all} $n=1$ KK particles.
The motivation for such a tour de force is twofold. First, 
recall that the importance of coannihilations is mostly
determined by the degeneracy of the corresponding particle 
with the dark matter candidate. In the Minimal UED model, 
the KK mass splittings are due almost entirely to radiative corrections.
In MUED, therefore, one might expect that, since the
corrections to KK particles other than the KK leptons 
are relatively large, their coannihilations can be safely neglected. 
However, the Minimal UED model makes an ansatz~\cite{Cheng:2002iz}
about the cut-off scale values of the so called boundary terms, 
which are not fixed by known SM physics, and are in principle arbitrary.  
In this sense, the UED scenario should be considered as a low energy 
effective theory with a multitude of parameters, just like the MSSM, 
and the MUED model should be treated as nothing more than a simple 
toy model with a limited number of parameters, just like the ``minimal
supergravity'' version of supersymmetry, for example. 
If one makes a different assumption 
about the inputs at the cut-off scale, both the KK spectrum and
its phenomenology can be modified significantly. In particular,
one could then easily find regions of this more general 
parameter space where other coannihilation processes become active.
On the other hand, even if we choose to restrict ourselves to MUED,
there is still a good reason to consider the coannihilation
processes which were omitted in the analysis of \cite{Servant:2002aq}.
While it is true that those coannihilations are more Boltzmann 
suppressed, their cross-sections will be larger, since they are
mediated by weak and/or strong interactions. Without an explicit 
calculation, it is impossible to estimate the size of the net 
effect, and whether it is indeed negligible compared to 
the purely hypercharge-mediated processes which have already 
been considered.
\item We will keep the exact value of each KK mass in our formulas 
for all annihilation cross-sections. 
This will render our analysis self-consistent. All
calculations of the LKP relic density available so 
far~\cite{Servant:2002aq,Kakizaki:2005en,Kakizaki:2005uy},
have computed the annihilation cross-sections in the limit when
all level 1 KK masses are the same. This approximation is
somewhat contradictory in the sense that {\em all} 
KK masses at level one are taken to be degenerate with LKP, 
yet only {\em a limited number} of coannihilation 
processes were considered. In reality, a completely 
degenerate spectrum would require the inclusion
of all possible coannihilations. Conversely, if some
coannihilation processes are being neglected, this
is presumably because the masses of the corresponding
KK particles are {\em not} degenerate with the LKP,
and are Boltzmann suppressed. However, the masses 
of these particles may still enter the formulas for 
the relevant coannihilation cross-sections, and
using approximate values for those masses would
lead to a certain error in the final answer.
Since we are keeping the exact mass dependence 
in the formulas, within our approach 
heavy particles naturally decouple, 
coannihilations are properly weighted, and all relevant
coannihilation cross-sections behave properly. 
Notice that the assumption of exact mass degeneracy 
overestimates the corresponding cross-sections and 
therefore underestimates the relic density. This 
expectation will be confirmed in our numerical analysis 
in Section~\ref{sec:mued}.
\item We will try to improve the numerical accuracy 
of the analysis by taking into account some minor
corrections which were neglected or approximated 
in~\cite{Servant:2002aq}. For example, we will use 
a temperature-dependent $g_\ast$ (the total number 
of effectively massless degrees of freedom, given 
by eq.~(\ref{gstar}) below) and include subleading 
corrections (\ref{bcorr}) in the velocity expansion 
of the annihilation cross-sections.
\end{itemize}

The availability of the calculation of the remaining coannihilation processes 
is important also for the following reason. Coannihilations with $SU(2)_W$-singlet 
KK leptons were found to reduce the effective annihilation cross-section, and 
therefore increase the LKP relic density. This has the effect of lowering the range 
of cosmologically preferred values of the LKP mass, or equivalently, the 
scale of the extra dimension. However, one could expect that coannihilations with 
the other $n=1$ KK particles would have the opposite effect, since they have
stronger interactions compared to the $SU(2)_W$-singlet KK leptons and the $B_1$ LKP.
As a result, the preferred LKP mass range could be pushed back up. 
For both collider and astroparticle searches for dark matter,
a crucial question is whether there is an upper limit on the WIMP mass
which could guarantee discovery, and if so, what is its precise numerical value.
To this end, one needs to consider the effect of all coannihilation processes
which have the potential to enhance the LKP annihilations. We will see 
that the lowering of the preferred LKP mass range in the case 
of coannihilations with $SU(2)_W$-singlet KK leptons is more of an 
exception rather than the rule, and the inclusion of {\em all} remaining 
processes is needed in order to derive an absolute upper bound on the LKP mass.

The paper is organized as follows. In Section~\ref{sec:ued} we review the 
Minimal model of Universal Extra Dimensions (MUED). In Section~\ref{sec:calc1}
(\ref{sec:calc2}) we review the standard calculation of the WIMP relic density 
in the absence (presence) of coannihilations. Then in Section~\ref{sec:mued}
we present our results for the LKP relic abundance in MUED.
In Section~\ref{sec:coann} we extend our analysis beyond MUED and
investigate the size of the coannihilation effects for each class of KK 
particles at level 1. We summarize and conclude in Section~\ref{sec:conclusions}.
The Appendix contains a list of our formulas for all relevant annihilation 
cross-sections, in the limit of equal masses. Those are given only for reference and 
contact with previous work~\cite{Servant:2002aq,BurKri}, since in our numerical 
code we use the corresponding different values for the individual masses.

\section{Universal Extra Dimensions}
\label{sec:ued}

The simplest version of UED has all the SM particles
propagating in a single extra dimension of size $R$,
which is compactified on an $S_1/Z_2$ orbifold.
More complicated models have also been considered,
motivated by ideas about electroweak symmetry breaking
and vacuum stability~\cite{Arkani-Hamed:2000hv,Bucci:2003fk,Bucci:2004xw},
neutrino masses~\cite{Appelquist:2002ft,Mohapatra:2002ug}, 
proton stability~\cite{Appelquist:2001mj}, the number of 
generations~\cite{Dobrescu:2001ae} or fermion chirality
\cite{Dobrescu:2004zi,Burdman:2005sr}.
A peculiar feature of UED is the conservation of
Kaluza-Klein number $n$ at tree level, which is a simple consequence
of momentum conservation along the extra dimension.
However, bulk and brane radiative effects 
\cite{Georgi:2000ks,vonGersdorff:2002as,Cheng:2002iz}
break KK number down to a discrete conserved quantity,
the so called KK parity, $(-1)^n$.
KK parity ensures that the lightest KK partners --
those at level one -- are always pair-produced
in collider experiments, similar to the case of supersymmetry models
with conserved $R$-parity. On the one hand, this leads to 
rather weak bounds on KK partner masses from direct searches 
at colliders~\cite{Appelquist:2000nn,Cheng:2002ab}.
On the other hand, the collider signatures of UED are very
similar to those of supersymmetry, and discriminating between 
the two scenarios at lepton~\cite{Cheng:2002rn,Battaglia:2005zf,Bhattacharyya:2005vm,Riemann:2005es,Bhattacherjee:2005qe} or 
hadron~\cite{Rizzo:2001sd,Macesanu:2002db,Barr:2004ze,Smillie:2005ar,Battaglia:2005ma,Datta:2005zs,Datta:2005vx,Barr:2005dz}
colliders is currently an active field of study.
KK parity conservation also implies
that the contributions to various precisely measured low-energy
observables \cite{Agashe:2001ra,Agashe:2001xt,Appelquist:2001jz,%
Petriello:2002uu,Appelquist:2002wb,Chakraverty:2002qk,Buras:2002ej,%
Oliver:2002up,Buras:2003mk,Iltan:2003tn,Khalil:2004qk}
only arise at loop level and are small.
As a result, the limits on the scale of the extra dimension, from 
precision electro-weak data, are rather weak, constraining  $R^{-1}$ to 
be larger than approximately 250~GeV \cite{Appelquist:2002wb}. An attractive feature of
UED is the presence of a stable massive particle which can be
a cold dark matter candidate \cite{Dienes:1998vg,Cheng:2002iz,Servant:2002aq,Cheng:2002ej}.
The lightest KK partner (LKP) at level one 
is also the lightest particle with negative KK parity and 
is stable on cosmological scales.
The identity of the LKP is a delicate issue, however, as it
depends on the interplay between the one-loop radiative corrections
to the KK mass spectrum and the brane terms generated by unknown physics 
at high scales~\cite{Cheng:2002iz}.
In the Minimal UED model defined in \cite{Cheng:2002ab}, 
the LKP turns out to be the KK partner $B_1$ of the
hypercharge gauge boson \cite{Cheng:2002iz} and
its relic density is typical  of a WIMP candidate:
in order to explain all of the dark matter, the $B_1$ mass
should be in the range $600-800$ GeV, depending on the rest of the KK 
spectrum~\cite{Servant:2002aq,Majumdar:2003dj,Kakizaki:2005en,Kakizaki:2005uy}. 
The experimental signals of Kaluza-Klein dark matter have also 
been discussed and it has been realized that it offers excellent prospects 
for direct~\cite{Cheng:2002ej,Servant:2002hb,Majumdar:2002mw}
or indirect detection~\cite{Cheng:2002ej,Hooper:2002gs,Bertone:2002ms,Hooper:2004xn,%
Bergstrom:2004cy,Baltz:2004ie,Bergstrom:2004nr,Bringmann:2005pp,Barrau:2005au,Birkedal:2005ep}.

In the Minimal UED model, the bulk interactions of the KK modes readily follow 
from the SM Lagrangian and contain no unknown parameters other
than the mass, $m_h$, of the SM Higgs boson. In contrast, the boundary 
interactions, which are localized on the orbifold fixed points, are in 
principle arbitrary, and their coefficients represent new free 
parameters in the theory. Since the boundary terms are renormalized
by bulk interactions, they are scale dependent \cite{Georgi:2000ks}
and cannot be completely ignored since they will be generated by 
renormalization effects. 
Therefore, we need an ansatz for their values at a particular scale.
As with any higher dimensional Kaluza-Klein theory, 
the UED model should be treated only as an effective
theory which is valid up to some high scale $\Lambda$, at which
it matches to some more fundamental theory. 
The Minimal UED model therefore has only two input parameters: 
the size of the extra dimension, $R$, and the cutoff scale, $\Lambda$. 
The number of KK levels present in the effective theory is 
simply $\Lambda R$ and may vary between a few and $\sim 40$, 
where the upper limit comes from the breakdown of perturbativity 
already below the scale $\Lambda$. Unless specified otherwise,
for our numerical results below, we shall always choose the 
value of $\Lambda$ so that $\Lambda R=20$, although analyses 
of unitarity constraints in the higher dimensional Standard Model 
typically yield a lower bound on $\Lambda$ \cite{Chivukula:2003kq,Muck:2004br}. 
Changing the value 
of $\Lambda$ will have very little impact on our results since the 
$\Lambda$ dependence of the KK mass spectrum is only logarithmic.
For the SM Higgs mass $m_h$ we shall adopt the value $m_h=120$ GeV.

\section{The Basic Calculation of the Relic Density} 
\label{sec:rd}

\subsection{The standard case}
\label{sec:calc1}

We first summarize the standard calculation for the relic abundance 
of a particle species $\chi$ which was in thermal equilibrium 
in the early universe and decoupled when it became nonrelativistic 
\cite{Srednicki:1988ce,Kolb:1990vq,Servant:2002aq}.
The relic abundance is found by solving the Boltzmann equation 
for the evolution of the $\chi$ number density $n$
\begin{equation}
\frac{d n}{ d t} = -3 Hn - \langle \sigma v \rangle ( n^2 - n^2_{eq})\ ,
\end{equation}
where $H$ is the Hubble parameter, $v$ is the relative velocity between two $\chi$'s, 
$\langle \sigma v \rangle$ is the thermally 
averaged total annihilation cross-section times relative velocity, 
and $n_{eq}$ is the equilibrium number density. 
At high temperature ($T \gg m$), $n_{eq} \sim T^3$ 
(there are roughly as many $\chi$ particles as photons).
At low temperature ($T \ll m$), in the nonrelativistic 
approximation, $n_{eq}$ can be written as 
\begin{equation}
n_{eq} = g \left ( \frac{m T}{ 2 \pi} \right )^{\frac{3}{2}} e^{-m/T}\ ,
\label{neq}
\end{equation}
where $m$ is the mass of the relic $\chi$, $T$ is the temperature and 
$g$ is the number of internal degrees of freedom of $\chi$ such as spin, 
color and so on. We see from eq.~(\ref{neq}) that the density 
$n_{eq}$ is Boltzmann-suppressed. 
At high temperature, $\chi$ particles are abundant and rapidly 
convert to lighter particles and vice versa. But shortly 
after the temperature $T$ drops below $m$, 
the number density decreases exponentially and 
the annihilation rate $\Gamma = \langle \sigma v \rangle n$ 
drops below the expansion rate $H$. 
At this point, $\chi$'s stop annihilating and escape 
out of the equilibrium and become thermal relics.
$\langle \sigma v \rangle$ is often approximated by the 
nonrelativistic expansion\footnote{Note, however, that the 
method fails near $s$-channel resonances and 
thresholds for new final states~\cite{Griest:1990kh}.
In the interesting parameter region of UED, 
we are always sufficiently far from thresholds, 
while for the treatment of resonances, see \cite{Kakizaki:2005en,Kakizaki:2005uy}.}
\begin{equation}
\langle \sigma v \rangle = a + b \langle v^2 \rangle + {\cal O}(\langle v^4 \rangle) 
\approx a+ 6 b /x + {\cal O}\left( \frac{1}{x^2}\right) \ ,
\end{equation}
where 
\beq
x = \frac{m}{T}\ .
\eeq
By solving the Boltzmann equation analytically with appropriate 
approximations \cite{Srednicki:1988ce,Kolb:1990vq,Servant:2002aq}, 
the abundance of $\chi$ is given by
\begin{equation}
\Omega_\chi h^2 \approx \frac{1.04 \times 10^9}{M_{Pl}}\frac{x_F}{\sqrt{g_\ast(x_F})} \frac{1}{a+3 b/x_F }\ ,
\end{equation}
where the Planck mass $M_{Pl} = 1.22\times 10^{19}$ GeV and
$g_\ast$ is the total number of effectively massless degrees of freedom,
\begin{equation}
g_\ast(T) = \sum_{i=bosons} g_i + \frac{7}{8} \sum_{i=fermions}g_i\ .
\label{gstar}
\end{equation}
The freeze-out temperature, $x_F$, is found iteratively from 
\begin{equation}
x_F = \ln \left ( c(c+2) \sqrt{\frac{45}{8}} \frac{g}{2\pi^3} \frac{m M_{Pl} (a+6b/x_F)}{\sqrt{g_\ast(x_F) x_F}}  \right )\ ,
\end{equation}
where the constant $c$ is determined  empirically by comparing to numerical
solutions of the Boltzmann equation and here we take $c=\frac{1}{2}$ as usual. 
The coefficient $\frac{7}{8}$ in the right hand side of (\ref{gstar}) 
accounts for the difference in Fermi 
and Bose statistics. Notice that $g_\ast$ is a function of the temperature $T$,
as the thermal bath quickly gets depleted of the heavy species with masses 
larger than $T$.

\subsection{The case with coannihilations}
\label{sec:calc2}

When the relic particle $\chi$ is nearly degenerate with other particles
in the spectrum, its relic abundance is determined not only by its own 
self-annihilation cross-section, but also by annihilation processes involving
the heavier particles. The previous calculation can be generalized 
to this ``coannihilation'' case in a straightforward 
way~\cite{Griest:1990kh,Kolb:1990vq,Servant:2002aq}.
Assume that the particles $\chi_i$ are labeled according to their masses, 
so that $m_i < m_j$ when $i < j$. 
The number densities $n_i$ of the various species $\chi_i$ obey
a set of Boltzmann equations. It can be shown that
under reasonable assumptions \cite{Griest:1990kh},
the ultimate relic density $n$ of the lightest species $\chi_1$ 
(after all heavier particles $\chi_i$ have decayed into it)
obeys the following simple Boltzmann equation
\begin{equation}
\frac{d n}{ d t} = -3 Hn - \langle \sigma_{eff} v \rangle ( n^2 - n^2_{eq})\ ,
\end{equation}
where
\begin{eqnarray}
\sigma_{eff}(x) &=& \sum_{ij}^N \sigma_{ij} \frac{g_i g_j}{g_{eff}^2} 
                 (1 + \Delta_i)^{3/2} (1 + \Delta_j)^{3/2} \exp(-x ( \Delta_i + \Delta_j ))\ ,
\label{sigmaeff}\\
g_{eff}(x)   &=& \sum_{i=1}^N g_i (1+\Delta_i)^{3/2} \exp(-x \Delta_i)\ , 
\label{geff}\\
\Delta_i     &=& \frac{m_i - m_1}{m_1}\ .
\end{eqnarray}
Here $\sigma_{ij}\equiv \sigma(\chi_i\chi_j\to SM)$,
$g_i$ is the number of internal degrees of freedom of particle $\chi_i$ and 
$n = \sum_{i=1}^N n_i$ is the density of $\chi_1$ we want to calculate.
This Boltzmann equation can be solved in a similar way 
\cite{Griest:1990kh, Servant:2002aq}, resulting in 
\begin{equation}
\Omega_\chi h^2 \approx \frac{1.04 \times 10^9}{M_{Pl}}
\frac{x_F}{\sqrt{g_\ast(x_F)}} \frac{1}{I_a+3 I_b/x_F }\ ,
\label{oh2coann}
\end{equation}
with 
\begin{eqnarray}
I_a &=& x_F \int_{x_F}^\infty a_{eff}(x) x^{-2} d x \ ,
\label{I_a}
\\
I_b &=& 2 x_F^2 \int_{x_F}^\infty b_{eff}(x) x^{-3} d x\ .
\label{I_b}
\end{eqnarray}
The corresponding formula for $x_F$ becomes
\begin{equation}
x_F = \ln \left ( c(c+2) \sqrt{\frac{45}{8}} \frac{g_{eff}(x_F)}{2\pi^3} \frac{m M_{Pl} (a_{eff}(x_F)+6b_{eff}(x_F)/x_F)}
{\sqrt{g_\ast(x_F) x_F}}  \right )\ .
\label{xfcoann}
\end{equation}
Here $a_{eff}$ and $b_{eff}$ are the first two terms 
in the velocity expansion of $\sigma_{eff}$
\begin{equation}
\sigma_{eff}(x)\,v = a_{eff}(x) + b_{eff}(x)\, v^2 + {\cal O}(v^4)\ .
\label{sigmaeffv}
\end{equation}
Comparing eqs.~(\ref{sigmaeff}) and (\ref{sigmaeffv}), one gets
\begin{eqnarray}
a_{eff}(x) &=& \sum_{ij}^N a_{ij} \frac{g_i g_j}{g_{eff}^2} 
                 (1 + \Delta_i)^{3/2} (1 + \Delta_j)^{3/2} \exp(-x ( \Delta_i + \Delta_j ))\ , 
\label{aeff}\\
b_{eff}(x) &=& \sum_{ij}^N b_{ij} \frac{g_i g_j}{g_{eff}^2} 
                 (1 + \Delta_i)^{3/2} (1 + \Delta_j)^{3/2} \exp(-x ( \Delta_i + \Delta_j ))\ ,
\label{beff}
\end{eqnarray}
where $a_{ij}$ and $b_{ij}$ are obtained from $\sigma_{ij}v = a_{ij} + b_{ij} v^2 + {\cal O}(v^4)$.

Considering relativistic corrections \cite{Srednicki:1988ce,Wells:1994qy,Lahanas:1999uy}
to the above treatment results in an additional subleading 
term which can be accounted for by the simple replacement 
\beq
b\to b-\frac{1}{4}a
\label{bcorr}
\eeq
in the above formulas.

\section{Relic Density in Minimal UED}
\label{sec:mued}

For the purposes of our study we have implemented the relevant 
features of the Minimal UED model in the {\tt CompHEP} 
event generator \cite{Pukhov:1999gg}.
We incorporated all  $n=1$ and $n=2$ KK modes as new particles, 
with the proper interactions and one-loop corrected 
masses~\cite{Cheng:2002iz}.
Similar to the SM case, the neutral gauge bosons at level~1, 
$Z_1$ and $\gamma_1$, are mixtures of the KK modes of the
hypercharge gauge boson and the neutral $SU(2)_W$ gauge boson.
However, as shown in~\cite{Cheng:2002iz}, the radiatively corrected
Weinberg angle at level~1 and higher is very small.
For example, $\gamma_1$, which is the LKP in the minimal
UED model, is mostly the KK mode of the hypercharge gauge boson.
Therefore, for simplicity, in the code we neglected neutral gauge boson mixing
for $n = 1$. We then use our UED implementation in CompHEP to derive 
analytic expressions for the (co)annihilation cross-sections
between any pair of $n=1$ KK particles. Our code has been 
subjected to numerous tests and cross-checks. For example, 
we reproduced all results from \cite{Servant:2002aq}.
We have also used the same code for independent studies of the
collider and astroparticle signatures of UED 
\cite{Battaglia:2005zf,Birkedal:2004xn,Birkedal:2005ep,Battaglia:2005ma} 
and thus have tested it from a different angle as well.

The mass spectrum of the $n=1$ KK partners in Minimal UED 
can be found, for example, in Fig.~1 of \cite{Cheng:2002ab}.
In MUED the next-to-lightest KK particles are the singlet 
KK leptons and their fractional mass difference from the LKP 
is\footnote{In this paper we follow the notation of \cite{Servant:2002aq}
where the two types of $n=1$ Dirac fermions are distinguished by 
an index corresponding to the chirality of their zero mode partner.
For example, $\ell_{R1}$ stands for an $SU(2)_W$-singlet Dirac fermion,
which has in principle both a left-handed and a right-handed component.}
\beq
\Delta_{\ell_{R1}} \equiv \frac{ m_{\ell_{R1}} - m_{\gamma_1} }{ m_{\gamma_1} } \sim 0.01 \ .
\eeq 
Notice that the Boltzmann suppression 
$$
e^{-\Delta_{\ell_{R1}}x_F}\sim e^{-0.01\cdot 25}=e^{-0.25}
$$
is not very effective and coannihilation processes with $\ell_{R1}$
are definitely important, hence they were considered in \cite{Servant:2002aq}.
What about the other, heavier particles in the $n=1$ KK spectrum in MUED?
Since their mass splittings from the LKP
\beq
\Delta_i\equiv \frac{m_i-m_{\gamma_1}}{m_{\gamma_1}}
\label{Delta_i}
\eeq
are larger, their annihilations suffer from a larger Boltzmann suppression.
However, the couplings of all $n=1$ KK partners other than $\ell_{R1}$
are larger compared to those of $\gamma_1$ and $\ell_{R1}$. 
For example, $SU(2)_W$-doublet
KK leptons $\ell_{L1}$ couple weakly, and the KK quarks $q_1$ 
and KK gluon $g_1$ have strong couplings. Therefore, their corresponding
annihilation cross-sections are expected to be larger than the
cross-section of the main $\gamma_1\gamma_1$ channel.

We see that for the other KK particles, there is a competition between the 
increased cross-sections and the larger Boltzmann suppression. An explicit 
calculation is therefore needed in order to evaluate the net effect of 
these two factors, and judge the importance of the coannihilation processes 
which have been neglected so far. One might expect that coannihilations with 
$SU(2)_W$-doublet KK leptons might be numerically significant, since their
mass splitting in MUED is $\sim 3\%$ and the corresponding Boltzmann 
suppression factor is only $e^{-0.03\cdot25}\sim e^{-0.75}$.

In our code we keep all KK masses different while we neglect all 
the masses of the Standard Model particles.
As an illustration, let us show the 
$a$ and $b$ terms for $\gamma_1\gamma_1$ annihilation only.
For fermion final states we find the $a$-term and $b$-term
of $\sigma(\gamma_1\gamma_1\to f\bar{f})v$ as follows
\begin{eqnarray}
a &=& \sum_{f} \frac{32 \pi \alpha_1^2 N_c m_{\gamma_1}^2}{9} \left ( \frac{Y_{f_L}^4}{(m_{\gamma_1}^2 + m_{f_{L1}}^2 )^2} 
+ \frac{Y_{f_R}^4}{(m_{\gamma_1}^2 + m_{f_{R1}}^2 )^2} \right ) 
\label{aff}\\
&\approx& \sum_{f}\frac{8 \pi \alpha_1^2}{9 m_{\gamma_1}^2} N_c \left ( Y_{f_L}^4 +  Y_{f_R}^4 \right ) 
= \frac{8 \pi \alpha_1^2}{9 m_{\gamma_1}^2} \left ( \frac{95}{18} \right ) \ ,
\label{affapprox}
\end{eqnarray}
\begin{eqnarray}
b &=& - \sum_{f} \frac{4 \pi \alpha_1^2 N_c m_{\gamma_1}^2 }{27} \Biggl(
   Y_{f_L}^4 \frac{11 m_{\gamma_1}^4 + 14 m^2 m_{f_{L1}}^2 - 13 m_{f_{L1}}^4 }{(m_{\gamma_1}^2 + m_{f_{R1}}^2 )^4} 
\nonumber 
\label{bff}
\\
&&\qquad\qquad\qquad\qquad
 +Y_{f_R}^4 \frac{11 m_{\gamma_1}^4 + 14 m^2 m_{f_{L1}}^2 - 13 m_{f_{L1}}^4 }{(m_{\gamma_1}^2 + m_{f_{R1}}^2 )^4}
\Biggr) \\
&\approx& - \sum_{f} \frac{\pi \alpha_1^2}{9 m_{\gamma_1}^2} N_c \left ( Y_{f_L}^4 +  Y_{f_R}^4 \right ) 
= - \frac{\pi \alpha_1^2}{9 m_{\gamma_1}^2} \left ( \frac{95}{18} \right )\ ,
\label{bffapprox}
\end{eqnarray}
where $g_1$ is the gauge coupling of the hypercharge $U(1)_Y$ gauge group,
$\alpha_1 = \frac{g_1^2}{4\pi}$ and $N_c=3$ for $f=q$ and $N_c=1$ for $f=\ell$.
$Y_f$ is the hypercharge of the fermion $f$.

\FIGURE[t]{
\epsfig{file=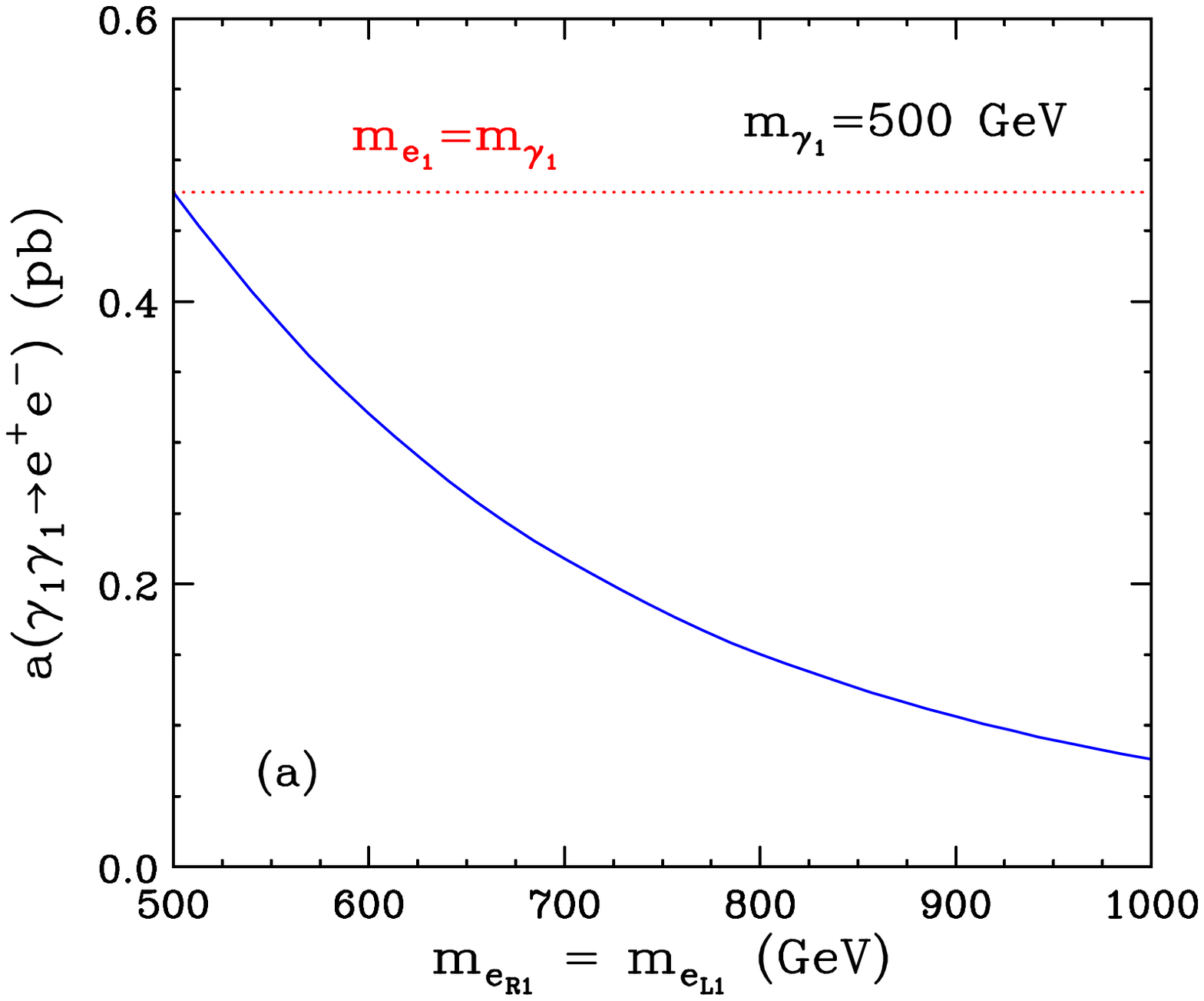,width=7.4cm}
\epsfig{file=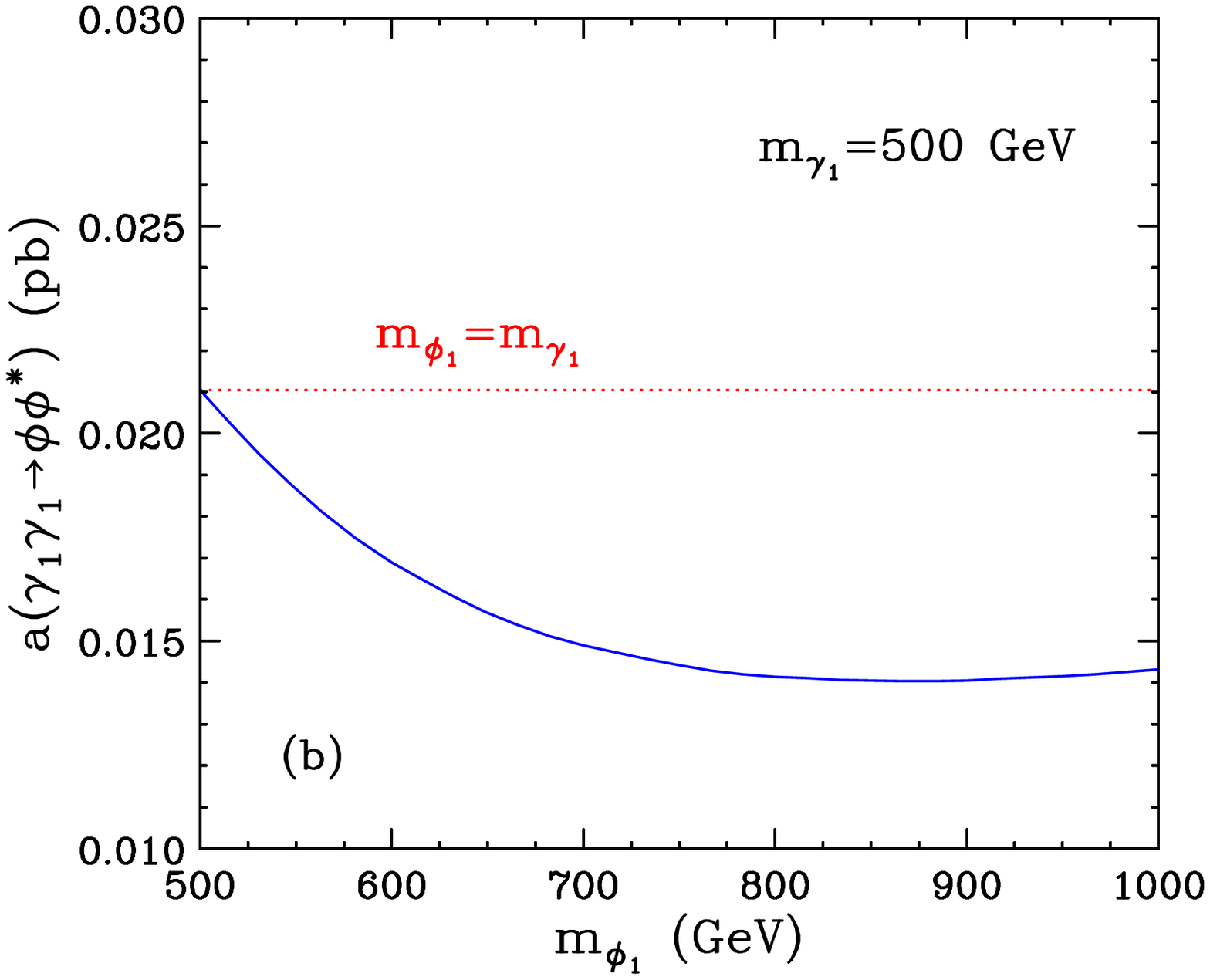,width=7.4cm}
\caption{\sl The $a$-term of the annihilation cross-section for 
(a) $\gamma_1 \gamma_1 \rightarrow e^+ e^-$ and 
(b) $\gamma_1 \gamma_1 \rightarrow \phi \phi^\ast$, as a function of
the mass of the $t$-channel particle(s). We fix the LKP mass at
$m_{\gamma_1}=500$ GeV and vary (a) the KK lepton mass $m_{e_{R1}}=m_{e_{L1}}$
or (b) the KK Higgs boson mass $m_{\phi_1}$. The blue solid lines are the
exact results (\ref{aff}) and (\ref{ahh}), while the red dotted lines 
correspond to the approximations (\ref{affapprox}) and (\ref{ahhapprox}).}
\label{fig:aterms}}

For the Higgs boson final states we get
\begin{eqnarray}
a &=& \sum_{i} \frac{2 \pi \alpha_1^2 Y_{\phi_i}^4 }{9 m_{\gamma_1}^2} \left ( 
\frac{11 m_{\gamma_1}^4 - 2m_{\gamma_1}^2 m_{\phi_i}^2 + 3m^4_{\phi_i} }{(m_{\gamma_1}^2 + m_{\phi_i}^2 )^2} 
 \right ) 
\label{ahh}\\
&\approx& \sum_{i} \frac{2 \pi \alpha_1^2 Y_{\phi_i}^4}{3m_{\gamma_1}^2} 
= \frac{4 \pi \alpha_1^2 Y_{\phi}^4}{3m_{\gamma_1}^2}\ , 
\label{ahhapprox}\\
b &=& - \sum_{i} \frac{\pi \alpha_1^2 Y_{\phi_i}^4}{108 m_{\gamma_1}^2} \left (
\frac{ 121 m_{\gamma_1}^8 + 140 m_{\gamma_1}^6 m_{\phi_i}^2 - 162 m_{\gamma_1}^4 m_{\phi_i}^4 
+ 60 m_{\gamma_1}^2 m_{\phi_i}^6 - 15 m_{\phi_i}^8  }{(m_{\gamma_1}^2 + m_{\phi_i}^2 )^4}
 \right )  \\
&\approx& - \sum_i \frac{\pi \alpha_1^2 Y^2_{\phi_i}}{12 m_{\gamma_1}^2} 
= - \frac{\pi \alpha_1^2 Y^2_{\phi}}{6 m_{\gamma_1}^2}\ .
\end{eqnarray}
In the limit where all KK masses are the same (the second line in each formula above), 
we recover the result of \cite{Servant:2002aq}. 
Notice the tremendous simplification which arises as a result 
of the mass degeneracy assumption. In Fig.~\ref{fig:aterms}
we show the $a$ terms of the annihilation cross-section for 
two processes: (a) $\gamma_1 \gamma_1 \rightarrow e^+ e^-$ and 
(b) $\gamma_1 \gamma_1 \rightarrow \phi \phi^\ast$, as a function of
the mass of the $t$-channel particle(s). We fix the LKP mass at
$m_{\gamma_1}=500$ GeV and vary (a) the KK lepton mass $m_{e_{R1}}=m_{e_{L1}}$
or (b) the KK Higgs boson mass $m_{\phi_1}$. The blue solid lines are the
exact results (\ref{aff}) and (\ref{ahh}), while the red dotted lines 
correspond to the approximations (\ref{affapprox}) and (\ref{ahhapprox})
in which the mass difference between the $t$-channel particles 
and the LKP has been neglected. We see that the 
approximations (\ref{affapprox}) and (\ref{ahhapprox}) can 
result in a relatively large error, whose size depends on 
the actual mass splitting of the KK particles. This is why in 
our code we keep all individual mass dependencies.

Another difference between our analysis and that of Ref.~\cite{Servant:2002aq}
is that here we shall use a temperature-dependent $g_\ast$ function as 
defined in (\ref{gstar}). The relevant value of $g_\ast$ which enters the 
answer for the LKP relic density (\ref{oh2coann}) is $g_\ast(T_F)$, where
$T_F=m_{\gamma_1}/x_F$ is the freeze-out temperature. 
In Fig.~\ref{fig:gstar}a we show a plot of $g_\ast(T_F)$ 
as a function of $R^{-1}$ in MUED, while in Fig.~\ref{fig:gstar}b
we show the corresponding values of $x_F$. In Fig.~\ref{fig:gstar}a
one can clearly see the jumps in $g_\ast$ when crossing the 
$b\bar{b}$, $W^+W^-$, $ZZ$ and $hh$ thresholds (from left to right).
The $t\bar{t}$ threshold is further to the right, outside the plotted range.
As we shall see below, cosmologically interesting values of 
$\Omega h^2$ are obtained for $R^{-1}$ below 1 TeV, where 
$g_\ast(T_F)=86.25$, since we are below the $W^+W^-$ threshold. The analysis 
of Ref.~\cite{Servant:2002aq} assumed a constant value of $g_\ast=92$, which
is only valid between the $W^+W^-$ and $ZZ$ thresholds. 
\FIGURE[t]{
\epsfig{file=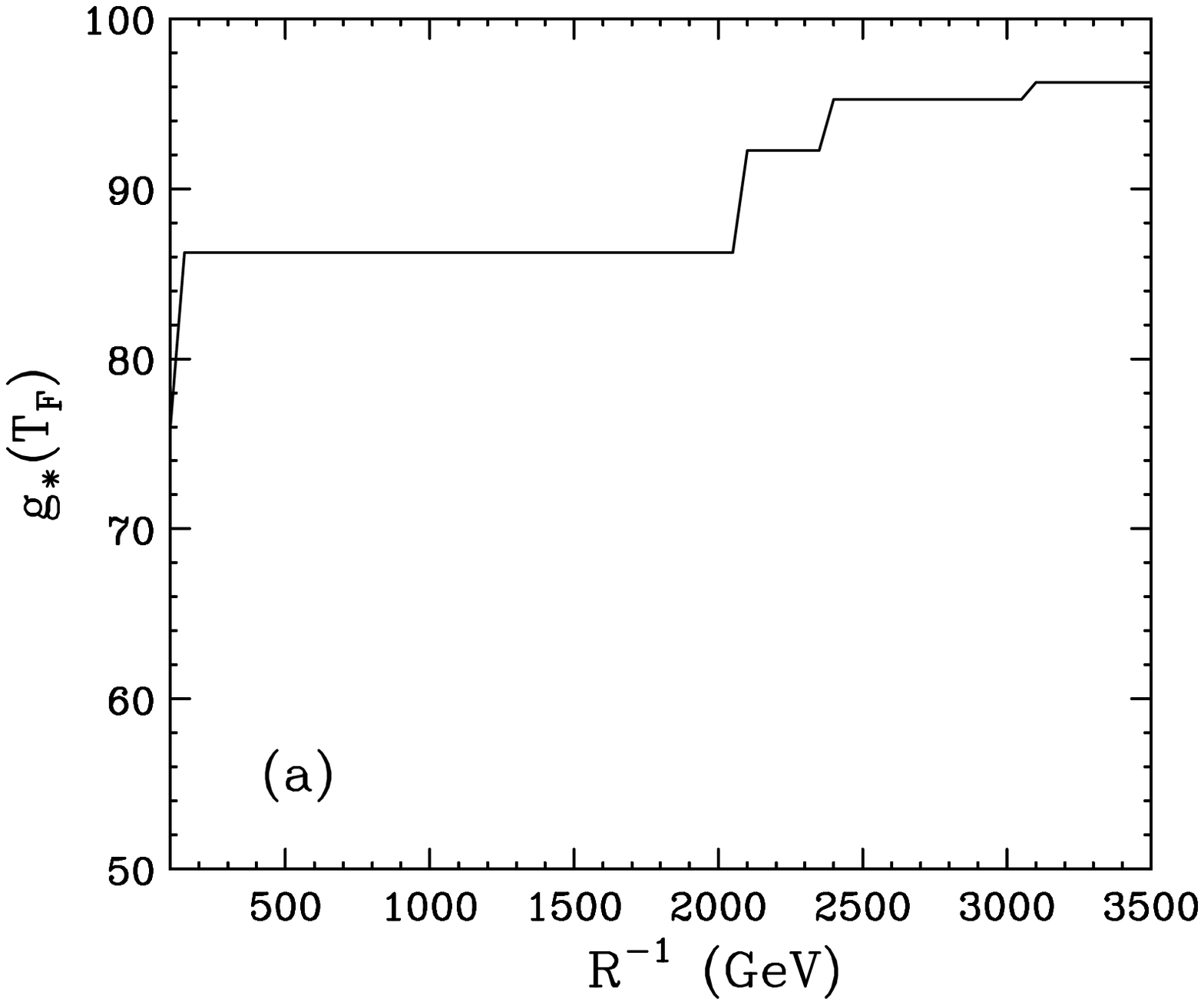,width=7.4cm}
\epsfig{file=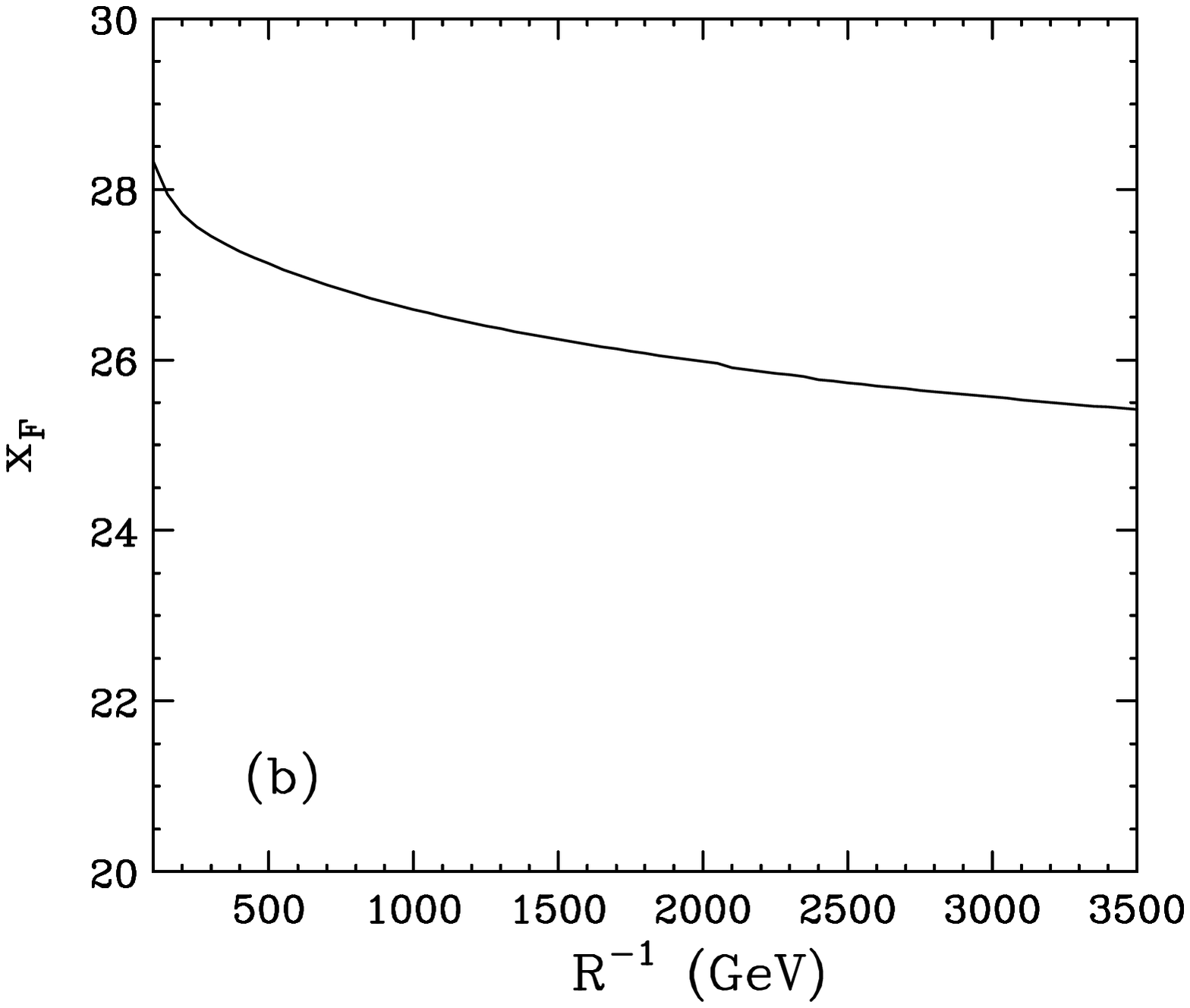,width=7.4cm}
\caption{\sl The quantities (a) $g_\ast(T_F)$ and (b) $x_F$ as a function 
of $R^{-1}$ in MUED.}
\label{fig:gstar}
}

The expert reader has probably noticed from Fig.~\ref{fig:gstar}b 
that the values of $x_F$ which we 
obtain in MUED are somewhat larger than the $x_F$ values one would have in
typical SUSY models. This is due to the effect of coannihilations, which 
increase $g_{eff}$ (see Fig.~\ref{fig:aeff}c below) and therefore $x_F$,
in accordance with (\ref{xfcoann}).

\FIGURE[t]{
\epsfig{file=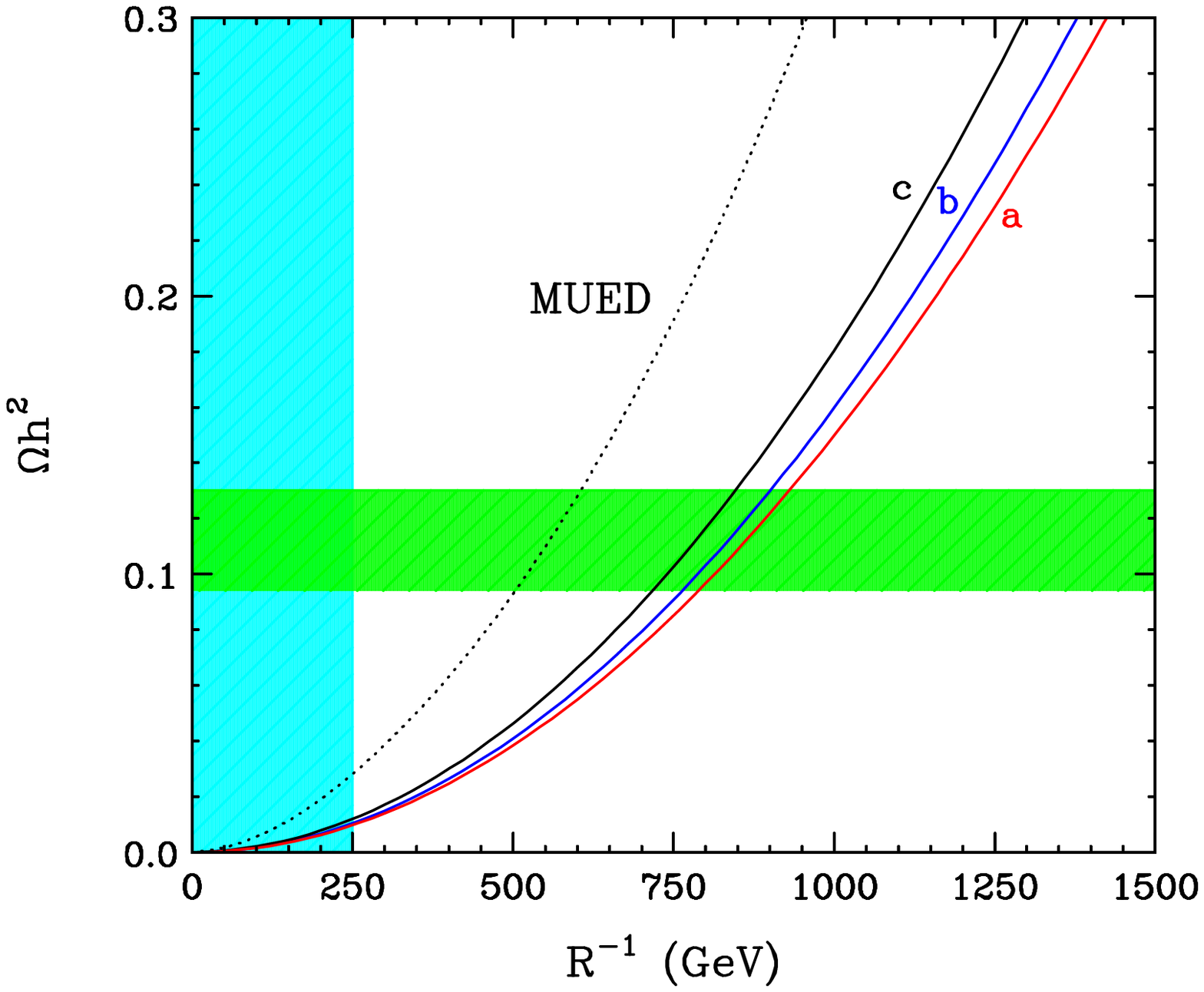,height=10.0cm}
\caption{\sl Relic density of the LKP as a function of $R^{-1}$ 
in the Minimal UED model. The (red) line marked ``a'' is the result
from considering $\gamma_1\gamma_1$ annihilation only, following
the analysis of Ref.~\cite{Servant:2002aq}, 
assuming a degenerate KK mass spectrum. The 
(blue) line marked ``b'' repeats the same analysis, but
uses $T$-dependent $g_\ast$ according to 
(\ref{gstar}) and includes the relativistic correction to the $b$-term
(\ref{bcorr}). The (black) line marked ``c'' relaxes the assumption of 
KK mass degeneracy, and uses the actual MUED mass spectrum. The dotted line is
the result from the full calculation in MUED, including all coannihilation 
processes, with the proper choice of masses. The green horizontal band
denotes the preferred WMAP region for the relic density 
$0.094<\Omega_{CDM}h^2<0.129$. The cyan vertical band delineates
values of $R^{-1}$ disfavored by precision data.}
\label{fig:MUED}}

We are now in a position to discuss our main result in MUED.
In Fig.~\ref{fig:MUED} we show the LKP relic density
as a function of $R^{-1}$ in the Minimal UED model. We show the
results from several analyses, each under different assumptions, 
in order to illustrate the effect of each assumption. 
We first show several calculations for the academic case 
of no coannihilations. The three solid lines in Fig.~\ref{fig:MUED}
account only for the $\gamma_1\gamma_1$ process.
The (red) line marked ``a'' recreates the
analysis of Ref.~\cite{Servant:2002aq}, 
assuming a degenerate KK mass spectrum. The 
(blue) line marked ``b'' repeats the same analysis, but
uses $T$-dependent $g_\ast$ according to 
(\ref{gstar}) and includes the relativistic correction to the $b$-term
(\ref{bcorr}). The (black) line marked ``c'' further 
relaxes the assumption of KK mass degeneracy, and 
uses the actual MUED mass spectrum. 

Comparing lines ``a'' and ``b'', we see that, as 
already anticipated from Fig.~\ref{fig:gstar}a,
accounting for the $T$ dependence in $g_\ast$
has the effect of lowering $g_\ast(x_F)$, $\sigma_{eff}(x_F)$,
and correspondingly, increasing the prediction for $\Omega h^2$.
This, in turns, lowers the preferred mass range for $\gamma_1$.
Next, comparing lines ``b'' and ``c'', we see that dropping the
mass degeneracy assumption has a similar effect on $\sigma_{eff}(x_F)$
(see Fig.~\ref{fig:aterms}), and further 
increases the calculated $\Omega h^2$. This can be easily understood 
from the $t$-channel mass dependence exhibited in 
(\ref{aff}) and (\ref{ahh}). The $t$-channel masses appear 
in the denominator, and they are by definition larger than the 
LKP mass. Therefore, using their actual values can only
decrease $\sigma_{eff}$ and increase $\Omega h^2$.

The dotted line in Fig.~\ref{fig:MUED} is the result from the 
full calculation in MUED, 
including all coannihilation processes, with the proper 
choice of masses. The green horizontal band
denotes the preferred WMAP region for the relic density 
$0.094<\Omega_{CDM}h^2<0.129$. The cyan vertical band delineates
values of $R^{-1}$ disfavored by precision data~\cite{Appelquist:2002wb}.
We see that according to the full calculation, the
cosmologically ideal mass range is $m_{\gamma_1}\sim 500-600$ GeV,
when $\gamma_1$ accounts for {\em all} of the 
dark matter in the Universe. This range is somewhat lower 
than earlier studies have indicated, mostly due to the  
effects discussed above. Since the MUED model will be our 
reference point for the investigations in Section~\ref{sec:coann},
the dotted line from Fig.~\ref{fig:MUED} will be appearing
in all subsequent plots in Section~\ref{sec:coann} below.

\section{Relative Weight of Different Coannihilation Processes}
\label{sec:coann}

As we already explained in the Introduction, the assumptions
behind the MUED model can be easily relaxed by allowing 
nonvanishing boundary terms at the scale $\Lambda$.
This would modify the KK spectrum and correspondingly 
change our prediction for the KK relic density from the
previous section. Our code is able to handle such more
general cases with ease, since we use as inputs the 
physical KK masses. In order to gain some insight into
the cosmology of such non-minimal scenarios, 
we have studied the effects of varying the $n=1$ 
KK masses one at a time. The change in any given KK mass 
will not only enhance or suppress the related 
coannihilation processes, but also impact any other 
cross-sections which happen to have a dependence on the
mass parameter being varied. Thus the results 
in this section may allow one to judge the importance 
of each individual coannihilation process, and
anticipate the answer for $\Omega h^2$ in non-minimal models.

We have classified the discussion in this section by particle types.
Section~\ref{sec:leptons} contains our results for 
the annihilation processes with KK leptons. 
Many of our results have already appeared in Ref.~\cite{Servant:2002aq}.
The new element here is the discussion of 
$\ell_{L1}$ coannihilations. The results presented in Sections
\ref{sec:quarks} and \ref{sec:EW} are completely new --
there we investigate the coannihilation effects with 
strongly interacting KK modes and electroweak gauge bosons 
and/or Higgs bosons, respectively.

\subsection{Effects due to coannihilations with KK leptons}
\label{sec:leptons}

We begin with a discussion of $\gamma_1$ coannihilations
with the $n=1$ $SU(2)_W$-singlet leptons $\ell_{R1}$ and
the $n=1$ $SU(2)_W$-doublet leptons $\ell_{L1}$. One might
expect that those processes will be important, since the
KK leptons receive relatively small one-loop mass corrections.
For example, in the Minimal UED model
$\Delta_{\ell_{R1}}\sim 1\%$ and 
$\Delta_{\ell_{L1}}\sim 3\%$. It is natural to expect that 
this degeneracy might persist in non-minimal models as well.

Our approach is as follows. Since we keep separate values for 
the KK masses, when we start varying any one of them, 
we have to somehow fix the remainder of the KK mass spectrum.
We choose to use MUED as our reference model, hence the
masses which are not being varied, will be fixed according 
to their MUED values. We shall still show results for $\Omega h^2$
as a function of $R^{-1}$, but for various fixed values of the
corresponding mass splitting $\Delta_i$ defined in 
eq.~(\ref{Delta_i}). We shall also
always display the reference MUED model line, for which, 
of course, $\Delta_i$ takes its MUED value. 

\FIGURE[t]{
\epsfig{file=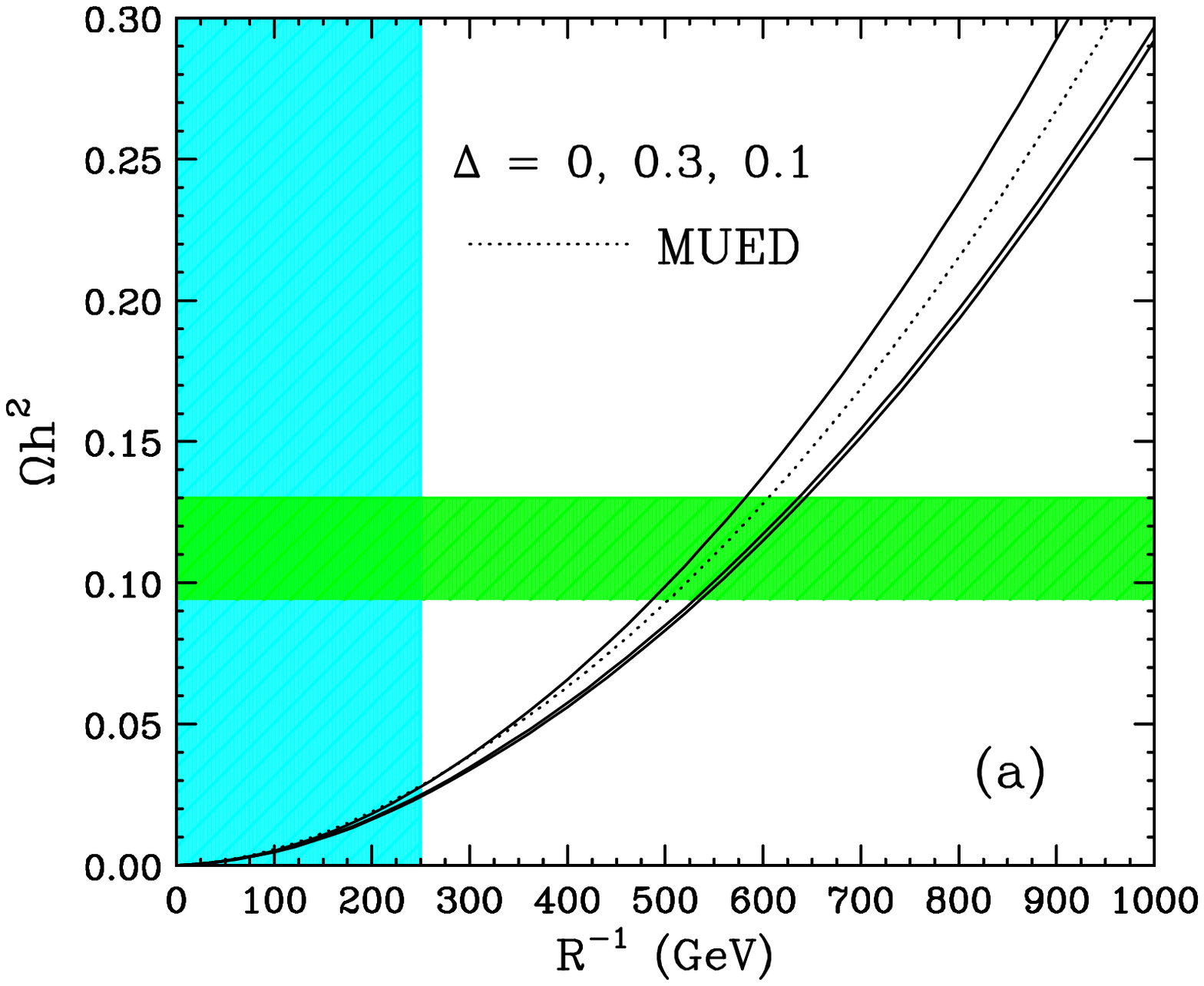,width=7.4cm}
\epsfig{file=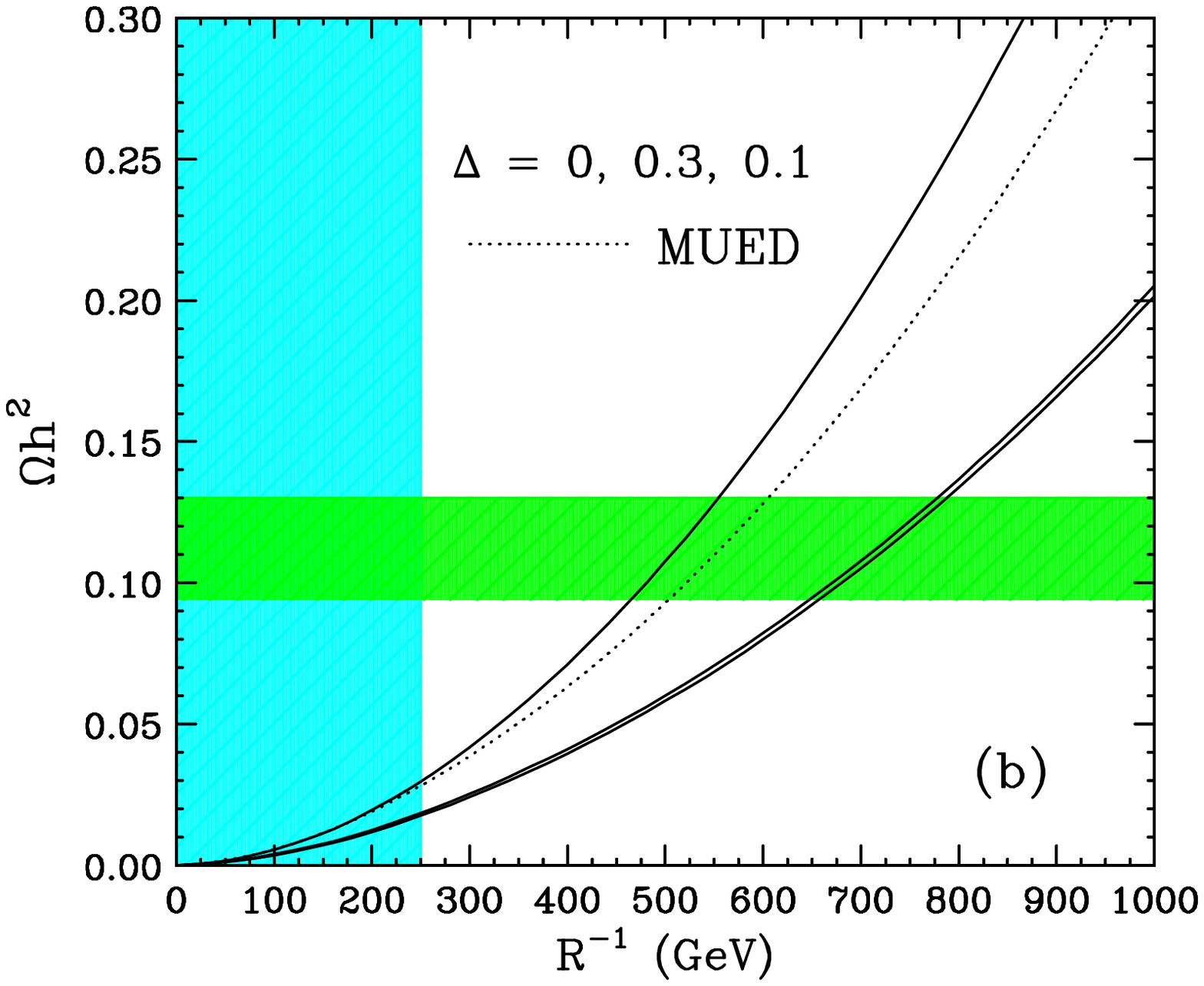,width=7.4cm}
\caption{\sl Coannihilation effects of (a) 1 generation or 
(b) 3 generations of singlet KK leptons. 
The lines show the LKP relic density as a function of $R^{-1}$, 
for several choices of the mass splitting $\Delta_{\ell_{R1}}$
between the LKP and the $SU(2)_W$-singlet KK fermions $\ell_{R1}$.
In each case we use the MUED spectrum to fix the masses of the
remaining particles, and then vary the $\ell_{R1}$ mass by hand.
The solid lines from top to bottom in both (a) and (b) correspond to
$\Delta_{e_{R1}}=0, 0.3, 0.1$.
The dotted line is the nominal UED case from Fig.~\ref{fig:MUED}.
\label{fig:lR}
}
}

Our first example is shown in Fig.~\ref{fig:lR}, where we 
illustrate the size of the coannihilation effects for
(a) 1 generation or (b) 3 generations of degenerate singlet KK leptons $\ell_{R1}$. 
The lines show the LKP relic density 
as a function of $R^{-1}$, for several choices of the mass splitting 
$\Delta_{\ell_{R1}}$
between the LKP and the $SU(2)_W$-singlet KK fermions $\ell_{R1}$.
The solid lines from top to bottom in both (a) and (b) correspond to
$\Delta_{e_{R1}}=0, 0.3, 0.1$, and the
dotted line is the nominal UED case from Fig.~\ref{fig:MUED},
for which $\Delta_{\ell_{R1}}=0.01$.
As expected, all lines follow the general trend of Fig.~\ref{fig:MUED}. 
In accord with the observations of Ref.~\cite{Servant:2002aq},
we see that $\ell_{R1}$ coannihilations {\em increase} the prediction 
for $\Omega h^2$. Such a behavior may seem peculiar
at first sight, since in supersymmetry one finds the opposite 
phenomenon --- coannihilations with sleptons tend to {\em reduce} the
SUSY WIMP relic density. The difference between the two cases
can be intuitively understood as follows. In SUSY, the
cross-section for the main annihilation channel 
($\tilde\chi^0_1\tilde\chi^0_1\to f\bar{f}$) 
is helicity suppressed, but the coannihilation processes 
are not. Adding coannihilations
therefore can only increase the effective cross-section
(\ref{sigmaeff}) and correspondingly decrease $\Omega h^2$. 
In contrast, in UED the main annihilation 
channel ($\gamma_1\gamma_1\to f\bar{f}$) is already
of normal strength. The effect of coannihilations can be 
easily guessed only if the additional processes have either 
much weaker or much stronger interactions. In the case of 
$\ell_{R1}$, however, the additional processes are of the 
same order (both $\gamma_1$ and $\ell_{R1}$ have hypercharge 
interactions only) and the sign of the coannihilation effect
depends on the detailed balance of numerical factors, which
will be illustrated in Fig.~\ref{fig:aeff} and discussed
in more detail below.

The spread in the lines in Fig.~\ref{fig:lR} 
is indicative of the importance of the coannihilations. 
Comparing Fig.~\ref{fig:lR}a and Fig.~\ref{fig:lR}b,
we see that in the case of three generations, the effects 
are magnified correspondingly. A similar conclusion was reached 
in Ref.~\cite{Servant:2002aq}.

Notice the peculiar ordering of the lines corresponding to different 
$\Delta_{\ell_{R1}}$. With respect to variations of
$\Delta_{\ell_{R1}}$, the maximum possible value of $\Omega h^2$
is obtained for $\Delta_{\ell_{R1}}\to0$, where 
the effect of coannihilations is maximal.
Then, as we increase the mass splitting between 
$\ell_{R1}$ and $\gamma_1$, at first $\Omega h^2$
decreases (see the sequence of $\Delta_{\ell_{R1}}=0$,
$\Delta_{\ell_{R1}}=0.01$ and $\Delta_{\ell_{R1}}=0.1$)
but then starts increasing again and the $\Omega h^2$ 
values that we get for $\Delta_{\ell_{R1}}=0.3$
are slightly larger than those for $\Delta_{\ell_{R1}}=0.1$. 
This behavior can be seen more clearly from 
Fig.~\ref{fig:aeff}a, where
we vary the mass of the $SU(2)_W$-singlet KK electron $e_{R1}$
and plot $\Omega h^2$ versus $\Delta_{e_{R1}}$ for a fixed
$R^{-1}=500$ GeV. 

\FIGURE[ht]{
\epsfig{file=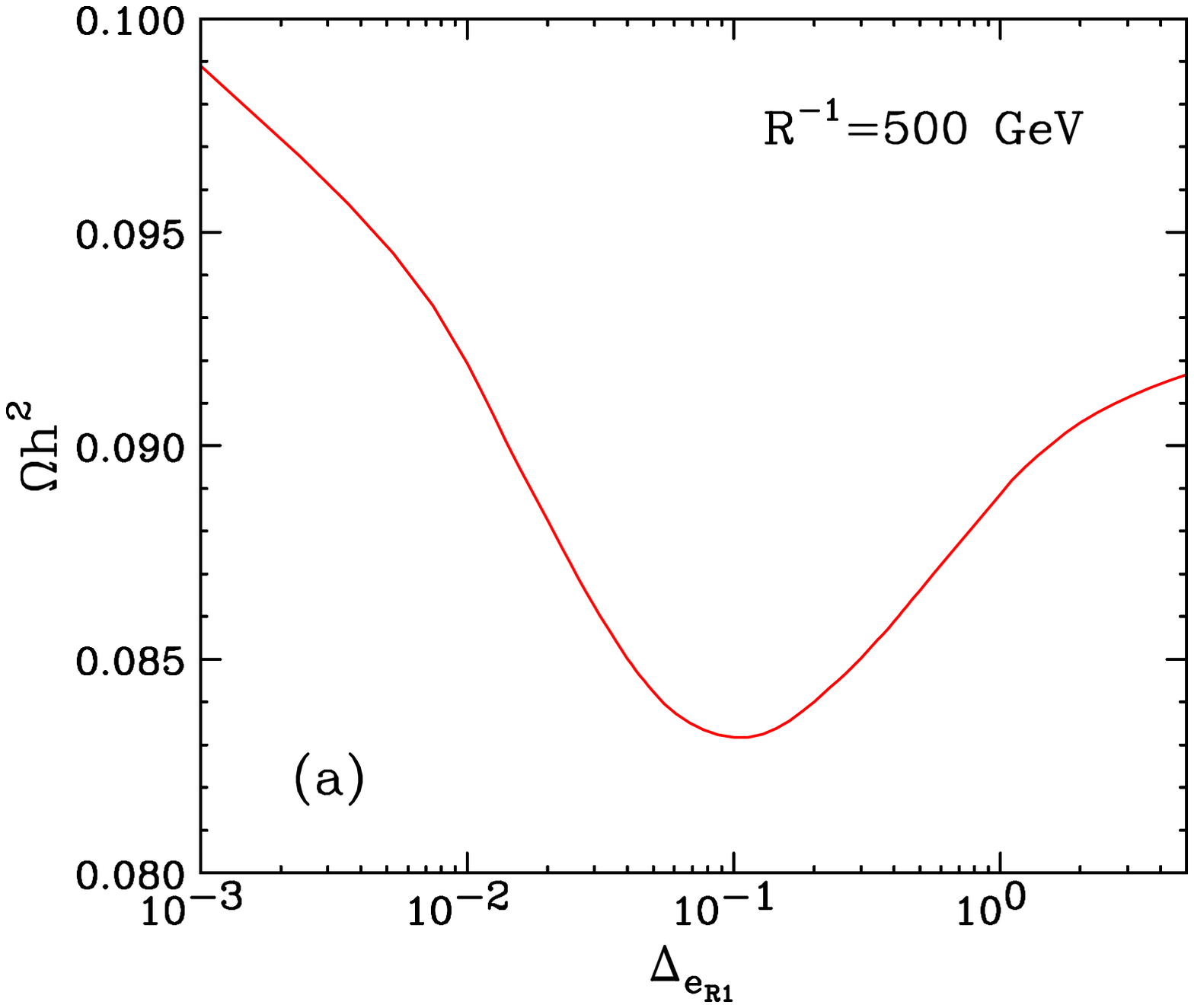,width=7.2cm}
\epsfig{file=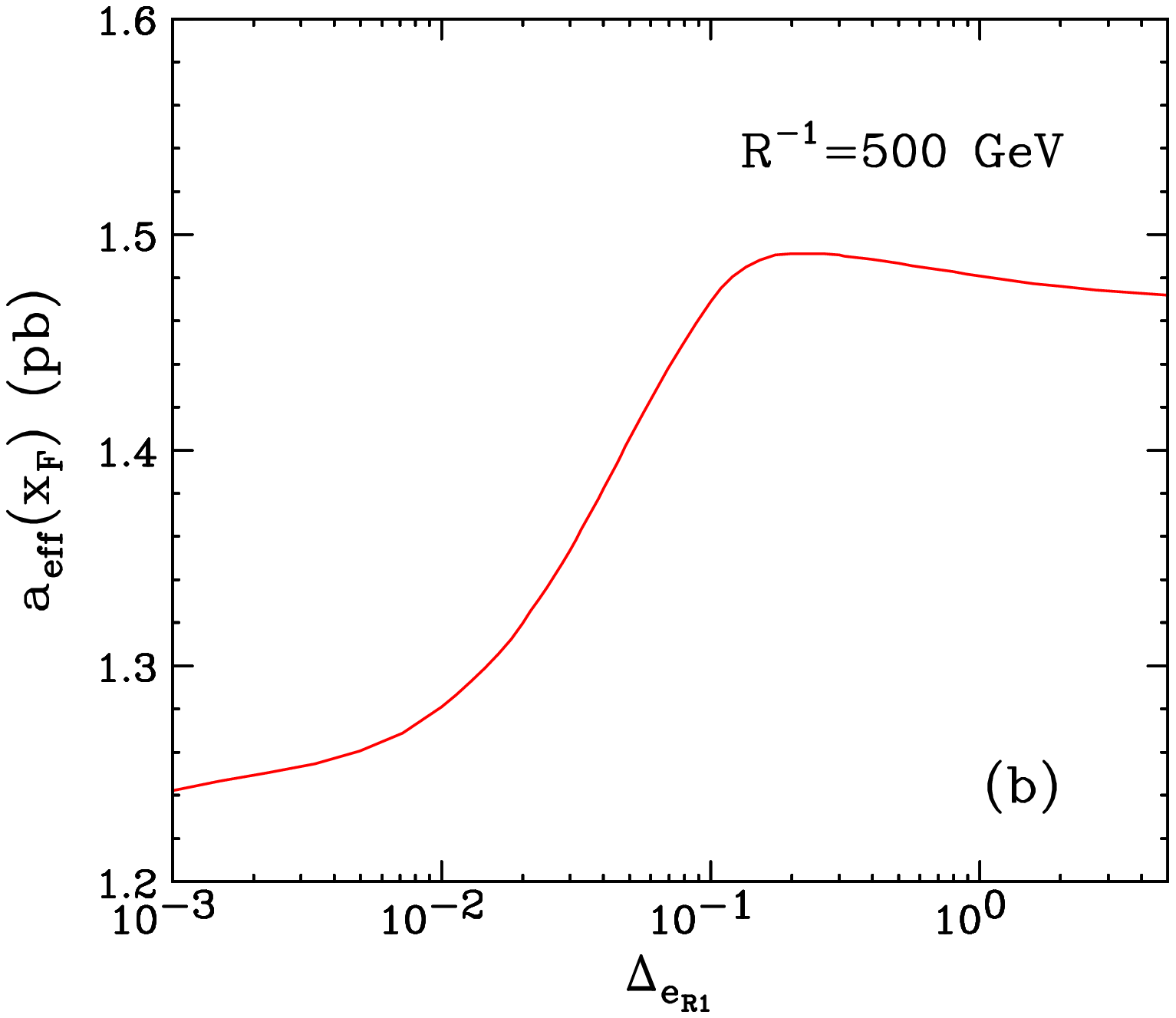,width=7.2cm}
\epsfig{file=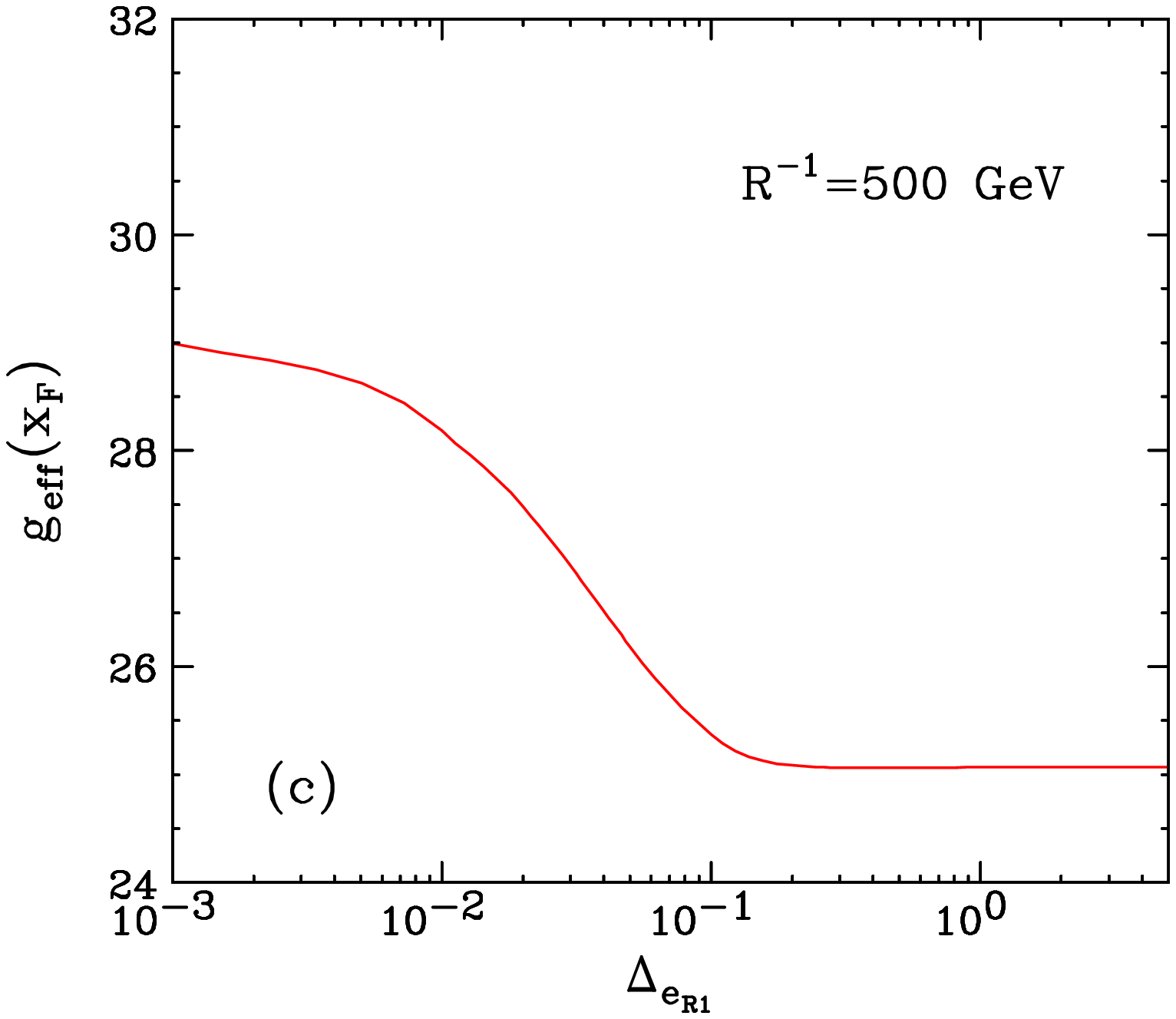,width=7.2cm}
\epsfig{file=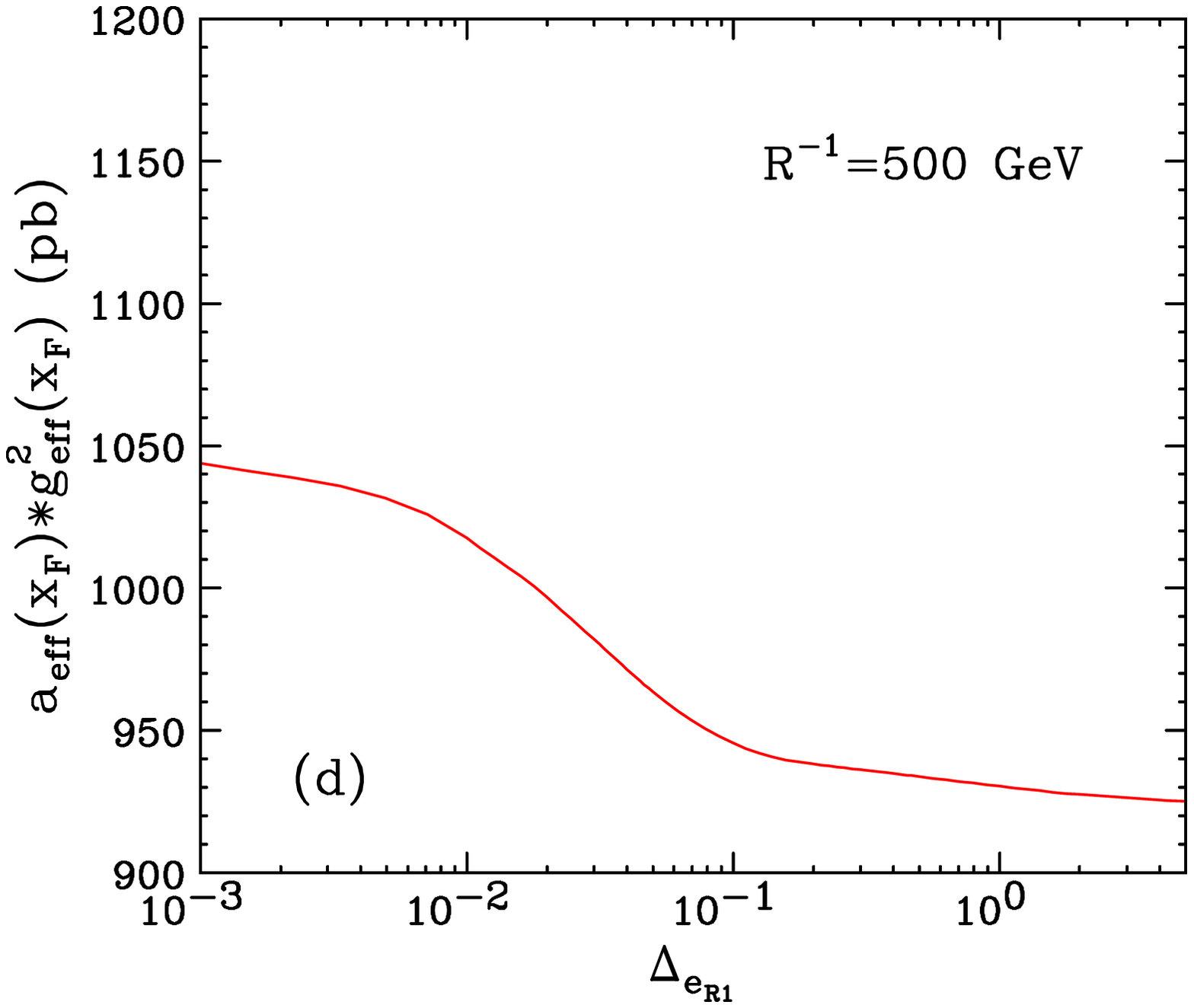,width=7.2cm}
\caption{\sl Plots of various quantities
entering the LKP relic density computation,
as a function of the mass splitting $\Delta_{e_{R1}}$ 
between the LKP and the $SU(2)_W$-singlet KK electron,
for $R^{-1}=500$ GeV in MUED.
(a) Relic density, (b) $a_{eff}(x_F)$,
(c) $g_{eff}(x_F)$ and 
(d) $a_{eff}(x_F)g^2_{eff}(x_F)$.
}
\label{fig:aeff}
}

The interesting behavior of $\Omega h^2$ exhibited in Fig.~\ref{fig:aeff}a
can be understood in terms of the $m_{\ell_{R1}}$ dependence of the
effective annihilation cross-section (\ref{sigmaeff}) which is
dominated by its $a$-term (\ref{aeff}). Both $\sigma_{eff}$ and
$a_{eff}$ are functions of $x$, but for the purposes of our discussion 
here it is sufficient to concentrate on the fixed value $x=x_F$
which dominates the integrals (\ref{I_a}) and (\ref{I_b}).
We plot $a_{eff}(x_F)$ as a function of $\Delta_{e_{R1}}$ 
in Fig.~\ref{fig:aeff}b. We see that $a_{eff}(x_F)$
exhibits exactly the opposite dependence to $\Omega h^2$,
and in particular, has an analogous local extremum at 
$\Delta_{e_{R1}}\sim 0.1$. Therefore, in order to 
understand qualitatively the behavior of $\Omega h^2$, 
we only need to concentrate on $a_{eff}(x_F)$.

Let us start with the large $\Delta_{e_{R1}}$ region
in Fig.~\ref{fig:aeff}b. The $SU(2)_W$-singlet KK electron 
$e_{R1}$ is then too heavy to participate in any 
relevant coannihilation processes. The effective cross-section
(\ref{sigmaeff}) then receives no contributions from
processes with $e_{R1}$. Nevertheless, the mass of
$e_{R1}$ enters $\sigma_{eff}$ through the cross-section
for the process $\gamma_1\gamma_1\to e^+e^-$
(see eqs.~(\ref{aff}) and (\ref{bff})).
Then as we lower $m_{e_{R1}}$, $\sigma(\gamma_1\gamma_1\to e^+e^-)$
is increased and this leads to a corresponding increase in 
$a_{eff}$ as seen in Fig.~\ref{fig:aeff}b.
This trend continues down to $\Delta_{e_{R1}}\sim 0.1$,
where coannihilations with $e_{R1}$ start becoming relevant.
This can be seen in Fig.~\ref{fig:aeff}c, where we plot
$g_{eff}(x_F)$ as a function of $\Delta_{e_{R1}}$.
From its defining equation (\ref{geff}) we see that
$g_{eff}(x)$ starts to deviate from a constant only when the
exponential terms (which signal the turning on of coannihilations)
become non-negligible. The exponential terms are all positive 
and increase $g_{eff}$. At the same time, there are new 
cross-section terms entering the sum for $\sigma_{eff}$, 
so we expect the numerator in (\ref{sigmaeff}) to increase 
as well. This is confirmed in Fig.~\ref{fig:aeff}d, where
we plot the numerator of (\ref{sigmaeff}) simply as
$a_{eff}(x_F) g^2_{eff}(x_F)$. From Figs.~\ref{fig:aeff}c
and \ref{fig:aeff}d we see that both the numerator and the 
denominator of (\ref{aeff}) increase at low $\Delta_{e_{R1}}$,
and so it is a priori unclear how their ratio will behave 
with $\Delta_{e_{R1}}$. In this particular case, $g_{eff}$ wins,
and $a_{eff}(x_F)$ is effectively decreased as a result 
of turning on the coannihilations with $e_{R1}$.
This feature was also observed in Ref.~\cite{Servant:2002aq}.

\FIGURE[ht]{
\epsfig{file=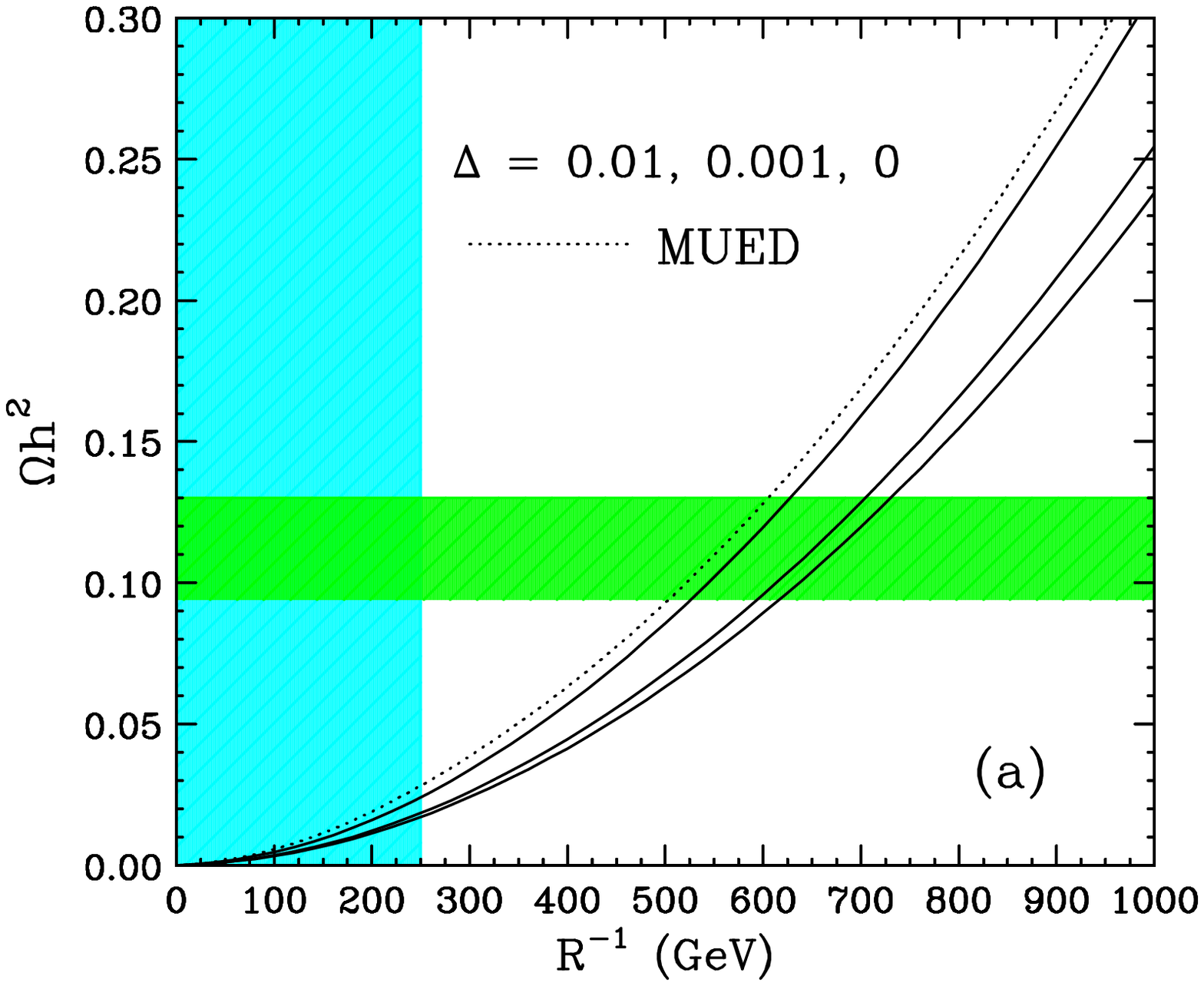,width=7.4cm}
\epsfig{file=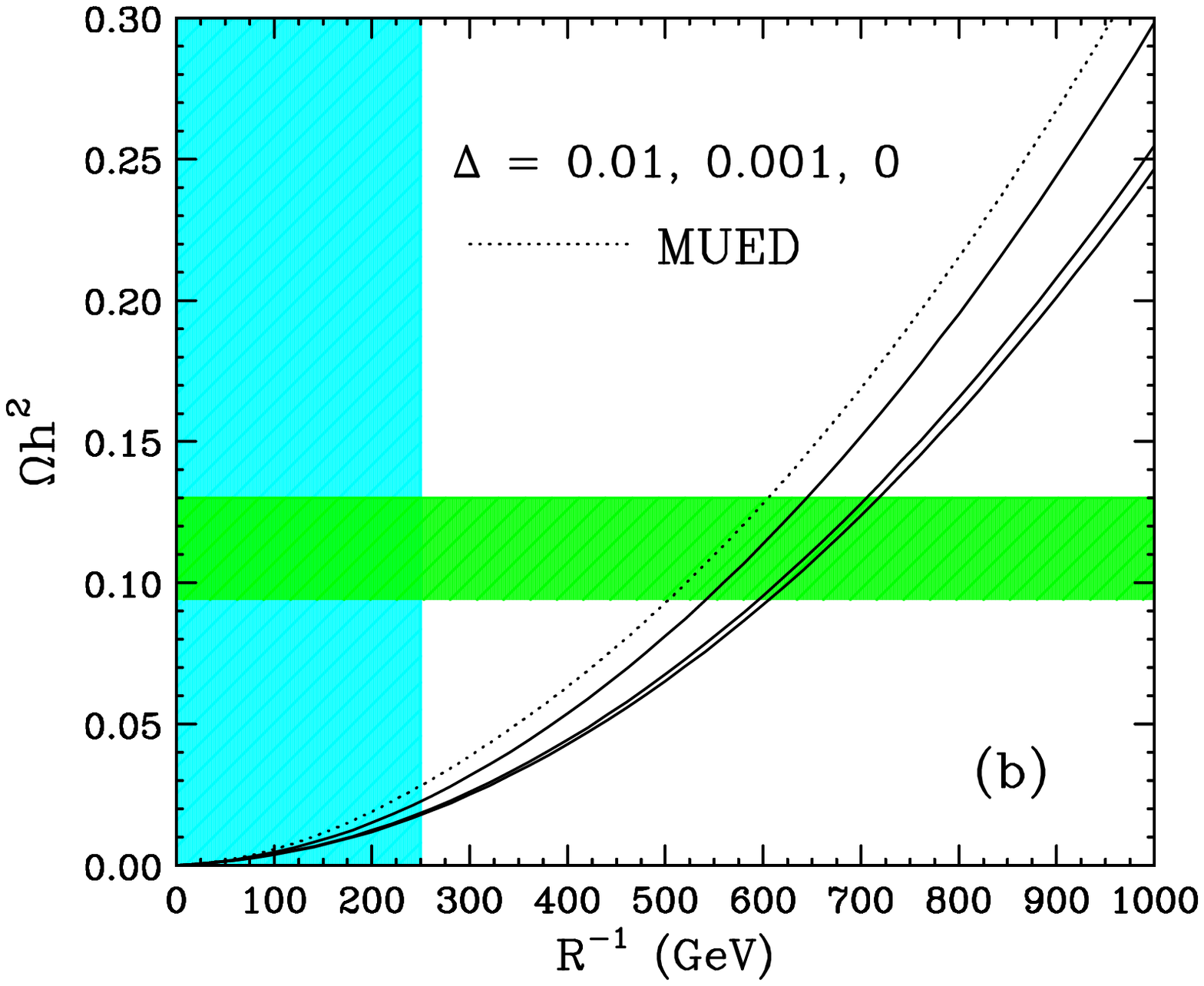,width=7.4cm}
\caption{\sl The same as Fig.~\ref{fig:lR}
but illustrating the effects of varying the $SU(2)_W$-doublet
KK electron mass. From top to bottom, the solid lines show $\Omega h^2$
as a function of $R^{-1}$, for $\Delta_{e_{L1}}=0.01, 0.001, 0$.
The dotted line is the nominal UED case from Fig.~\ref{fig:MUED}.
\label{fig:lL}
}
}

We are now in position to repeat the same analysis, but 
for the case of the $SU(2)_W$-doublet KK leptons $\ell_{L1}$.
In Fig.~\ref{fig:lL}, in complete analogy to Fig.~\ref{fig:lR},
we illustrate the effects on the relic density from
varying the $SU(2)_W$-doublet KK electron mass.
From top to bottom, the solid lines show $\Omega h^2$
as a function of $R^{-1}$, for $\Delta_{e_{L1}}=0.01, 0.001, 0$. 
The dotted line is again the MUED reference model.
We see that the case of $SU(2)_W$-doublet KK leptons is different.
Unlike $\ell_{R1}$, they have weak interactions, and 
the extra terms which they bring into the sum (\ref{sigmaeff})
are larger than the main annihilation channel. 
The increase in $g_{eff}$ is similar as before.
As a result, this time the increase in the numerator of
(\ref{sigmaeff}) wins, and the net effect is to increase
the effective annihilation cross-section. This leads 
to a reduction in the predicted value for the relic density, 
as evidenced from Fig.~\ref{fig:lL}. Notice how
the decrease in $\Omega h^2$ is monotonic with $\Delta_{e_{L1}}$.

Another difference between $\ell_{R1}$ and $\ell_{L1}$
coannihilations is revealed by comparing the case of 1 generation
(panels (a) in Figs.~\ref{fig:lR} and \ref{fig:lL})
and 3 generations (panels (b) in Figs.~\ref{fig:lR} and \ref{fig:lL}).
We see that for $SU(2)_W$ singlets, the coannihilations are more
prominent for the case of 3 generations, while for
$SU(2)_W$ doublets, it is the opposite.
This is due to the different number of degrees of freedom
contributed to $g_{eff}$ in each case, which shifts the 
delicate balance between the numerator and denominator of 
(\ref{sigmaeff}), as discussed above.

\subsection{Effects due to coannihilations with KK quarks and KK gluons}
\label{sec:quarks}

We will now consider coannihilation effects with colored 
KK particles (KK quarks and KK gluons). Since they
couple strongly, we expect on general grounds that the
effective annihilation cross-sections will be
enhanced, and the preferred range of the LKP mass 
will correspondingly be shifted higher.
\FIGURE[ht]{
\epsfig{file=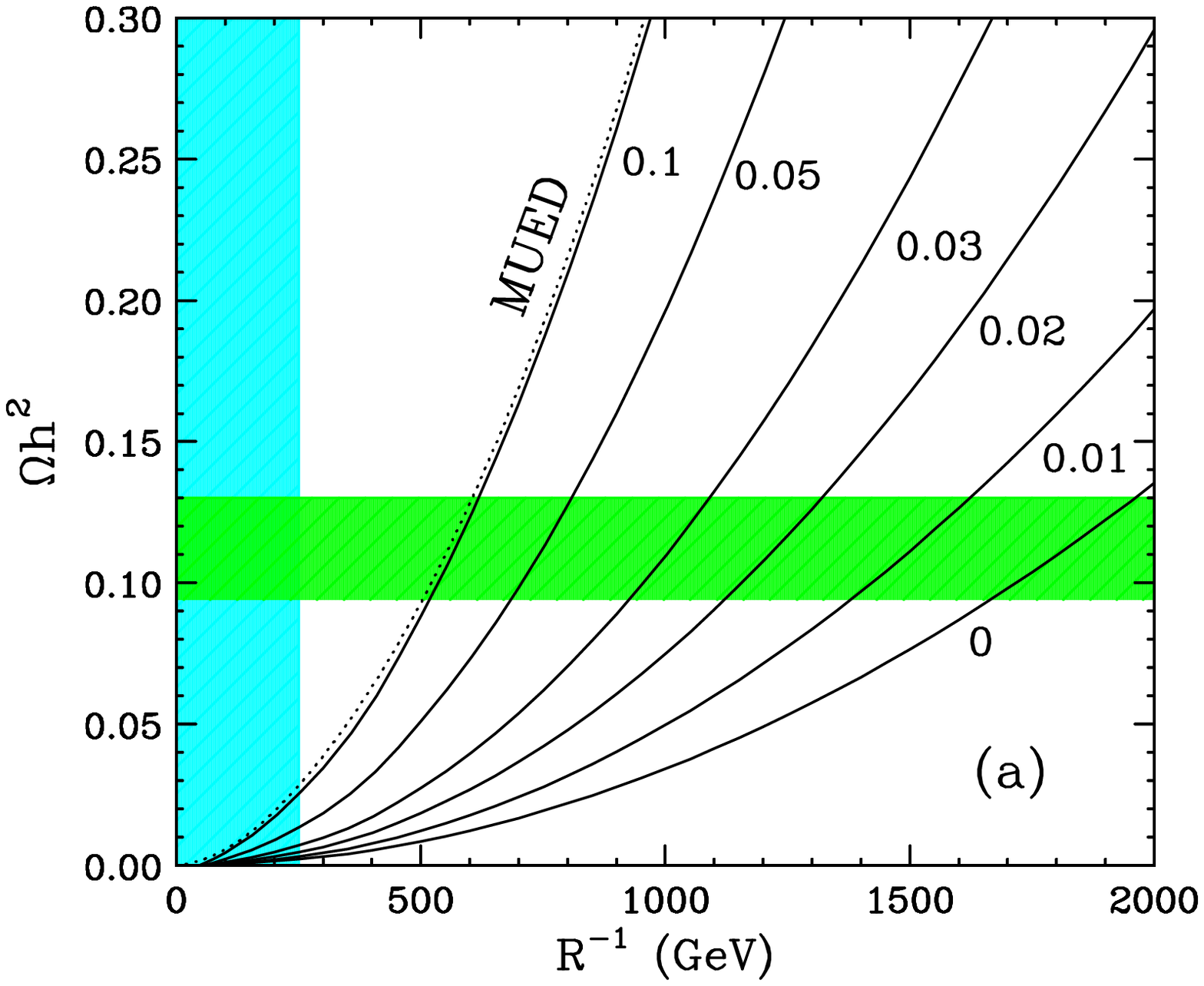,width=7.4cm}
\epsfig{file=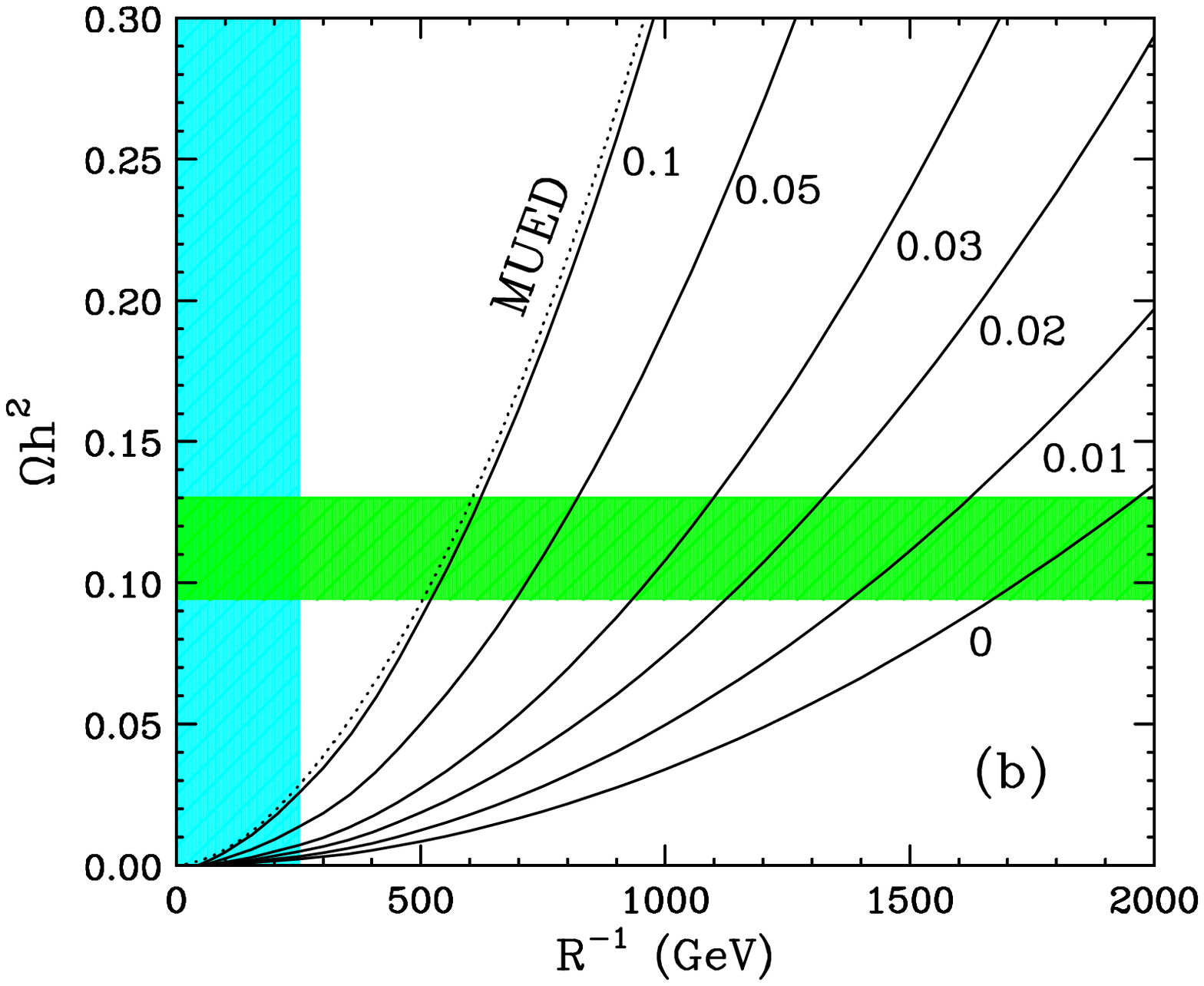,width=7.4cm}
\caption{\sl The same as Figs.~\ref{fig:lR}(b) and
\ref{fig:lL}(b), but for the case of KK quarks.
Each solid line is labeled by the value of (a) $\Delta_{q_{R1}}$ 
or (b) $\Delta_{q_{L1}}$ used. 
The dotted line is the nominal UED case from Fig.~\ref{fig:MUED}.
\label{fig:q}
}
}
\FIGURE[ht]{
\epsfig{file=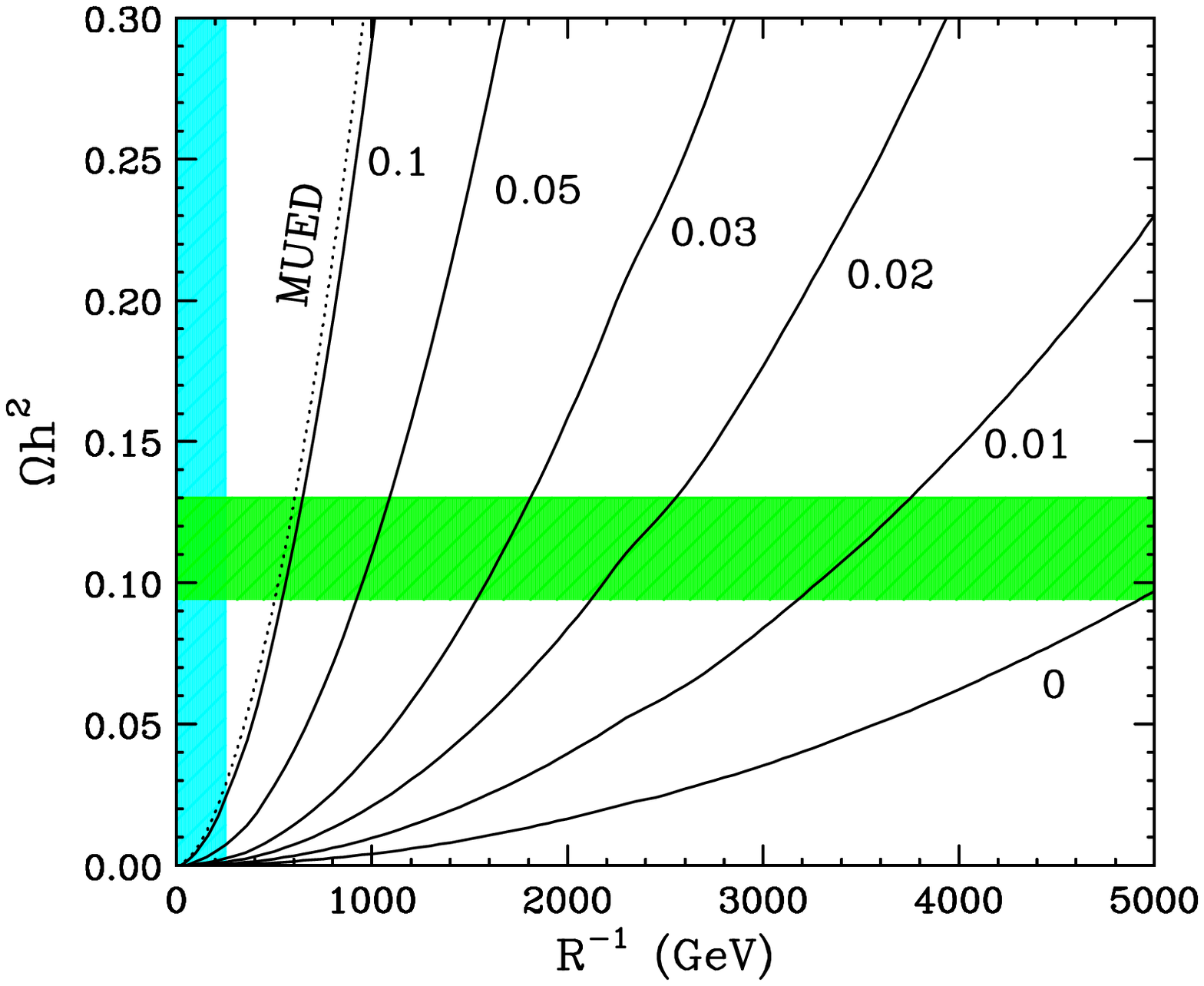,width=10cm}
\caption{\sl The same as Fig.~\ref{fig:q}, but for the case of varying 
the KK gluon mass. The lines are labeled by the value of $\Delta_{g_1}$.
The dotted line is the nominal UED case from Fig.~\ref{fig:MUED}.
\label{fig:gluon}
}
}
These expectations are confirmed by our explicit calculation
whose results are shown in Figs.~\ref{fig:q} and \ref{fig:gluon}. 
In Fig.~\ref{fig:q} we show
the effects on the relic density from varying the
masses of all three generations of (a) $SU(2)_W$-singlet KK 
quarks and (b) $SU(2)_W$-doublet KK quarks.
The solid lines show $\Omega h^2$ as a function of $R^{-1}$, 
and are labeled by the corresponding value
of $\Delta_{q_{R1}}$ or $\Delta_{q_{L1}}$ used.
As before, the dotted line is the MUED reference model.
Comparing the results in Figs.~\ref{fig:q}a and \ref{fig:q}b,
we find that the coannihilations with $q_{R1}$ and $q_{L1}$
have very similar effects, as they are both dominated by the
strong interactions, which are the same for $q_{R1}$ and $q_{L1}$.
Fig.~\ref{fig:gluon} shows the analogous result 
for the case of varying the KK gluon mass, where the
labels now show the values of $\Delta_{g_1}$.
There is a noticeable distortion of the lines
around $R^{-1}\sim 2300$ GeV, which is due to the change in
$g_\ast$ (see Fig.~\ref{fig:gstar}a).

From Figs.~\ref{fig:q} and \ref{fig:gluon} we see that
in non-minimal UED models where the colored KK modes
happen to exhibit some sort of degeneracy with 
the LKP, multi-TeV values for $m_{\gamma_1}$ are in 
principle possible. From that point of view, 
unfortunately, there is no ``no-lose'' theorem for
the LHC or ILC regarding a potential absolute upper bound on
the LKP mass.

\subsection{Effects due to coannihilations with electroweak KK bosons}
\label{sec:EW}

We finally show our coannihilation results for the case of
electroweak KK gauge bosons ($W^0_1$ and $W^\pm_1$) and KK Higgs bosons
($H^0_1$, $G^0_1$ and $G^\pm_1$). The results are displayed in 
Figs.~\ref{fig:WH}a and \ref{fig:WH}b, correspondingly.
Due to the $SU(2)_W$ symmetry, all three $n=1$ KK $W$-bosons
are very degenerate, and we have assumed a common
parameter $\Delta_{W_1}$ for all three. Similarly, the
masses of the $n=1$ KK Higgs bosons differ only by
electroweak symmetry breaking effects, which we neglect 
throughout the calculation. We have therefore assumed a 
common parameter $\Delta_H$ for them as well.

Since both the electroweak KK gauge bosons and the KK Higgs bosons
have weak interactions, we expect the results to be similar to the 
case of $SU(2)_W$-doublet leptons in the sense that
coannihilations would lower the predictions for $\Omega h^2$.
This is confirmed by Fig.~\ref{fig:WH}. We observe that
the effects from the KK $W$-bosons are actually quite significant, 
and can push the preferred LKP mass as high as 1.4 TeV.

\FIGURE[ht]{
\epsfig{file=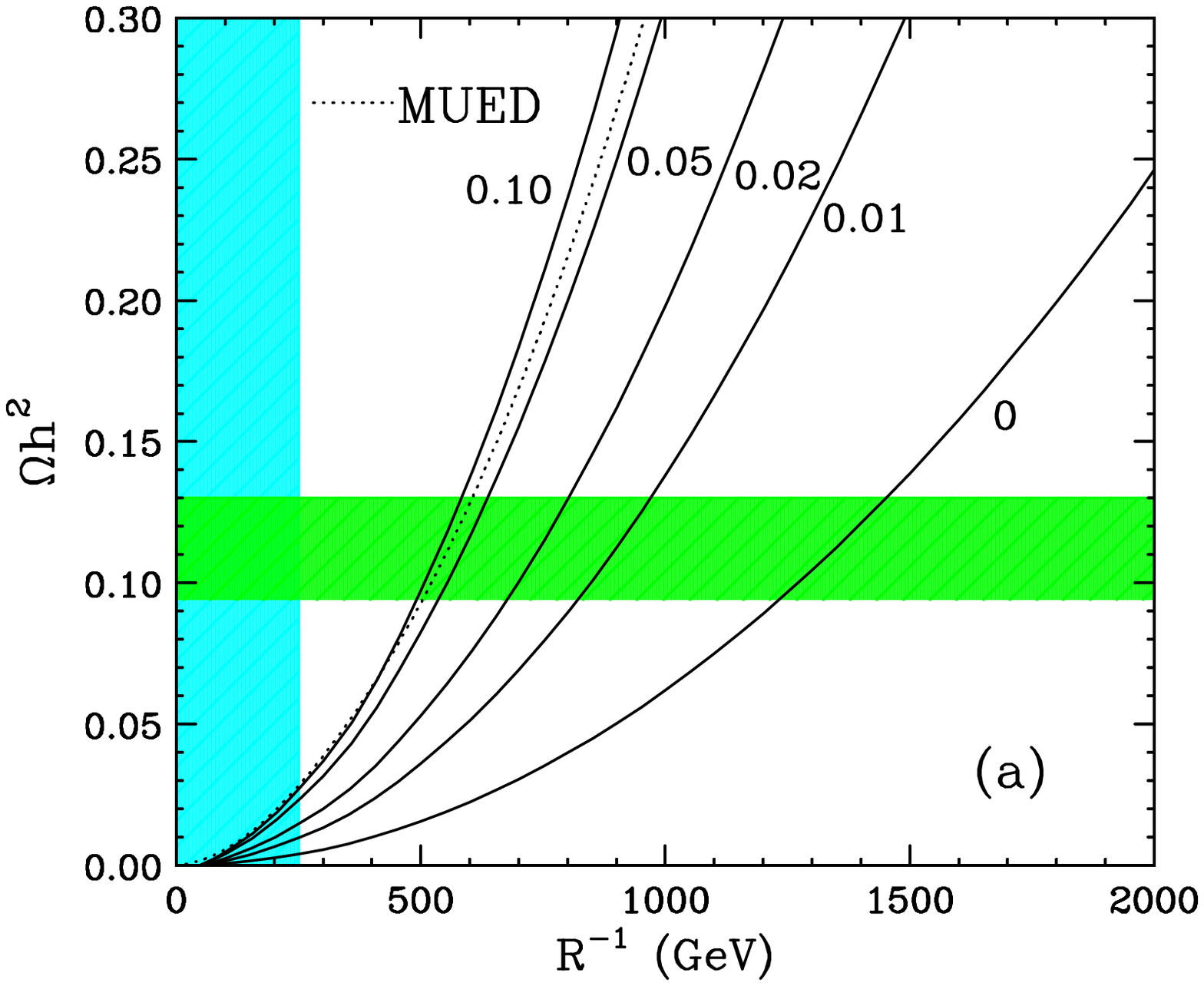,width=7.4cm}
\epsfig{file=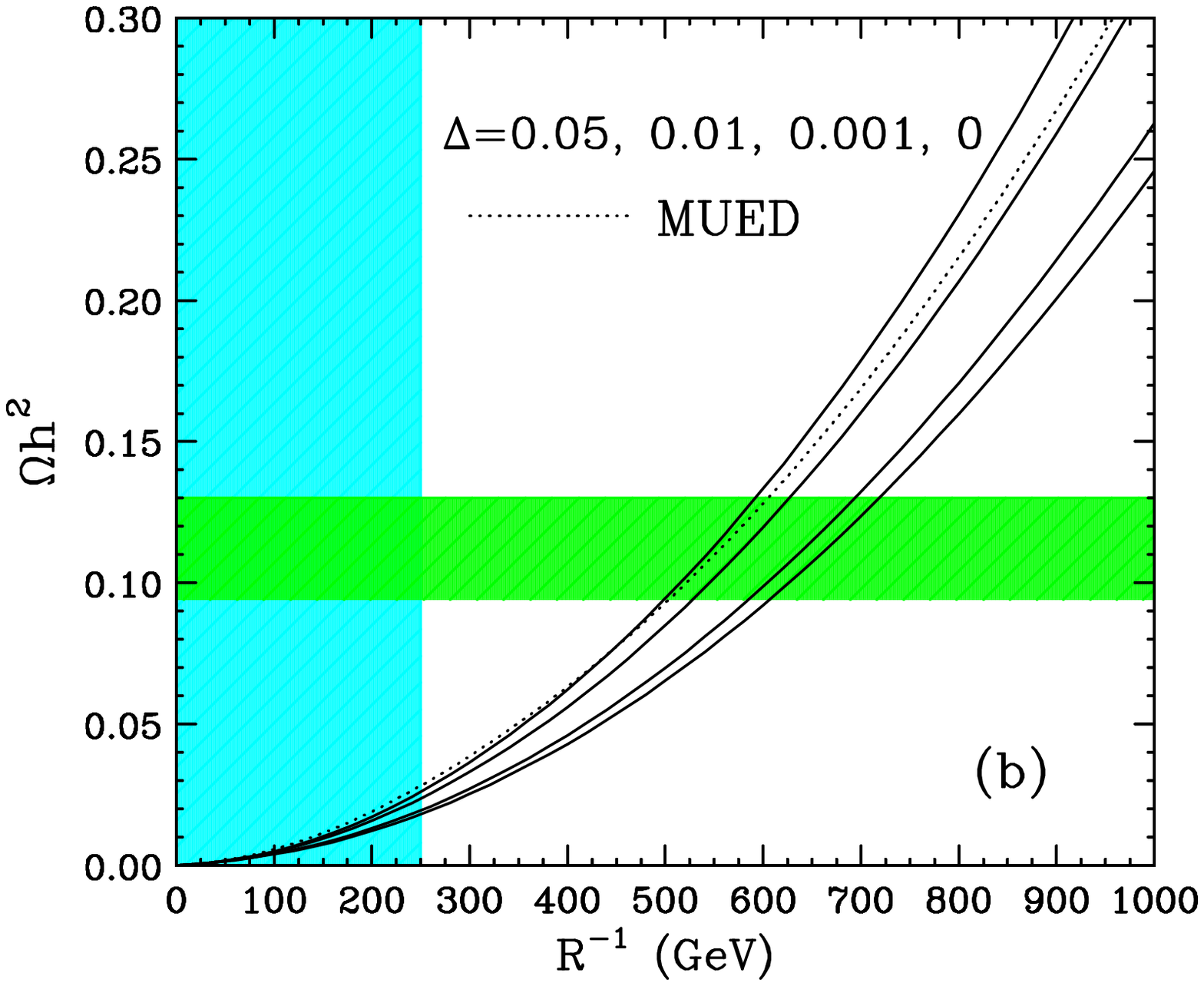,width=7.4cm}
\caption{\sl The same as Fig.~\ref{fig:gluon}, but
illustrating the effect of varying simultaneously 
the masses of all (a) $SU(2)$ KK gauge bosons
and (b) KK Higgs bosons. In (a) the lines are labeled by
the value of $\Delta_{W_1}$, while in (b) the values of $\Delta_H$
are (from top to bottom) $\Delta_H=0.05, 0.01, 0.001, 0$.
The dotted line is the nominal UED case from Fig.~\ref{fig:MUED}.
\label{fig:WH}
}
}

\section{Summary and Conclusions}
\label{sec:conclusions}

In this paper we revisited the calculation of the LKP relic density in 
the scenario of Universal Extra Dimensions. We extended the analysis of
Ref.~\cite{Servant:2002aq} to include {\em all} coannihilation processes
involving $n=1$ KK partners. This allowed us to predict reliably the 
preferred mass range for the KK dark matter particle in the Minimal UED model.
We found that in order to account for all of the dark matter in the universe, 
the mass of $\gamma_1$ should be within $500-600$ GeV, which is 
somewhat lower than the range found in \cite{Servant:2002aq}.
This is due to a combination of several factors. Among the
effects which caused our prediction for $\Omega h^2$ to go up are the following:
we used a lower value of $g_\ast$, we kept the individual KK masses 
in our formulas, and we accounted for the relativistic correction
(\ref{bcorr}). On the other hand, as we saw in Section~\ref{sec:coann},
including the effect of coannihilations with KK particles 
other than $SU(2)_W$-singlet KK leptons, always has the
effect of lowering the predicted $\Omega h^2$. Finally,
the cosmologically preferred range for $\Omega h^2$ itself
has shifted lower since the publication of \cite{Servant:2002aq}.

The lower range of preferred values for $R^{-1}$ 
is good news for collider and astroparticle searches 
for KK dark matter. It should be kept 
in mind that it is quite plausible, and in fact very likely, that 
the dark matter is made up of not one but several different components,
in which case the LKP could be even lighter.
We should mention that several collider studies 
\cite{Cheng:2002ab,Battaglia:2005zf,Smillie:2005ar,Battaglia:2005ma}
have already used an MUED benchmark point with $R^{-1}=500$ GeV,
a choice which we now see also happens to be relevant for cosmology.

In Section~\ref{sec:coann} we also investigated how each class 
of $n=1$ KK partners impacts the KK relic density. 
We summarize the observed trends in Fig.~\ref{fig:finger}, where
we fix $\Omega h^2=0.1$ and then show the required $R^{-1}$
for any given $\Delta_i$, for each class of KK particles. 
We show variations of the masses of
one (red dotted) or three (red solid) generations of $SU(2)_W$-singlet
KK leptons; three generations of $SU(2)_W$-doublet leptons (magenta);
three generations of $SU(2)_W$-singlet quarks (blue)
(the result for three generations of $SU(2)_W$-doublet quarks
is almost identical); KK gluons (cyan) and
electroweak KK gauge bosons (green). The circle on each
line denotes the MUED values of $\Delta$ and $R^{-1}$.

\FIGURE[ht]{
\epsfig{file=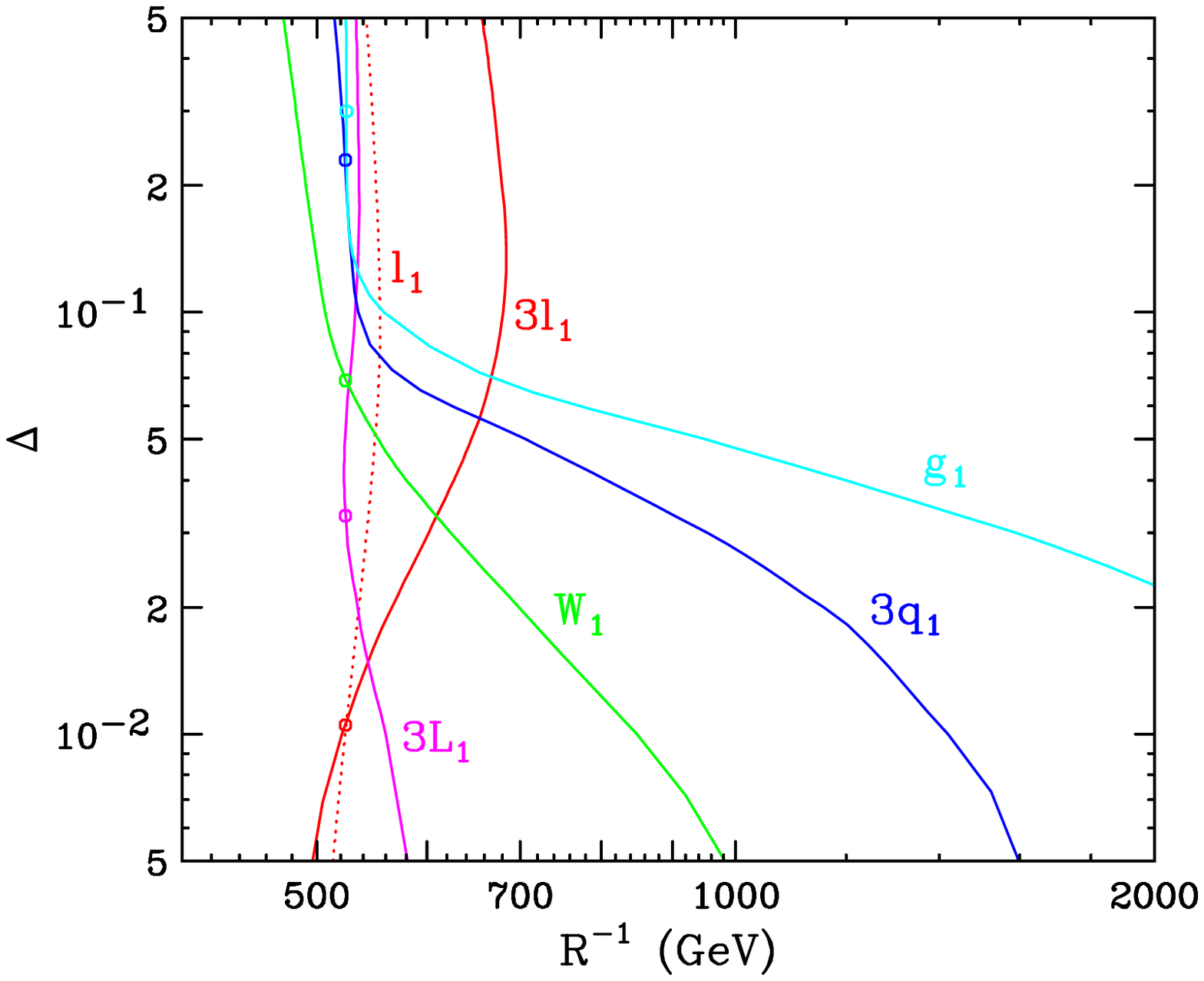,width=10.0cm}
\caption{\sl The change in the cosmologically preferred value for $R^{-1}$
as a result of varying the different KK masses away from their nominal 
MUED values. Along each line, the LKP relic density is $\Omega_\chi h^2=0.1$.
To draw the lines, we first fix the MUED spectrum, and then vary the 
corresponding KK mass and plot the value of $R^{-1}$ which is required
to give $\Omega_\chi h^2=0.1$. We show variations of the masses of
one (red dotted) or three (red solid) generations of $SU(2)_W$-singlet
KK leptons; three generations of $SU(2)_W$-doublet leptons (magenta);
three generations of $SU(2)_W$-singlet quarks (blue)
(the result for three generations of $SU(2)_W$-doublet quarks
is almost identical); KK gluons (cyan) and
electroweak KK gauge bosons (green). The circle on each
line denotes the MUED values of $\Delta$ and $R^{-1}$.
\label{fig:finger}
}
}

Fig.~\ref{fig:finger} summarizes our results from Section~\ref{sec:coann}.
It also provides a quick reference guide for
the expected variations in the predicted value of $\Omega h^2$
as we move away from the Minimal UED model. 
For example, it is clear that unlike the case of 
coannihilations with $\ell_{R1}$, which was
considered  in \cite{Servant:2002aq}, coannihilations with
all other KK particles will lower the prediction for $\Omega h^2$ 
and correspondingly increase the preferred range of $R^{-1}$. 
This is due to the larger couplings of those particles.
Fig.~\ref{fig:finger} can also be used to quantitatively estimate
the variations in the preferred value of $R^{-1}$ in non-minimal
models.

On a final note, in the non-minimal UED model, other neutral KK particles 
such as $Z_1$ can also be dark matter candidates. 
On dimensional grounds, the relic density is inversely proportional 
to the square of the LKP mass,
\begin{eqnarray}
\Omega {h^2} &\sim&  \frac{g_1^4}{m_{\gamma_1}^2}\ , \\
\Omega {h^2} &\sim&  \frac{g_2^4}{m_{Z_1}^2}\ .
\end{eqnarray}
Due to the larger coupling $g_2$ of the $SU(2)_W$ gauge interactions, 
we expect the upper bound on $m_{Z_1}$, consistent with WMAP, 
to be larger than the bound on $m_{\gamma_1}$ roughly
by a factor of $g^2_2/g^2_1 \sim 3$. However, in the $Z_1$ LKP case, 
$SU(2)_W$ symmetry implies that the charged $W_1$ states are almost 
degenerate with $Z_1$, and therefore coannihilations with $W_1^\pm$ 
will be very important and will need to be considered. 
The analysis of the cases of $Z_1$ and $H_1$ LKP and their detection 
prospects is currently in progress~\cite{KM}.
The results presented in this paper are also relevant for the case
of KK graviton superwimps~\cite{Feng:2003xh,Feng:2003uy,Feng:2003nr},
whose relic density is still determined by the freeze-out of the 
next-to-lightest KK particle.

In conclusion, dark matter candidates from theories with extra dimensions 
should be considered on an equal footing with more conventional 
candidates such as SUSY dark matter or axions. The framework of 
Universal Extra Dimensions provides a useful playground for
gaining some experience about the signals one could expect
from extra dimensional dark matter. If extra dimensions have indeed
something to do with the dark matter problem, the explicit realization of 
that idea may look quite differently (see for 
example~\cite{Byrne:2003sa,Agashe:2004ci,Agashe:2004bm}),
especially if one wants to resolve the radion stabilization problem
\cite{Kolb:2003mm,Mazumdar:2003vg}.
Nevertheless, we believe that the methods and insight we developed 
in this paper will prove useful in more general contexts.

\bigskip

{\bf Note added:} During the completion of this work we became aware of
an analogous calculation done independently by another group~\cite{BurKri}.
We have compared extensively the formulas for the annihilation cross-sections 
involving KK quarks, KK leptons and KK gauge bosons, in the limit of degenerate 
KK masses, as listed in the Appendix. In all considered cases we found perfect 
agreement.

\bigskip

\acknowledgments
We are grateful to F.~Burnell and G.~Kribs for extensively comparing their results
with ours.
We are grateful to A.~Birkedal, G.~Servant and T.~Tait for extensive discussions 
and/or correspondence regarding the calculation of annihilation cross-sections in UED.
This work is supported in part by a US Department of Energy Outstanding Junior 
Investigator award under grant DE-FG02-97ER41209.

\appendix
%%%%%%%%%%%%%%%%%%%%%%%%%%%%%%%%%%%%%%%%%%%%%%%%%%%%%%%%%%%%%%%
\section{Annihilation Cross-Sections}
\label{app:xsec}
%\addcontentsline{toc}{section}{Appendix:  \ Annihilation Cross-Sections}
\renewcommand{\theequation}{A.\arabic{equation}}
\setcounter{equation}{0}

In this section, we summarize the annihilation cross 
sections of any pair of $n=1$ KK particles into SM 
fields in the limit of no electroweak symmetry breaking,
as in \cite{Servant:2002aq}. In order to render 
the formulas manageable for publication, in this Appendix we
list our results in the limit of equal KK masses. 
However, in our numerical calculation, we kept
different masses for all KK particles, which often
leads to enormously complicated analytical expressions.
We also assume all SM particles to be massless, since we
are working in the limit where we neglect electroweak 
symmetry breaking (EWSB) effects of order $vR$, where
$v$ is the Higgs vacuum expectation value in the SM.
All cross-sections are calculated at tree level. 
All vertices satisfy KK-number conservation and KK-parity 
since KK-number violating interactions are only induced 
at the loop level \cite{Cheng:2002iz}. 
Some of the cross-sections have already appeared in \cite{Servant:2002aq} 
and we find perfect agreement with those results.
We define a few constants below which are commonly used in our 
formulas for the cross-sections.
\begin{eqnarray}
g_1 &=& \frac{e}{c_w} \ , \\
g_2 &=& \frac{e}{s_w} \ , \\
g_Z &=& \frac{e}{2 s_w c_w} \ , \\
\beta &=& \sqrt{1-\frac{4 m^2}{s}} \ , \\
L &=& \log \left ( \frac{1-\beta}{1+\beta} \right ) = -2 \tanh^{-1}\beta \ .
\end{eqnarray}
Here $g_1$, $g_2$ and $g_3$ are the gauge couplings of $U(1)_Y$, $SU(2)_W$ and $SU(3)$. 
$\beta$ is the velocity of the incoming KK particle in the annihilation process. 
Notice that $L$ is negative since $0 < \beta < 1$. $m$ is the KK mass 
which for the purposes of this appendix is the same for all KK particles, 
$e$ is the electric charge and $c_w$ and $s_w$ are the cosine and sine
of the Weinberg angle in the SM. Table~\ref{tab:xsection}
provides a quick reference guide for the different process types.

\TABULAR{|c|c|c|c|c|}{
\hline 
               &    Gauge bosons     &    Leptons      &   Quarks       &   Higgses     \\  \hline \hline
Gauge bosons   &        A.2          &                 &                &               \\  \hline
Leptons        &        A.3          &       A.1       &                &               \\  \hline
Quarks         &        A.3          &       A.5       &      A.4       &               \\  \hline
Higgses        &        A.7          &       A.8       &      A.8       &       A.6     \\  \hline
}{\label{tab:xsection} A guide to the formulas in the Appendix. Each box in the table corresponds to
a particular type of an initial state. The entry points to the section in the Appendix where the 
corresponding annihilation cross-sections can be found.
Here ``gauge bosons'' include EW KK gauge bosons ($W_1^\pm$, $Z_1$ and $\gamma_1$) and 
the KK gluon ($g_1$). 
``Higgses'' stands for the KK Higgs ($H_1$) and KK Goldstone bosons ($G_1^\pm$ and $G_1$).
Leptons contain both $SU(2)_W$-singlet KK leptons ($\ell_{R1}$) and 
$SU(2)_W$-doublet KK leptons ($\ell_{L1}$ and $\nu_{\ell_1}$).
Quarks include both $SU(2)_W$-doublet KK quarks ($q_{R1}$) and 
$SU(2)_W$-singlet KK quarks ($q_{L1}$).}

\subsection{Leptons}
Coannihilations with $SU(2)_W$-singlet KK leptons 
$\ell_{R1}$ are important since they are expected 
to be the next-to-lightest KK particles in the 
Minimal UED model~\cite{Servant:2002aq, Cheng:2002iz}.
For fermion final states with $f\ne \ell$, the cross-section is 
\beq
\sigma ( \ell_{R1}^{+} {\ell}_{R1}^{-} \rightarrow f \bar{f} )
\ =\ \frac{N_c g_1^4 Y_{\ell}^2 ( Y_{f_L}^2 + Y_{f_R}^2 )(s+2m^2)}{24 \pi\beta s^2} \ ,
\label{llff}
\eeq
where $N_c$ is 3 for quarks and 1 for leptons. For cases with
the same lepton flavor in the initial and final state, we have
\begin{eqnarray}
\sigma ( \ell_{R1}^{+} \ell_{R1}^{-} \rightarrow \ell^+ \ell^- )
&=& \frac{g_1^4 Y_{\ell_{R1}}^4 (5\beta s + 2(2s+3m^2)L)}{32 \pi \beta^2 s^2} \nonumber \\
&+& \frac{g_1^4 Y_{\ell_{R1}}^4 (\beta (4s+9m^2) + 8m^2 L)}{64 \pi m^2 \beta^2 s}  \\
&+& \frac{g_1^4 Y_{\ell_{R1}}^2 (Y_{\ell_R}^2+Y_{\ell_L}^2) (s+2m^2 )}{24 \pi \beta s^2} \ , \nonumber
\end{eqnarray}
\beq
\sigma ( \ell_{R1}^{\pm} \ell_{R1}^{\pm} \rightarrow \ell^\pm \ell^\pm )
\ =\ \frac{g_1^4 Y_{\ell}^4 ( -m^2(4s-5m^2)L - \beta s (2s-m^2))}{32 \pi \beta^2 s^2 m^2} \ ,
\eeq
\beq
\sigma ( \ell_{R1}^\pm {\ell'}_{R1}^\pm \rightarrow \ell^\pm {\ell'}^{\pm} )
\ =\ \frac{g_1^4 Y_\ell^4 (4s-3m^2)}{64 \pi \beta s m^2}  \ ,
\eeq
\beq
\sigma ( \ell_{R1}^\pm {\ell'}_{R1}^\mp \rightarrow \ell^\pm {\ell'}^\mp )
\ =\ \frac{g_1^4 Y_\ell^4 ( \beta (4s+9m^2) + 8m^2 L )}{64 \pi \beta^2 s m^2} \ , 
\eeq
where $\ell$ and $\ell'$ are the leptons from different families.
For the remaining final states we get
\beq
\sigma ( \ell_{R1}^{+} {\ell}_{R1}^{-} \rightarrow \phi \phi^* )
\ =\ \frac{g_1^4 Y_{\ell}^2 Y_{\phi}^2 (s+2m^2)}{48\pi \beta s^2} \ ,
\eeq
\beq
\sigma ( \ell_{R1}^{+} {\ell}_{R1}^{-} \rightarrow B_0 B_0 )
\ =\ \frac{g_1^4 Y_{\ell}^4 ( 2(s^2+4m^2 s-8m^4) - \beta s (s+4m^2))}{8 \pi \beta^2 s^3} \ .
\label{llbb}
\eeq
Our results, (\ref{llff}-\ref{llbb}), exactly agree with (C.1) - (C.8) from \cite{Servant:2002aq}.

The cross-sections among left handed fermions are somewhat complicated since 
they involve $SU(2)_W$ gauge bosons as well as $U(1)_Y$ gauge bosons. For KK neutrinos we find
\beq
\sigma ( \nu_{\ell 1} \bar{\nu}_{\ell 1} \rightarrow f \bar{f} )
\ = \ \frac{N_c g_Z^2 (\bar{g}_L^2 + \bar{g}_R^2) (s+2m^2)}{24 \pi \beta s^2} \ ,
\label{nunuff}
\eeq
\beq
\sigma ( \nu_{\ell 1} \bar{\nu}_{\ell 1} \rightarrow \phi \phi^* )
\ = \ \frac{g_\phi^2 g_Z^2 (s+2m^2)}{48 \pi \beta s^2}  \ ,
\eeq
\beq
\sigma ( \nu_{\ell 1} \bar{\nu}_{\ell 1} \rightarrow Z Z  )
\ = \ \frac{g_Z^4 ((8m^4-s^2-4m^2 s)L - \beta s (s+4m^2))}{8\pi \beta^2 s^3} \ ,
\eeq
\begin{eqnarray}
\sigma ( \nu_{\ell 1} \bar{\nu}_{\ell 1} \rightarrow W^+ W^-  )
&=& - \frac{5 g_2^4 (s+2m^2)}{96 \pi \beta s^2} 
    + \frac{g_2^4 (\beta s -2m^2 L)}{32\pi\beta^2 s^2} \nonumber \\
&-& \frac{g_2^4 (\beta(s+4m^2) + (s+2m^2)L)}{32\pi\beta^2 s^2} \ , 
\end{eqnarray}
\beq
\sigma ( \nu_{\ell 1} \nu_{\ell 1} \rightarrow \nu_\ell \nu_\ell   )
\ = \ \frac{g_Z^4 ( \beta s (2s -m^2) - m^2 (4s-5m^2) L) }{32 \pi \beta^2 s^2 m^2} \ ,
\eeq
\beq
\sigma ( \nu_{\ell 1} \nu_{\ell' 1} \rightarrow \nu_\ell \nu_{\ell '}  )
\ = \ \frac{g_Z^4 (4s - 3m^2)}{64 \pi \beta s m^2} \ , 
\eeq
\beq
\sigma ( \nu_{\ell 1} \bar{\nu}_{\ell' 1} \rightarrow \nu_\ell \bar{\nu}_{\ell '}  )
\ = \ \frac{g_Z^4 ( \beta (4s + 9m^2) + 8m^2 L)}{64 \pi \beta^2 s m^2} \ , 
\eeq
\beq
\sigma ( \nu_{\ell 1} \bar{\nu}_{\ell' 1} \rightarrow \ell^- {\ell '}^+  )
\ = \ \frac{g_2^4 ( \beta (4s + 9m^2) + 8m^2 L)}{256 \pi \beta^2 s m^2} \ , 
\eeq
\begin{eqnarray}
\sigma ( \nu_{\ell 1} \bar{\nu}_{\ell 1} \rightarrow \ell^+ {\ell}^-  )
&=& \frac{g_Z \hat{g}_L^2 \bar{g}_L (5\beta s + 2(2s+3m^2)L)}{32\pi \beta^2 s^2} \nonumber \\
&& \hspace{-2cm}+ \frac{g_Z^2 (\bar{g}_L^2 + \bar{g}_R^2 ) (s+2m^2)}{24\pi \beta s^2} 
     + \frac{ \hat{g}_L^4 ( \beta (4s+9m^2)+ 8m^2 L)}{64\pi m^2 \beta^2 s} \ .
\end{eqnarray}
Here $\bar{g}_{L(R)} = \frac{e}{s_w c_w} \left ( T^3 - Q_f s_w^2 \right )$, $\hat{g}_L = g_Z$ for neutrinos and
$\hat{g}_L = g_2/\sqrt{2} $ for charged leptons. 
$g_\phi = \frac{e}{s_w c_w} \left ( T^3 - Q_\phi s_w^2 \right )$ with $Q_\phi = 1$ 
for the upper entry in the Higgs doublet and $Q_\phi = 0$ for the lower entry. 
Since we ignore EWSB, all gauge bosons have transverse polarizations only.
$\phi$ represents either a charged Higgs boson ($\phi_u$, isospin $\frac{1}{2}$) or 
a neutral Higgs boson ($\phi_d$, isospin $-\frac{1}{2}$).

The previous results allow us to immediately obtain
\begin{eqnarray}
\sigma ( \nu_{\ell 1} \bar{\nu}_{\ell 1} \rightarrow \phi \phi^* )
&=& \sigma ( \ell_{L1}^+ \ell_{L1}^- \rightarrow \phi \phi^* ) \nonumber \ , \\
\sigma ( \nu_{\ell 1} \bar{\nu}_{\ell 1} \rightarrow Z Z  )
&=& \sigma ( \ell_{L1}^+ {\ell}_{L1}^- \rightarrow Z Z + Z\gamma + \gamma\gamma  ) \nonumber \ , \\
\sigma ( \nu_{\ell 1} \bar{\nu}_{\ell 1} \rightarrow W^+ W^-  )
&=& \sigma ( \ell_{L1}^+ \ell_{L1}^- \rightarrow W^+ W^-  )\nonumber \ , \\
\sigma ( \nu_{\ell 1} \nu_{\ell 1} \rightarrow \nu_\ell \nu_\ell   )
&=& \sigma ( \ell_{L1}^\pm \ell_{L1}^\pm  \rightarrow \ell^\pm \ell^\pm   ) \ , \\
\sigma ( \nu_{\ell 1} \nu_{\ell' 1} \rightarrow \nu_\ell \nu_{\ell '}  )
&=& \sigma ( \nu_{\ell 1} \ell'_{L1} \rightarrow \nu_\ell \ell '  ) \nonumber  \\
&=& \sigma ( \ell_{L1}^\pm {\ell'}_{L1}^\pm \rightarrow \ell^\pm {\ell '}^\pm  ) \nonumber \ ,\\
\sigma ( \nu_{\ell 1} \bar{\nu}_{\ell' 1} \rightarrow \nu_\ell \bar{\nu}_{\ell '}  )
&=& \sigma ( \ell_{L1}^\pm {\ell'}_{L1}^\mp \rightarrow \ell^\pm {\ell '}^\mp  ) \nonumber \ .
\end{eqnarray}

For at least one charged KK lepton in the initial state we get
\beq
\sigma ( \ell_{L1}^+ {\ell}_{L1}^- \rightarrow f \bar{f} \,\, or \,\, \ell^+_R \ell^-_R )
\ = \ \frac{N_c g^4 (s+2m^2)}{24 \pi \beta s^2} \ ,
\eeq
\begin{eqnarray}
\sigma ( \ell_{L1}^+ {\ell}_{L1}^- \rightarrow  \nu_\ell \bar{\nu}_\ell \,\, or \,\, \ell^+_L \ell^-_L)
&=&  \frac{\hat{g}_L^2 g^2 (5\beta s + 2(2s+3m^2)L)}{32 \pi \beta^2 s^2}  \nonumber \\
&+& \frac{\hat{g}_L^4 (\beta (4s+9m^2) + 8m^2 L)}{64 \pi m^2 \beta^2 s}
   + \frac{g^4 (s+2m^2)}{24 \pi \beta s^2} \ ,
\end{eqnarray}
\beq
\sigma ( \ell_{L1}^- \bar{\nu}_{L1} \rightarrow f \bar{f}')
\ = \ \frac{N_c g_2^4 (s+2m^2)}{96\pi \beta s^2} \ ,
\eeq
\beq
\sigma ( \ell_{L1}^- \bar{\nu}_{L1} \rightarrow \phi^*_u \phi_d)
\ = \ \frac{g_2^4 (s+2m^2)}{192 \pi \beta s^2} \ ,
\eeq
\beq
\sigma ( \ell_{L1}^- \bar{\nu}_{L1} \rightarrow W^- B_0 )
\ = \ \frac{g_1^2 g_2^2 ((8m^4-s^2-4m^2 s)L - \beta s (s+4m^2))}{32 \pi \beta^2 s^3} \ ,
\eeq
\begin{eqnarray}
\sigma ( \ell_{L1}^- \bar{\nu}_{L1} \rightarrow W^- W^0_3) 
&=& - \frac{5 g_2^4 (s+2m^2)}{48 \pi s^2 \beta} + \frac{g_2^4 (\beta s - 2m^2 L)}{32 \pi s^2 \beta^2} \nonumber \\
&-&  \frac{g_2^4 (\beta (s+4m^2) + (s+2m^2)L)}{64 \pi s^2 \beta} + \frac{g_2^4 m^2 L}{16 \pi s^2} \ ,
\end{eqnarray}
\begin{eqnarray}
\sigma ( \ell_{L1}^- \bar{\nu}_{L1} \rightarrow \ell^- \bar{\nu}_\ell ) 
&=&  \frac{g_1 g_2^3 (5\beta s + 2(2s+3m^2)L)}{64 \pi \beta^2 s^2 (2 s_w^2 -1)}  \nonumber \\
&+& \frac{g_1^2 g_2^2 (\beta (4s+9m^2) + 8m^2 L)}{64 \pi m^2 \beta^2 s (2 s_w^2 -1)^2}
   + \frac{g_2^4 (s+2m^2)}{96 \pi \beta s^2} \ ,
\end{eqnarray}
\begin{eqnarray}
\sigma ( \ell_{L1}^- \nu_{L1} \rightarrow \ell^- \nu_\ell) 
&=& \frac{g_1 g_2^3 (- 2 m^2 (4s-5m^2)L + m^2 \beta s)}{64 \pi \beta^2 s^2 m^2 (2 s_w^2 -1)} \nonumber \\
&+& \frac{\beta s (4s-3m^2)}{64\pi\beta^2 s^2 m^2} \left ( \frac{g_1^2 g_2^2}{(2s_w^2-1)^2} 
             + \frac{g_2^4}{4} \right ) \ , 
\end{eqnarray}
\beq
\sigma ( \nu_{L1} \ell'_{L1} \rightarrow \nu_{\ell '} \ell^- ) 
\ = \ \frac{g_2^4 (4s-3m^2)}{256 \pi m^2 s \beta} \ , 
\eeq
\beq
\sigma ( \nu_{L1} \ell'_{L1} \rightarrow \nu_{\ell} \ell'^- ) 
\ = \ \frac{g_1^2 g_2^2 (4s-3m^2)}{64 \pi m^2 s \beta (2 s_w^2 -1)^2} \ , 
\eeq
\beq
\sigma ( \bar{\nu}_{L1} \ell'_{L1} \rightarrow \bar{\nu}_{\ell} \ell'^- ) 
\ = \ \frac{g_1^2 g_2^2 ( \beta (4s+9m^2) + 8m^2 L)}{64 \pi m^2 s \beta^2 (2 s_w^2-1)^2} \ ,
\eeq
\beq
\sigma ( \ell_{L1} \bar{\ell}'_{L1} \rightarrow \nu_\ell \bar{\nu}_{\ell '} ) 
\ = \ \frac{g_2^4 ( \beta (9m^2+4s) + 8m^2 L)}{256 \pi m^2 s \beta^2} \ ,
\label{llpnunup}
\eeq
where $g^2 = g_1^2 Y_f Y_{\ell_{R}} + g_2^2 T_f^3 T^3_{\ell_{L}} $.
The above cross-sections, (\ref{nunuff}-\ref{llpnunup}) 
are consistent with (B.48) - (B.62) and (B.71) - (B.74) in \cite{Servant:2002aq}.

For one $SU(2)_W$-singlet KK lepton and one $SU(2)_W$-doublet KK lepton we get
\beq
\sigma ( \ell_{R1} \ell_{L1} \rightarrow \ell \ell )
\ = \ \frac{g_1^4 Y^2_{e_L} Y^2_{e_R}}{64 \pi m^2 s \beta^2} \left ( 8m^2 L + \beta (9m^2+4s) \right ) \ ,
\eeq
\beq
\sigma ( \ell_{R1} \bar{\ell}_{L1} \rightarrow \ell \bar{\ell} )
\ = \ \frac{g_1^4 Y^2_{e_L} Y^2_{e_R}}{64 \pi m^2 s \beta} \left ( 4s-3m^2  \right ) \ .
\eeq
These two formulas have the same structure as 
$\sigma ( \ell_{R1}^\pm {\ell'}_{R1}^\pm \rightarrow \ell^\pm {\ell'}^{\pm} )$
and $\sigma ( \ell_{R1}^\pm {\ell'}_{R1}^\mp \rightarrow \ell^\pm {\ell'}^\mp )$.

\subsection{Gauge bosons}

The self-annihilation cross-sections of $\gamma_1$ are
\beq
\sigma ( \gamma_1 \gamma_1 \rightarrow f \bar{f} )
\ = \ \frac{N_c g_1^4 (Y_{f_L}^4 + Y_{f_R}^4)}{72 \pi s^2 \beta^2 } \left ( -5s (2m^2+s)L-7s\beta \right ) \ ,
\eeq
\beq
\sigma ( \gamma_1 \gamma_1 \rightarrow \phi \phi^* )
\ =  \frac{g_1^4 Y_\phi^4}{12 \pi s \beta} \ .
\eeq
These two cross-sections are identical to (A.44) and (A.47) in \cite{Servant:2002aq}. 
For $Z_1$ self-annihilation into fermions and Higgs bosons,
\beq
\sigma ( Z_1 Z_1 \rightarrow f \bar{f} )
\ = \ \frac{N_c g_2^4}{1152 \pi s^2 \beta^2} \left ( -5 (2m^2 + s) L - 7 s \beta \right ) \ ,
\eeq
\beq
\sigma ( Z_1 Z_1 \rightarrow \phi \phi^* )
\ = \ \frac{g_2^4}{192 \pi s \beta} \ .
\eeq
The cross-section for the above two processes are obtained from 
$\sigma ( \gamma_1 \gamma_1 \rightarrow f \bar{f} )$
and $\sigma ( \gamma_1 \gamma_1 \rightarrow \phi \phi^* )$ 
by replacing $g_1 Y$ with $g_2/2$, which corresponds to 
the $Z$ couplings to SM fermions and Higgs bosons. 
For the coannihilations of $SU(2)_W$ KK bosons into SM gauge bosons, we get
\beq
\sigma ( Z_1 Z_1 \rightarrow W^+ W^- )
\ = \ \frac{g_2^4}{18\pi m^2 s^3 \beta^2} \left ( 12m^2 (s-2m^2 )L + s \beta (12m^4+3s m^2+4s^2) \right )  ,
\eeq
\beq
\sigma ( W^+_1 W^+_1 \rightarrow W^+ W^+ )
\ = \ \frac{g_2^4}{36 \pi m^2 s^3 \beta^2} \left ( 12m^4 (s-2m^2)L + s\beta (12m^4+3sm^2+4s^2) \right )  ,
\eeq
\beq
\sigma ( W^+_1 W^-_1 \rightarrow \gamma\gamma )
\ = \ \frac{e^4}{36 \pi m^2 s^3 \beta^2} \left ( 12m^4 (s-2m^2)L + s\beta (12m^4+3sm^2+4s^2) \right ) \  ,
\eeq
\beq
\sigma ( W^+_1 W^-_1 \rightarrow \gamma Z )
\ =\ \frac{g_2^2 e^2 c_w^2}{18 \pi m^2 s^3 \beta^2} \left ( 12m^4 (s-2m^2)L + s\beta (12m^4+3sm^2+4s^2) \right )\  ,
\eeq
\beq
\sigma ( W^+_1 W^-_1 \rightarrow Z Z )
\ = \ \frac{g_2^4 c_w^4}{36 \pi m^2 s^3 \beta^2} \left ( 12m^4 (s-2m^2)L + s\beta (12m^4+3sm^2+4s^2) \right ) \  .
\eeq
We see that the above five cross-sections contain similar 
expressions up to overall factors due to the gauge structure of 
$SU(2)_W \times U(1)_Y$. For $W_1^\pm$ annihilation into other final states, we have
\beq
\sigma ( W^+_1 W^-_1 \rightarrow f \bar{f} )
\ = \ \frac{-N_c g_2^4}{576 \pi s^2 \beta^2} \left ( (12m^2+5s)L + 2\beta (4m^2+5s) \right ) \ ,
\eeq
\beq
\sigma ( W^+_1 W^-_1 \rightarrow W^+W^- )
\ = \ \frac{g_2^4}{18\pi m^2 s^2 \beta^2} \left ( 2m^2 (3m^2+2s)L + \beta (11m^4+5s m^2+2s^2) \right ) \ ,
\eeq
\beq
\sigma ( W^+_1 W^-_1 \rightarrow \phi \phi^* )
\ = \ \frac{g_2^4 (s-m^2)}{144 \pi s^2 \beta}\ .
\eeq
These three cross-sections are different since they involve $s$-channel $Z$ diagrams. 
For $\gamma_1 Z_1$ and $\gamma_1 W^-_1$ into fermions we can recycle
$\sigma ( \gamma_1 \gamma_1 \rightarrow f \bar{f} )$ and obtain
\beq
\sigma ( \gamma_1 Z_1 \rightarrow f \bar{f} )
\ = \ \frac{-N_c g_1^2 g_2^2 Y_f^2}{288 \pi s^2 \beta^2} \left ( 5 (2m^2+s)L+7s\beta \right ) \ ,
\eeq
\beq
\sigma ( \gamma_1 W^-_1 \rightarrow f \bar{f'} )
\ = \ \frac{-N_c g_1^2 g_2^2 Y_f^2}{144 \pi s^2 \beta^2} \left ( 5 (2m^2+s)L+7s\beta \right ) \ .
\eeq
For the annihilation of two different KK gauge bosons into Higgs bosons we have
\beq
\sigma ( \gamma_1 Z_1 \rightarrow \phi \phi^* ) 
\ = \ \frac{g_1^2 g_2^2}{192 \pi s \beta} \ ,
\eeq
\beq
\sigma ( \gamma_1 W^-_1 \rightarrow \phi_d \phi_u^* ) 
\ = \ \frac{g_1^2 g_2^2}{96 \pi s \beta} \ ,
\eeq
\beq
\sigma  ( Z_1 W^-_1 \rightarrow \phi_d \phi_u^* ) 
\ = \ \frac{g_2^4 \beta}{288\pi s}         \ ,
\eeq
which can be obtained from $\sigma ( \gamma_1 \gamma_1 \rightarrow \phi \phi^* )$.
The cross-section for $Z_1 W^-_1$ into fermions
\beq
\sigma  ( Z_1 W^-_1 \rightarrow f \bar{f'} )
\ = \ \frac{-N_c g_2^4}{576\pi s^2 \beta^2} \left ( (14m^2+5s) L + \beta (16m^2+13s) \right ) \ ,
\eeq
has a different structure compared to other fermion final states due to 
an s-channel $W$ diagram. The cross-sections for 
$Z_1 W^-_1$ into gauge boson final states
\beq
\sigma  ( Z_1 W^-_1 \rightarrow Z W^- )
\ = \ \frac{g_2^4 c_w^2}{18\pi m^2 s^2 \beta^2} \left ( 2m^2 (3m^2+2s)L + \beta (11m^4 +5 s m^2 +2s^2) \right ) \ ,
\eeq
\beq
\sigma  ( Z_1 W^-_1 \rightarrow \gamma W^- )
\ = \ \frac{e^2 g_2^2 }{18\pi m^2 s^2 \beta^2} \left ( 2m^2 (3m^2+2s)L + \beta (11m^4 +5 s m^2 +2s^2) \right ) \ ,
\eeq
can be obtained from $\sigma ( W^+_1 W^-_1 \rightarrow W^+W^- )$. For KK gluons we get 
\beq
\sigma ( g_1 g_1 \rightarrow g g )
\ = \ \frac{g_3^4}{64\pi m^2 s^3\beta^2} \left ( 8m^2 (s^2+3sm^2-3m^4)L + s\beta (34m^2 +13sm^2+8s^2) \right ) \ ,
\eeq
\beq
\sigma ( g_1 g_1 \rightarrow q\bar{q} )
\ = \ \frac{-g_3^4}{3456\pi s^2\beta^2} \left ( 2 (20s+49m^2)L + \beta (72m^2+83s) \right ) \ ,
\eeq
for which there are no analogous processes. 
The cross-sections associated with one gluon and one electroweak 
gauge bosons in the initial state
\beq
\sigma ( g_1 \gamma_1 \rightarrow q\bar{q} )
\ = \ \frac{g_1^2 g_3^2 (Y^2_{q_L} + Y^2_{q_R})}{144\pi s^2 \beta^2} \left ( -5(2m^2+s)L-7s\beta \right ) \ ,
\eeq
\beq
\sigma ( g_1 Z_1 \rightarrow q\bar{q} )
\ = \ \frac{g_2^2 g_3^2 }{576\pi s^2 \beta^2} \left ( -5(2m^2+s)L-7s\beta \right ) \ ,
\eeq
\beq
\sigma ( g_1 W^-_1 \rightarrow q\bar{q'} )
\ = \ \frac{g_2^2 g_3^2 }{288\pi s^2 \beta^2} \left ( -5(2m^2+s)L-7s\beta \right ) \ ,
\eeq
are obtained from 
$\sigma ( \gamma_1 \gamma_1 \rightarrow f \bar{f} )$, $\sigma ( \gamma_1 Z_1 \rightarrow f \bar{f} )$ and 
$\sigma ( \gamma_1 W_1^- \rightarrow f \bar{f'} )$ by simple coupling replacements
and accounting for the additional color factors.

\subsection{Fermions and Gauge Bosons}

Note that in UED, the Dirac KK fermions are constructed out 
of two Weyl fermions with the same $SU(2)_W \times U(1)_Y$ quantum numbers 
while a Dirac fermion in the Standard Model is made up of
two Weyl fermions of different $SU(2)_W \times U(1)_Y$ quantum numbers. 
Therefore the couplings of KK fermions to zero mode gauge bosons are vector-like. 
This difference shows up in processes involving gauge-boson couplings with fermions.
For the vertices which involve $n=1$ gauge bosons, we need one KK fermion and one SM fermion 
in order to conserve KK number. In this case, there is always a projection operator associated 
with the subscript ($L/R$) of the KK fermion. The annihilation cross-sections with 
$SU(2)_W$-singlet KK fermions and $SU(2)_W$ KK gauge bosons ($Z_1$ and $W_1^{\pm}$) are 
zero:
\begin{eqnarray}
\sigma ( f_{R1} Z_1 \rightarrow SM )
&=& 0\ , \nonumber \\
\sigma ( f_{R1} W^{\pm}_1 \rightarrow SM )
&=& 0\ .
\end{eqnarray}
The cross-section for $SU(2)_W$-singlet leptons and $\gamma_1$ is
\beq
\sigma ( \gamma_1 \ell_{R1}^\pm \rightarrow B_0 \ell^\pm )
\ =\ \frac{Y_{\ell}^4 g_1^4 (-6L-\beta)}{96 \pi s \beta^2} \ .
\eeq
For a KK quark and $\gamma_1$ we have
\beq
\sigma ( q_1 \gamma_1 \rightarrow q g )
\ = \ \frac{g_1^2 g_3^2 Y_{q_1}^2}{72 \pi s\beta^2} \left ( -6L-\beta \right ) \ .
\eeq
This cross-section is basically the same as $\sigma ( \gamma_1 \ell_{R1}^\pm \rightarrow B_0 \ell^\pm )$,
up to a group factor. In $\sigma ( q_1 \gamma_1 \rightarrow q g )$, the vertex associated with 
the SM gluon $g$ contains a Gell-Mann matrix $t_{ij}^a$. 
In the squared matrix element, we then get \cite{Peskin:1995ev}
\beq
\sum_{a=1}^{8} \frac{1}{3}\sum_{i,j = 1}^{3} t^a_{ij} t^a_{ji} 
          = \frac{1}{3} \times \frac{4}{3} \times 3 = \frac{4}{3} \ .
\eeq
Similarly, for the $SU(2)_W$-doublet quarks with $Z_1$, we get
\beq
\sigma ( q_{L1} Z_1 \rightarrow g q )
\ = \ \frac{g_2^2 g_3^2}{288\pi s \beta^2} \left ( -6 L -\beta \right ) \ ,
\eeq
by replacing the $g_1$ coupling in $\sigma ( q_1 \gamma_1 \rightarrow q g )$
with $g_2/2$. For $W_1^\pm$, one should use the coupling $g_2/\sqrt{2}$ instead:
\beq
\sigma ( q_{L1} W_1 \rightarrow g q' )
\ = \ \frac{g_2^2 g_3^2}{144\pi s \beta^2} \left ( -6 L -\beta \right ) \ .
\eeq
For the $SU(2)_W$-doublet KK leptons and $\gamma_1$ or $Z_1$, we get
\beq
\sigma ( \ell_{L1} \gamma_1 \rightarrow \ell_L \gamma / Z )
\ = \ \frac{g_1^4}{1536 \pi s \beta^2 s_w^2} \left ( -6 L -\beta \right ) \ ,
\eeq
\beq
\sigma ( \ell_{L1} Z_1 \rightarrow \ell_L \gamma / Z )
\ = \ \frac{g_2^4}{1536 \pi s \beta^2 c_w^2} \left ( -6 L -\beta \right ) \ ,
\eeq
\beq
\sigma ( \ell_{L1} \gamma_1 \rightarrow \nu_{\ell} W )
\ = \ \frac{g_1^2 g_2^2}{768 \pi s \beta^2} \left ( -6 L -\beta \right ) \ .
\eeq
The last 7 cross-sections have a similar structure 
since they all have $s$- and $t$-channel diagrams only.
For the cross-sections associated with $SU(2)_W$-doublet KK leptons and 
electroweak KK gauge bosons into other final states, we have
\beq
\sigma ( \ell_{L1} Z_1 \rightarrow \nu_{\ell} W )
\ = \ \frac{g_2^4}{768 \pi m^2 s \beta^2} \left ( 26 m^2 L + \beta(23m^2 +32s) \right ) \ ,
\eeq
\begin{eqnarray}
\sigma ( \ell_{L1} W_1 \rightarrow \nu_{\ell '} (\gamma + Z) )
&=& \frac{g_2^4}{768 \pi m^2 s \beta^2 c_w^2} \left ( m^2 (32c_w^2-6)L  \nonumber \right . \\ 
&& \hspace{1cm }\left . +\beta m (24c_w^2-1) +32s\beta c_w^2 \right ) \ ,
\end{eqnarray}
\beq
\sigma ( \ell_{L1} W_1^- \rightarrow \ell_L W^- )
\ = \ \frac{g_2^4}{192 \pi m^2 s \beta^2} \left ( -3m^2 L + 4s\beta  \right ) \ ,
\eeq
\beq
\sigma ( \ell_{L1} W_1^+ \rightarrow \ell_L W^+ )
\ = \ \frac{g_2^4}{384 \pi m^2 s \beta^2} \left ( 16m^2 L + \beta (11m^2 + 8s) \right ) \ ,
\eeq
\begin{eqnarray}
\sigma ( \ell_{L1} W_1^- \rightarrow \ell_L W^- )
&=& \sigma ( \nu_{\ell1} W_1^+ \rightarrow \nu_\ell W^+ )\ , \\
\sigma ( \ell_{L1} W_1^+ \rightarrow \ell_L W^+ )
&=& \sigma ( \nu_{\ell1} W_1^- \rightarrow \nu_\ell W^- ) \ .
\end{eqnarray}
For KK gluon - KK quark annihilation, we obtain
\beq
\sigma ( g_1 q_1 \rightarrow g q )
\ = \ \frac{g_3^4}{846 \pi m^2 s \beta^2} \left ( 24 m^2 L + \beta (25m^2 + 36s) \right ) \ .
\eeq

\subsection{Quarks}

The annihilation cross-section of two KK quarks into SM quarks of different 
flavor
\beq
\sigma ( q_1 \bar{q}_1 \rightarrow q' \bar{q}' )
\ = \ \frac{g_3^4 (s+2m^2)}{54 \pi s^2 \beta}
\eeq
can be obtained from $\sigma ( \ell_{R1}^{+} {\ell}_{R1}^{-} \rightarrow f \bar{f} )$ 
by multiplying with the following group factor \cite{Peskin:1995ev}
\beq
\frac{1}{3} \frac{1}{3} tr \left ( t^a t^b \right ) tr \left ( t^b t^a \right ) 
= \frac{1}{9} C(r)^2 \delta^{ab} \delta^{ab} = \frac{1}{9} \cdot \frac{1}{4} \cdot 8 = \frac{2}{9} .
\eeq
Here $C(r) = 1/2$ is the quadratic Casimir operator for the fundamental representation of $SU(3)$.
KK quark annihilation into same flavor SM quarks is given by
\beq
\sigma ( q_1 q_1 \rightarrow q q )
\ = \ \frac{g_3^4}{432 \pi m^2 s^2 \beta^2} \left ( 2 m^2 (4s-5m^2)L + (6s-5m^2)s\beta \right ) \ .
\eeq
In this process, there are three terms in the squared matrix element 
since we have both $t$- and $u$-channel diagrams. 
This process also has an analogy with 
$\sigma ( \ell_{R1}^{\pm} \ell_{R1}^{\pm} \rightarrow \ell^\pm \ell^\pm )$. 
However, each term gets a different group factor.  The squares of the $t$- and $u$-channel
diagrams get the same factor of $2/9$ but for the cross term we obtain \cite{Peskin:1995ev}
\begin{eqnarray}
\left ( \frac{1}{3} \right )^2 tr\left ( t^a t^b t^a t^b\right ) 
&=& \frac{1}{9} \left ( C_2 (2) - \frac{1}{2} C_2 (G) \right ) tr \left ( t^a t^a \right ) \nonumber \\
&=& \frac{1}{9} \left ( \frac{4}{3} - \frac{3}{2} \right ) \frac{4}{3} = - \frac{2}{9}
= -\frac{2}{27} \ .
\end{eqnarray}
For $q_1 \bar{q}_1$ annihilation into same flavor SM quarks we get
\beq
\sigma ( q_1 \bar{q}_1 \rightarrow q \bar{q} )
\ = \ \frac{g_3^4}{864 \pi m^2 s^2 \beta^2} \left ( 4m^2 (4s-3m^2)L + \beta (32m^4 + 33s m^2 +12s^2) \right ) \ ,
\eeq
which can also be obtained using the analogy to 
$\sigma ( \ell_{R1}^{+} \ell_{R1}^{-} \rightarrow \ell^+ \ell^- )$ 
and taking into account group factors. For the final state with gluons we have
\beq
\sigma ( q_1 \bar{q}_1 \rightarrow g g )
\ = \ \frac{-g_3^4}{54 \pi s^3 \beta^2} \left ( 4 (m^4+4sm^2+s^2) L + s\beta (31m^2 +7s) \right ) \ .
\eeq
For different quark flavors in the initial state, we have
\beq
\sigma ( q_1 q'_1 \rightarrow q q' )
\ = \ \frac{g_3^4 (4s-3m^2)}{288 \pi m^2 s \beta} \ ,
\eeq
\beq
\sigma ( q_1 \bar{q}'_1 \rightarrow q \bar{q}' )
\ = \ \frac{g_3^4}{288 \pi m^2 s \beta^2} \left ( 8m^2 L + \beta (9m^2 +4s) \right ) \ .
\eeq
The above two cross-sections can be obtained from 
$\sigma ( \ell_{R1}^\pm {\ell'}_{R1}^\pm \rightarrow \ell^\pm {\ell'}^{\pm} )$
and $\sigma ( \ell_{R1}^\pm {\ell'}_{R1}^\mp \rightarrow \ell^\pm {\ell'}^\mp )$,
correspondingly, by multiplying with the group factor $2/9$. 
The remaining cross-sections are
\begin{eqnarray}
\sigma ( q_1 \bar{q}'_1 \rightarrow q \bar{q}' )
&=& \sigma ( u_{R1} \bar{d}_{R1} \rightarrow u \bar{d} ) \nonumber  \ ,\\
&=& \sigma ( u_{R1} d_{L1} \rightarrow u d ) \ , \\
&=& \sigma ( u_{R1} u_{L1} \rightarrow u u  ) \nonumber \ ,
\end{eqnarray}
and
\begin{eqnarray}
\sigma ( q_1 q'_1 \rightarrow q q' )
&=& \sigma ( u_{R1} d_{R1} \rightarrow u d ) \nonumber  \ , \\
&=& \sigma ( u_{R1} \bar{d}_{L1} \rightarrow u \bar{d} )  \ , \\
&=& \sigma ( u_{R1} \bar{u}_{L1} \rightarrow u \bar{u}  ) \ . \nonumber
\end{eqnarray}

\subsection{Quarks and Leptons}

The cross-sections listed below are mediated by $t$- or $u$-channel diagrams with 
KK gauge bosons. For one KK lepton and one KK quark in the initial state, we get
\beq
\sigma ( \ell_{L1} u_{L1} \rightarrow \ell u )
\ = \ \frac{(4g_1^2 Y_{e_L} Y_{u_L} -g_2^2)^2 (4s-3m^2)}{1024 \pi m^2 s \beta} \ ,
\eeq
\beq
\sigma ( \nu_1 u_{L1} \rightarrow \nu u )
\ = \ \frac{(4g_1^2 Y_{e_L} Y_{u_L} +g_2^2)^2 (4s-3m^2)}{1024 \pi m^2 s \beta} \ ,
\eeq
\beq
\sigma ( \ell_{L1} u_{L1} \rightarrow \nu d )
\ = \  \frac{g_2^4 (4s-3m^2)}{256 \pi m^2 s \beta}  \ .
\eeq
These three cross-sections can be obtained from 
$\sigma ( \ell_{R1}^\pm {\ell'}_{R1}^\pm \rightarrow \ell^\pm {\ell'}^{\pm} )$. 
For one KK lepton and one KK anti-quark in the initial state, we have
\beq
\sigma ( \nu_1 \bar{u}_{L1} \rightarrow \ell \bar{d} )
\ = \ \frac{g_2^4}{256 \pi m^2 s \beta^2} \left ( 8m^2 L + \beta (9m^2 + 4s) \right ) \ ,
\eeq
\beq
\sigma ( \ell_{L1} \bar{u}_{L1} \rightarrow \ell \bar{u} )
\ = \ \frac{(4g_1^2 Y_{e_L} Y_{u_L} -g_2^2)^2 }{1024 \pi m^2 s \beta^2} \left ( 8m^2 L + \beta (9m^2 +4s) \right ) \ ,
\eeq
\beq
\sigma ( \nu_1 \bar{u}_{L1} \rightarrow \nu \bar{u} )
\ = \ \frac{(4g_1^2 Y_{e_L} Y_{u_L} +g_2^2)^2 }{1024 \pi m^2 s \beta^2} \left ( 8m^2 L + \beta (9m^2 +4s) \right ) \ ,
\eeq
which can be obtained from
$\sigma ( \ell_{R1}^\pm {\ell'}_{R1}^\mp \rightarrow \ell^\pm {\ell'}^\mp )$.
If one of the particles in the initial state is an $SU(2)_W$-singlet fermion, 
only $\gamma_1$ can mediate the process and
the cross-sections can be obtained from our previous results:
\begin{eqnarray}
\sigma ( \ell_{R1} u_{L1} \rightarrow \ell u ) 
&=& \sigma ( \ell_{L1} u_{R1} \rightarrow \ell u ) 
= \sigma ( \ell_{R1} \bar{u}_{R1}     \rightarrow \ell \bar{u} ) \nonumber  \ , \\
&=& \sigma ( \ell_{R1} \bar{\ell'}_{R1} \rightarrow \ell \bar{\ell'}  ) 
= \sigma ( \ell_{R1} \ell_{L1}  \rightarrow \ell \ell )  \ , 
\end{eqnarray}
\begin{eqnarray}
\sigma ( \ell_{R1} \bar{u}_{L1} \rightarrow \ell \bar{u} ) 
&=& \sigma ( \ell_{L1} \bar{u}_{R1} \rightarrow \ell \bar{u} ) 
= \sigma ( \ell_{R1} u_{R1} \rightarrow \ell u )  \nonumber  \ ,\\
&=& \sigma ( \ell_{R1} \ell'_{R1} \rightarrow \ell \ell' )
= \sigma ( \ell_{R1} \bar{e}_{L1} \rightarrow \ell \bar{\ell} )  \ . 
\end{eqnarray}

\subsection{Higgs Bosons}
\label{sec:hh}

The mass terms for the KK $SU(2)_W$-singlets appear with the 
wrong sign in the fermion Lagrangian. For example, the mass term 
for the KK quarks leads to the following structure for the mass matrix
at tree level
\begin{equation}
\begin{array}{cc}
( \bar{u}_{Ln} (x), \bar{u}_{Rn}(x) ) \\
 &
\end{array} \hspace{-0.3cm}
\left (
\begin{array}{cc}
\frac{n}{R} & m           \\
m           & -\frac{n}{R} \\
\end{array}
\right )
\left (
\begin{array}{cc}
u_{Ln} (x) \\
u_{Rn} (x)
\end{array}
\right ) \ .
\end{equation}
The corresponding mass eigenstates $u'_{Ln}$ and $u'_{Rn}$ have mass
\begin{equation}
M_n = \sqrt{\left ( \frac{n}{R} \right )^2 + m^2} \ .
\end{equation}
These mass eigenstates receive different radiative corrections which lift
the degeneracy~\cite{Cheng:2002iz}.
The interaction eigenstates are related to the mass eigenstates by
\begin{equation}
\left (
\begin{array}{c}
u_{Ln} \\
u_{Rn}
\end{array}
\right ) = \left (
\begin{array}{cc}
\cos\alpha & \gamma_5 \sin\alpha \\
\sin\alpha & -\gamma_5 \cos\alpha
\end{array}
\right )
\left (
\begin{array}{c}
u'_{Ln} \\
u'_{Rn}
\end{array}
\right )\ ,
\label{rotation}
\end{equation}
where $\alpha$ is the mixing angle between $SU(2)_W$-singlet and 
$SU(2)_W$-doublet fermions defined by
\begin{equation}
\tan 2\alpha = \frac{m}{n/R} .
\end{equation}
This mixing is very small except for the top quark. 
However, even with $\alpha\approx0$
the effect of the rotation (\ref{rotation}) is present
in the Yukawa couplings through the redefinition $u_{Rn} \rightarrow -\gamma^5 u_{Rn}$. 
It does not affect the gauge-fermion couplings.
We use the following notation for KK Higgs bosons, 
\begin{equation}
\left(
\begin{array}{c}
G_1^+ \\
\frac{H_1 + i G_1}{\sqrt{2}}
\end{array}
\right) .
\end{equation}
We keep only the top-Yukawa coupling and 
we also keep the Higgs self-coupling assuming $m_h=120$ GeV. 
Below we list the cross-sections associated with two KK Higgs bosons in the initial state.
\begin{eqnarray}
\sigma ( H_1 H_1 \rightarrow G^+ G^- )
\ &=& \  \frac{1}{64 \pi m^2 S \beta^2 s_w^4} \left ( 8e^2 m^2 \lambda^2 s_w^2 L 
      + \beta \{ (2s+m^2)e^4  \right . \nonumber \\
& & \left .  \hspace{2cm} +4 \lambda m^2 s_w^2 e^2 + 4\lambda^2 m^2 s_w^4    \} \right ) \ ,
\end{eqnarray}
\beq
\sigma ( H_1 H_1 \rightarrow H H )
\ = \ \frac{9\lambda^2}{32 \pi s \beta}     \ ,
\eeq
\begin{eqnarray}
\sigma ( H_1 H_1 \rightarrow G G )
\ &=& \ \frac{1}{128 \pi m^2 s \beta^2 s_w^4 c_w^4} \left ( 8 e^2 m^2 \lambda s_w^2 c_w^2 L 
    + \beta \{ (2s+m^2)e^4  \right . \nonumber \\
& & \left . \hspace{2cm} + 4 \lambda m^2 e^2 s_w^2 c_w^2 + 4 \lambda^2 m^2 s_w^4 c_w^4 \} \right ) \ ,
\end{eqnarray}
\beq
\sigma ( H_1 H_1 \rightarrow Z Z )
\ = \ \frac{g_2^4}{64 \pi s^3 \beta^2 c_w^4} \left ( s\beta (s+4m^2) + 4 m^2 (s-2m^2)L \right ) \ , 
\eeq
\beq
\sigma ( H_1 H_1 \rightarrow W^+ W^- )
\ = \ \frac{g_2^4}{32 \pi s^3 \beta^2 c_w^4} \left ( s\beta (s+4m^2) + 4 m^2 (s-2m^2)L \right ) \ ,
\eeq
\beq
\sigma ( H_1 H_1 \rightarrow t \bar{t} )
\ = \ \frac{3 y_t^4}{16 \pi s^2 \beta^2} \left ( -(s+2m^2)L - 2 s \beta \right ) \ ,
\eeq
\begin{eqnarray}
\sigma ( G_1^+ G_1^+ \rightarrow G^+ G^+ ) 
&=& \frac{1}{128 \pi m^2 s \beta^2 s_w^4 c_w^4} \left ( -16 e^2 m^2 \lambda s_w^2 c_w^2 L \right . \nonumber \\
&+& \left .     \beta \{ (2s+m^2)e^4 -8 \lambda m^2 e^2 s_w^2 c_w^2 + 16 \lambda^2 m^2 s_w^4 c_w^4 \} \right ) \ ,
\end{eqnarray}
\begin{eqnarray}
\sigma ( G_1^+ G_1^- \rightarrow t \bar{t} )
&=& \frac{1}{1152 \pi s^2 \beta^2 s_w^4 c_w^4}
 \left ( 72 s_w^4 c_w^2 y_t^2 (-3 s c_w^2 y_t^2 - 4 m^2 e^2) L - 432 s \beta s_w^4 c_w^4 y_t^4  \right . \nonumber \\
& &\hspace{0.5cm} + \left . s \beta^3 (20 s_w^4 - 12 s_w^2 +9)e^4  - 144 s \beta s_w^4 c_w^2 y_t^2 e^2 \right ) \ ,
\end{eqnarray}
\begin{eqnarray}
\sigma ( G_1^+ G_1^- \rightarrow b \bar{b} )
&=& \frac{1}{1152 \pi s^2 \beta^2 s_w^4 c_w^4} \left ( 72 s_w^2 c_w^2 y_t^2 
(-3s s_w^2 c_w^2 y_t^2 + e^2 m^2 (4s_w^2 -3))L  \right . \nonumber \\
& &\hspace{1cm} -432 s \beta s_w^4 c_w^4 y_t^4  +  s \beta^3 (20s_w^4-24s_w^2+9) e^4 \\
& & \left . \hspace{1cm} -36 s \beta s_w^2 y_t^2 e^2 (4s_w^4-7s_w^2+3) \right )  \ , \nonumber
\end{eqnarray}
\begin{eqnarray}
\sigma ( G_1^+ G_1^- \rightarrow G^+ G^- )
&=& \frac{-1}{192 \pi m^2 s^2 \beta^2 s_w^4 c_w^4} \left ( 6 m^2 e^2 s (e^2 + 4 \lambda s_w^2 c_w^2) L 
                   - 48 \lambda^2 m^2 s s_w^4 c_w^4 \right . \nonumber \\ 
& & \hspace{0.5cm}+ \left . \beta \{ e^4 (m^4-7sm^2-3s^2) -12 \lambda m^2 s s_w^2 c_w^2 e^2  \} \right )  \ ,
\end{eqnarray}
\begin{eqnarray}
\sigma ( G_1^+ G_1^- \rightarrow G H )
&=& \frac{1}{128 \pi m^2 s \beta^2 s_w^4} \left ( 8s_w^2 m^2 e^2 \lambda L + \beta \{ (2s+m^2)e^4 \right . \nonumber \\
&& \hspace{1cm} \left . + 4 \lambda s_w^2 m^2 e^2 + 4\lambda^2 m^2 s_w^4 \} \right ) \ ,
\end{eqnarray}
\begin{eqnarray}
\sigma ( G_1^+ G_1^- \rightarrow G H )
&=& \frac{g_Z^4}{48 \pi m^2 s^2 \beta^2} \left ( 24 m^2 c_w^2 s L 
-  \beta \{ 4 (1-2s_w^2)^2 m^4  \right . \nonumber\\
&+& \left . s (92 s_w^4 - 140 s_w^2 + 47) -24 s^2 c_w^4     \} \right )   \ ,
\end{eqnarray}
\beq
\sigma ( G_1^+ G_1^- \rightarrow Z Z )
\ = \ \frac{g_Z^4 (1-2s_w^2)^4}{4 \pi s^3 \beta^2} \left ( 4 m^2 (s-2m^2) L + s \beta (s+4 m^2) \right ) \ , 
\eeq
\beq
\sigma ( G_1^+ G_1^- \rightarrow \gamma Z )
\ = \  \frac{e^2 g_Z^2 (1-2s_w^2)^2}{2 \pi s^3 \beta^2} \left ( 4 m^2 (s-2m^2) L + s \beta (s+4 m^2) \right ) \ ,
\eeq
\beq
\sigma ( G_1^+ G_1^- \rightarrow \gamma \gamma )
\ = \ \frac{e^4}{4 \pi s^3 \beta^2} \left ( 4 m^2 (s-2m^2) L + s \beta (s+4 m^2) \right ) \ ,
\eeq
\beq
\sigma ( G_1^+ G_1^- \rightarrow W^+ W^- )
\ = \ \frac{g_2^4}{24 \pi s^2 \beta^2} \left ( 6 m^2 L + \beta (s + 11 m^2) \right ) \ ,
\eeq
\beq
\sigma ( G_1^+ G_1^- \rightarrow f \bar{f} )
\ = \ \frac{N_c g_Z^4 \beta}{24 \pi s} \left ( \hat{g}_L^2 + \hat{g}_R^2 \right )         \ ,
\eeq
where $f$ represents leptons and first two generations of quarks and 
$$\hat{g}_{L(R)}^2 = -e^2 Q_f - 2 g_Z^2 (1-2 s_w^2) (T^3_f - Q_f s_w^2).$$
A number of cross-sections can be simply related: 
\begin{eqnarray}
\sigma ( G_1 G_1 \rightarrow G^+ G^- )   &=& \sigma ( H_1 H_1 \rightarrow G^+ G^- ) \nonumber \ , \\
\sigma ( G_1 G_1 \rightarrow G G )       &=& \sigma ( H_1 H_1 \rightarrow H H ) \nonumber \ , \\
\sigma ( G_1 G_1 \rightarrow H H )       &=& \sigma ( H_1 H_1 \rightarrow G G ) \nonumber \ , \\ 
\sigma ( G_1 G_1 \rightarrow Z Z )       &=& \sigma ( H_1 H_1 \rightarrow Z Z ) \nonumber  \ , \\
\sigma ( G_1 G_1 \rightarrow W^+ W^- )   &=& \sigma ( H_1 H_1 \rightarrow W^+ W^- ) \ , \\
\sigma ( G_1 G_1 \rightarrow t \bar{t} ) &=& \sigma ( H_1 H_1 \rightarrow t \bar{t} ) \nonumber \ , \\
\sigma ( G_1^+ G_1^- \rightarrow G G )   &=& \sigma ( G_1^+ G_1^- \rightarrow H H ) \nonumber \ , \\
\sigma ( G_1^+ G_1^- \rightarrow H H )   &=& \sigma ( G_1^+ G_1^- \rightarrow G G ) \nonumber \ , \\
                                         &=&\frac{1}{2}\sigma ( H_1 H_1 \rightarrow G^+ G^- ) \nonumber \ .
\end{eqnarray}
The rest of the cross-sections are
\begin{eqnarray}
\sigma ( H_1 G_1 \rightarrow H G ) 
&=& \frac{1}{192 \pi m^2 s^2 \beta^2 s_w^4 c_w^4} \left ( 6 m^2 e^2 s (e^2 - 2\lambda s_w^2 c_w^2) L \right . \nonumber \\
& & - \beta \{ (m^4-7sm^2-3s^2)e^4  \\
& & \left . + 6 \lambda m^2 s s_w^2 c_w^2 e^4 - 12 \lambda^2 m^2 s s_w^4 c_w^4 \}  \right )  \nonumber \ ,
\end{eqnarray}
\begin{eqnarray}
\sigma ( H_1 G_1 \rightarrow G^+ G^- ) 
&=& \frac{g_Z^4}{48 \pi m^2 s^2 \beta^2} \left ( 24 m^2 c_w^2 s L + \beta \{ -4(1-2s_w^2 )^2 m^4 \right . \nonumber \\
& & \left . + s m^2 (-92 s_w^4 + 140 s_w^2 -47) + 24 s^2 c_w^4 \} \right ) \ ,
\end{eqnarray}
\beq
\sigma ( H_1 G_1 \rightarrow W^+ W^- ) 
\ = \ \frac{g_2^4}{96 \pi s^3 \beta^2} \left ( 12m^2 (2m^2 +s)L + s\beta (32m^2 +s) \right ) \ ,
\eeq
\begin{eqnarray}
\sigma ( H_1 G_1 \rightarrow t \bar{t} ) 
&=& \frac{1}{288 \pi s^2 \beta^2 c_w^2 s_w^2} \left ( -54 y_t^2 (e^2 m^2 + c_w^2 s_w^2 (s-2m^2) y_t^2) L \right . \nonumber \\
&+& \left . \beta \{ -54 s y_t^4 c_w^2 s_w^2 -27 e^2 y_t^2 s + e^2 g_Z^2 s \beta^2 (32 s_w^4 -24 s_w^2 +9) \} \right ) \ ,
\end{eqnarray}
\beq
\sigma ( H_1 G_1 \rightarrow f \bar{f} ) 
\ = \ \frac{N_c \beta g_Z^2}{24 \pi s} \left ( \bar{g}_L^2 + \bar{g}_R^2 \right )   \ ,
\eeq
\begin{eqnarray}
\sigma ( H_1 G_1^+ \rightarrow H G^+ ) 
&=& \frac{1}{192 \pi m^2 s^2 \beta^2 s_w^4} \left ( 6 m^2 e^2 s (e^2 - 2\lambda s_w^2 ) L  - 12 \lambda^2 m^2 s s_w^4 
 \right . \nonumber \\
&-& \left . \beta \{ (m^4-7sm^2-3s^2)e^4 + 6 \lambda m^2 s s_w^2 e^4 \}  \right )   \ ,
\end{eqnarray}
\begin{eqnarray}
\sigma ( H_1 G_1^+ \rightarrow G G^+ ) 
&=& \frac{g_Z^4}{48 \pi m^2 s^2 \beta^2} \left ( 12m^2(1-2s_w^2 ) (2-3_w^2)L \right . \nonumber \\
& & \hspace{1cm} - \beta  \{ (4m^4 +47s m^2 -24 s^2)c_w^4 \\
& & \hspace{1cm} \left . - 24 (m^2 -s)s s_w^2 c_w^2 - 3 s (m^2+4s)s_w^4  \} \right )   \nonumber \ ,
\end{eqnarray}
\begin{eqnarray}
\sigma ( H_1 G_1^+ \rightarrow Z W^+ ) 
&=& \frac{g_Z^4}{6 \pi s^3 \beta^2} \left ( 6m^2 L \{ (1-2s_w^2)(4m^2 -2s s_w^2 + s) + s \} \right . \nonumber \\
&+& \left . s \beta \{ 4 s_w^4 (s+11m^2) + (1-2s_w^2)(s+32m^2) \} \right )    \ ,
\end{eqnarray}
\beq
\sigma ( H_1 G_1^+ \rightarrow \gamma W^+ ) 
\ = \ \frac{e^2 g_2^2}{24 \pi s^2 \beta^2} \left ( 6m^2 L + \beta (11m^2 + s) \right )    \ ,
\eeq
\begin{eqnarray}
\sigma ( H_1 G_1^+ \rightarrow t \bar{b} ) 
&=& \frac{1}{64 \pi s^2 \beta^2 s_w^4} \left ( -6 s_w^2 (2 e^2 m^2 + s s_w^2 y_t^2) y_t^2 L \right . \nonumber \\
&+& \left . \beta \{ s \beta^2 e^4 - 6s s_w^4 y_t^2 e^2 - 12 s s_w^4 y_t^4 \}\right ) \ ,
\end{eqnarray}
\beq
\sigma ( H_1 G_1^+ \rightarrow f \bar{f}' ) 
\ = \ \frac{N_c g_2^4 \beta}{192 \pi s} \ ,
\eeq
\begin{eqnarray}
\sigma ( G_1 G_1^+ \rightarrow G G^+ )       &=& \sigma ( H_1 G_1^+ \rightarrow H G^+ ) \nonumber \ ,\\
\sigma ( G_1 G_1^+ \rightarrow H G^+ )       &=& \sigma ( H_1 G_1^+ \rightarrow G G^+ ) \nonumber  \ , \\
\sigma ( G_1 G_1^+ \rightarrow Z W^+ )       &=& \sigma ( H_1 G_1^+ \rightarrow Z W^+ ) \ , \\
\sigma ( G_1 G_1^+ \rightarrow \gamma W^+ )  &=& \sigma ( H_1 G_1^+ \rightarrow \gamma W^+ ) \nonumber \ , \\ 
\sigma ( G_1 G_1^+ \rightarrow f \bar{f}' )  &=& \sigma ( H_1 G_1^+ \rightarrow f \bar{f}' ) \nonumber \ .
\end{eqnarray}

\subsection{Higgs Bosons and Gauge Bosons}

The cross-sections involving one KK Higgs boson and one KK gauge boson are given below. 
They are rather simple compared to the cross-sections from Section~\ref{sec:hh}.
\beq
\sigma ( H_1 g_1 \rightarrow t \bar{t} )
\ = \ \frac{g_3^2 y_t^2}{48 \pi m^2 s^2 \beta^2} \left ( 2 m^2 (s-m^2) L + (2s-5m^2)s\beta \right )  \ ,
\eeq
\beq
\sigma ( H_1 Z_1 \rightarrow Z H )
\ = \ \frac{g_2^4}{96 \pi c_w^2 s\beta^2} \left ( L +4\beta \right )  \ ,
\eeq
\beq
\sigma ( H_1 Z_1 \rightarrow W^- G^+ )
\ = \ \frac{g_2^4}{96 \pi m^2 s^2 \beta^2} \left ( 4m^2 L +s\beta(4s+m^2) \right ) \ ,
\eeq
\beq
\sigma ( H_1 Z_1 \rightarrow t \bar{t} )
\ = \ \frac{g_2^2 y_t^2}{64 \pi m^2 s\beta^2} \left ( 2m^2 L + (4s-11m^2)\beta \right ) \ ,
\eeq
\beq
\sigma ( G_1^+ Z_1 \rightarrow Z G^+ )
\ = \ \frac{g_2^4 (1-2s_w^2)^2}{96 \pi c_w^2 s \beta^2} \left ( L +4\beta \right )  \ ,
\eeq
\beq
\sigma ( G_1^+ Z_1 \rightarrow \gamma G^+ )
\ = \ \frac{e^2 g_2^2}{24\pi s\beta^2} \left ( L + 4\beta \right )  \ ,
\eeq
\beq
\sigma ( G_1^+ Z_1 \rightarrow t \bar{b} )
\ = \ \frac{g_2^2 y_t^2}{64 \pi m^2 s\beta^2} \left ( 2m^2 L + (4s-11m^2)\beta \right ) \ ,
\eeq
\beq
\sigma ( H_1 \gamma_1 \rightarrow Z H )
\ = \ \frac{g_1^2 g_2^2}{96 \pi c_w^2 s \beta^2} \left ( L + 4\beta \right ) \ ,
\eeq
\beq
\sigma ( G_1 \gamma_1 \rightarrow W^- G^+ )
\ = \ \frac{g_1^2 g_2^2}{96 \pi s\beta^2} \left ( L + 4\beta \right ) \ ,
\eeq
\beq
\sigma ( H_1 \gamma_1 \rightarrow t \bar{t} )
\ = \ \frac{g_1^2 y_t^2}{576 \pi m^2 c_w^2 s^2 \beta^2} \left ( -2m^2 (7s+8m^2)L +(4s-43m^2)s\beta \right ) \ ,
\eeq
\beq
\sigma ( G_1^+ \gamma_1 \rightarrow Z G^+ )
\ = \ \frac{g_1^2 g_2^2 (1-2s_w^2)^2}{96 \pi c_w^2 s \beta^2} \left ( L + 4\beta \right )  \ ,
\eeq
\beq
\sigma ( G_1^+ \gamma_1 \rightarrow \gamma G^+ )
\ = \ \frac{e^2 g_1^2}{24 \pi s \beta^2} \left ( L + 4\beta \right ) \ ,
\eeq
\beq
\sigma ( G_1^+ W_1^+ \rightarrow G^+ W^+ )
\ = \ \frac{g_2^4}{96\pi m^2 s \beta^2} \left ( 12m^2 L + \beta (6m^2+5s)  \right ) \ ,
\eeq
\begin{eqnarray}
\sigma ( H_1 W_1^+ \rightarrow G^+ Z )
&=& \frac{g_2^4}{96 \pi m^2 s\beta^2 c_w^2} \left ( m^2 (2-s_w^2 -2s_w^4) L \right . \nonumber \\
& & \left . \hspace{0.5cm}  -\beta \{ m^2 (4s_w^4-s_w^2+1)+(3s_w^4-7s_w^2+4)  \} \right ) \ ,
\end{eqnarray}
\beq
\sigma ( H_1 W_1^+ \rightarrow G^+ \gamma )
\ = \ \frac{g_1^2 e^2}{96 \pi m^2 s\beta^2} \left ( -2m^2 L +\beta ( 4m^2 +3s)  \right ) \ ,
\eeq
\beq
\sigma ( H_1 W_1^+ \rightarrow H W^+ )
\ = \ \frac{g_2^4}{96 \pi s\beta^2} \left ( L +4\beta   \right ) \ ,
\eeq
\beq
\sigma ( G_1^- W_1^+ \rightarrow W^+ G^- )
\ = \ \frac{g_2^4}{96 \pi m^2 s \beta^2} \left ( -2m^2L +\beta (4m^2+3s)   \right ) \ , 
\eeq
\begin{eqnarray}
\sigma ( G_1^- W_1^+ \rightarrow Z G )
&=& \frac{g_2^4}{96 \pi m^2 s\beta^2 c_w^2} \left ( m^2 (12s_w^4 -15s_w^2 +4)L \right . \nonumber \\
& & \left . \hspace{0.5cm} +\beta \{ (6 s_w^4 -3s_w^2 +1)m^2 +s(5s_w^4-9s_w^2+4)  \}   \right )  \ ,
\end{eqnarray}
\beq
\sigma ( G_1^- W_1^+ \rightarrow \gamma G )
\ = \ \frac{g_2^2 e^2}{96 \pi m^2s\beta^2} \left ( 12m^2 L + \beta (6m^2 +5s)   \right ) \ ,
\eeq
\beq
\sigma ( G_1^- W_1^+ \rightarrow t \bar{t} )
\ = \ \frac{g_2^2}{64 \pi m^2 s \beta^2} \left ( -4m^2 y_t^2 L 
        + \beta \{ e^2 m^2 + 2 (4s-11m^2) y_t^2    \}  \right ) \ ,
\eeq
where $y_t$ is the top quark Yukawa coupling. 
The cross-sections listed below are obtained from our previous calculations. 
For one KK Higgs boson and one KK gluon we get
\begin{eqnarray}
\sigma ( H_1 g_1 \rightarrow t \bar{t} ) &=& \sigma ( G_1 g_1 \rightarrow t \bar{t} ) \nonumber  \ , \\
                                         &=& \sigma ( G_1^+ g_1 \rightarrow t \bar{b} )  \ .
\end{eqnarray}
For one KK Higgs boson and one $Z_1$ we have
\begin{eqnarray}
\sigma ( H_1 Z_1 \rightarrow Z H )       &=& \sigma ( G_1 Z_1 \rightarrow Z H ) \nonumber \ , \\
\sigma ( H_1 Z_1 \rightarrow W^- G^+ )   &=& \sigma ( G_1 Z_1 \rightarrow W^- G^+ ) \nonumber \ , \\
                                         &=& \sigma ( G_1^+ Z_1 \rightarrow W^+ G )\nonumber  \ , \\
                                         &=& \sigma ( G_1^+ Z_1 \rightarrow W^+ H )\nonumber \ , \\
                                         &=& \sigma ( W_1^+ H_1 \rightarrow W^+ G ) \ , \\
                                         &=& \sigma ( W_1^+ H_1 \rightarrow W^+ H ) \nonumber \ , \\
\sigma ( H_1 Z_1 \rightarrow t \bar{t} ) &=& \sigma ( G_1 Z_1 \rightarrow t \bar{t} ) \nonumber \ , \\
\sigma ( G_1^+ Z_1 \rightarrow t \bar{b} ) &=& \sigma ( H_1 W_1^+ \rightarrow t \bar{b} ) \nonumber \ , \\
                                           &=& \sigma ( G_1 W_1^+ \rightarrow t \bar{b} ) \nonumber \ .
\end{eqnarray}
For one KK Higgs boson and one $\gamma_1$ we obtain
\begin{eqnarray}
\sigma ( H_1 \gamma_1 \rightarrow Z H )         &=& \sigma ( G_1 \gamma_1 \rightarrow Z G )\nonumber \ , \\
\sigma ( G_1 \gamma_1 \rightarrow W^- G^+ )     &=& \sigma ( H_1 \gamma_1 \rightarrow W^- G^+ )\nonumber  \ ,\\
                                           &=& \sigma ( G_1^+ \gamma_1 \rightarrow W^+ G )\nonumber \ , \\
                                           &=& \sigma ( G_1^+ \gamma_1 \rightarrow W^+ H )   \ , \\
\sigma ( H_1 \gamma_1 \rightarrow t \bar{t} )   &=& \sigma ( G_1 \gamma_1 \rightarrow t \bar{t} ) \nonumber \ , \\
                                           &=& \sigma ( G_1^+ \gamma_1 \rightarrow t \bar{b} ) \nonumber \ .
\end{eqnarray}
For one KK Higgs boson and one $W^\pm_1$, we get
\begin{eqnarray}
\sigma ( H_1 W_1^+ \rightarrow G^+ Z )     &=& \sigma ( G_1 W_1^+ \rightarrow G^+ Z ) \nonumber \ , \\
\sigma ( H_1 W_1^+ \rightarrow G^+\gamma ) &=& \sigma ( G_1 W_1^+ \rightarrow G^+ \gamma ) \nonumber \ , \\
\sigma ( H_1 W_1^+ \rightarrow H W^+ ) &=& \sigma ( G_1 W_1^+ \rightarrow G W^+ ) \ , \\
\sigma ( G_1^- W_1^+ \rightarrow Z G ) &=& \sigma ( G_1^- W_1^+ \rightarrow Z H ) \nonumber \ , \\
\sigma ( G_1^- W_1^+ \rightarrow \gamma G ) &=& \sigma ( G_1^- W_1^+ \rightarrow \gamma H ) \nonumber \ .
\end{eqnarray}

\subsection{Higgs Bosons and Fermions}

For the cross-sections between one KK Higgs boson and one KK $SU(2)_W$-singlet fermion, we have
\beq
\sigma ( H_1 f_{R1} \rightarrow f G)
\ = \ \frac{g_1^4 Y_f^2}{32 \pi m^2 s \beta^2} \left ( m^2 L + s\beta \right )  \ ,
\eeq
\beq
\sigma ( H_1 t_{R1} \rightarrow g t)
\ = \ - \frac{g_3^2 y_t^2}{48 \pi s \beta^2} \left ( 2 L + 3\beta \right ) \ ,
\eeq
\begin{eqnarray}
\sigma ( G_1^+ t_{R1} \rightarrow t G^+)
&=& \frac{1}{288 \pi m^2 s^2 \beta^2 c_w^2} 
     \left ( 12 c_w^2 e^2 m^2 y_t^2 + m^2 L \{ + 12c_w^2 (m^2-s) y_t^2 e^2 \right . \nonumber \\
&+& \left . 9c_w^4 (m^2-s)y_t^2 +  4se^4  \} + s\beta \{ 4se^4 - 9c_w^4 m^2 y_t^4 \} \right )  \ ,
\end{eqnarray}
\begin{eqnarray}
\sigma ( H_1 f_{R1} \rightarrow f G) &=& \sigma ( G_1 f_{R1} \rightarrow f H) \nonumber \ , \\
                                     &=& \sigma ( G_1^+ f_{R1} \rightarrow f G^+ ) \nonumber \ , \\ 
                                     &=& \sigma ( G_1^- f_{R1} \rightarrow f G^- ) \nonumber \ , \\ 
\sigma ( H_1 t_{R1} \rightarrow g t) &=& \sigma ( H_1 t_{L1} \rightarrow g t) \ , \\
                                     &=& \sigma ( G_1 t_{R1} \rightarrow g t) \nonumber \ , \\
                                     &=& \sigma ( G_1^- t_{R1} \rightarrow g b) \nonumber \ , \\
                                     &=& \frac{1}{2} \sigma ( G_1^+ b_{L1} \rightarrow g t) \nonumber  \ .
\end{eqnarray}
For the cross-sections between one KK Higgs boson and one KK $SU(2)_W$-doublet fermion, we get
\beq
\sigma ( H_1 f_{L1} \rightarrow G f )
\ = \ \frac{e^2 (T^3_f c_w g_2 - 2g_1 s_w Y_f )^2}{128 \pi m^2 s_w^2 c_w^2 s \beta^2} \left ( m^2 L + s\beta \right ) \ ,
\eeq
\beq
\sigma ( H_1 f^+_1 \rightarrow f^- G^+ )
\ = \ \frac{g_2^4}{64 \pi m^2 s \beta^2} \left ( m^2 L + s\beta \right ) \ ,
\eeq
\beq
\sigma ( G_1^+ t_{L1} \rightarrow t G^+ )
\ =\ \frac{e^2 ( c_w g_2 - 2g_1 s_w Y_f )^2}{128 \pi m^2 s_w^2 c_w^2 s \beta^2} \left ( m^2 L + s\beta \right ) 
   +  \frac{y_t^4}{32\pi s^2\beta^2} \left ( m^2 L + s \beta \right ) \ ,
\eeq
\beq
\sigma ( G_1^+ t_{L1} \rightarrow t W^+ )
\ = \ -\frac{g_2^2 y_t^2 L}{32 \pi s \beta^2} \ ,
\eeq
\begin{eqnarray}
\sigma ( G_1^- b_{L1} \rightarrow b G^- )
&=& \frac{e^2 ( c_w g_2 - 2g_1 s_w Y_f )^2}{128 \pi m^2 s_w^2 c_w^2 s \beta^2} \left ( m^2 L + s\beta \right ) \nonumber\\
&+& \frac{1}{32\pi c_w^2 s_w^2 s^2 \beta^2} \left ( y_t^2 L \{ s e^2 (c_w^2-2s_w^2 Y_b) \right . \\
& & \left . \hspace{1cm} + c_w^2 s_w^2 (s-m^2) y_t^2 - s_w^2 c_w^2 s\beta^2 y_t^2 \} \right ) \nonumber \ ,
\end{eqnarray}
where $T^3_f$ denotes the fermion isospin.
\begin{eqnarray}
\sigma ( H_1 b_{L1} \rightarrow G f )    &=& \frac{1}{2}\sigma ( G_1^+ t_{L1} \rightarrow t W^+ ) \nonumber  \ , \\
\sigma ( H_1 f_{L1} \rightarrow G f )    &=& \sigma ( G_1 f_{L1} \rightarrow H f ) \nonumber  \ , \\
                                         &=& \sigma ( G_1^{\pm} f_{L1} \rightarrow G^{\pm} f ) \nonumber  \ , \\
\sigma ( H_1 f^+_1 \rightarrow f^- G^+ ) &=& \sigma ( H_1 f^-_1 \rightarrow f^+ G^- ) \nonumber  \ , \\
                                         &=& \sigma ( G_1 f^+_1 \rightarrow f^- G^+ ) \ , \\
                                         &=& \sigma ( G_1 f^-_1 \rightarrow f^+ G^- ) \nonumber  \ , \\
                                         &=& \sigma ( G_1^+ f^-_1 \rightarrow f^+ G ) \nonumber  \ , \\
                                         &=& \sigma ( G_1^- f^+_1 \rightarrow f^- G ) \nonumber \ ,
\end{eqnarray}
where $f$ stands for any lepton or quark, except $t_{L1}$ and $b_{L1}$,
and $f^+$ ($f^-$) denotes isospin $+1/2$ (isospin $-1/2$) fermions.

%%%%%%%%%%%%%%%%%%%%%%%%%%%%%%%%%%%%%%%%%%%%%%%%%%%%%%%%%%%%%%%
%\listoftables           % ONLY DRAFT
%\listoffigures          % ONLY DRAFT

%%%%%%%%%%%%%%%%%%%%%%%%%%%%%%%%%%%%%%%%%%%%%%%%%%%%%%%%%%%%%%%

\end{document}